\documentclass[prc,aps,superscriptaddress,twocolumn,amsmath,amssymb]{revtex4}
\usepackage[pdftex]{graphicx} 
\usepackage{epsfig}
\usepackage{subfigure}
\usepackage{multirow} %
\usepackage{rotating}
\newcommand\captionof[1]{\def\@captype{#1}\caption}
\newcommand{\myreac}{\gamma p \rightarrow \phi p}
\newcommand{\cmangle}{\cos \theta^\phi_{\mbox{\scriptsize c.m.}}}
\newcommand{\phith}{\theta^\phi_{\mbox{\scriptsize c.m.}}}
\newcommand{\Pom}{I$\!$P}
\newcommand{\sqrts}{\sqrt{s}}
\newcommand{\kkb}{K\overline{K}}
\newcommand{\kp}{K^+}
\newcommand{\km}{K^-}
\newcommand{\ks}{K^0_S}
\newcommand{\kl}{K^0_L}
\newcommand{\pip}{\pi^+}
\newcommand{\pim}{\pi^-}
\newcommand{\piz}{\pi^0}
\usepackage{relsize}
\def\babar{\mbox{\slshape B\kern-0.1em{\smaller A}\kern-0.1em
    B\kern-0.1em{\smaller A\kern-0.2em R}}}
\begin{document}
\newcommand*{\CMU}{Carnegie Mellon University, Pittsburgh, Pennsylvania 15213, USA}
\newcommand*{\CMUindex}{1}
\affiliation{\CMU}
\newcommand*{\WJ}{Washington \& Jefferson College, Washington, Pennsylvania 15301, USA}
\newcommand*{\WJindex}{2}
\affiliation{\WJ}
\newcommand*{\ANL}{Argonne National Laboratory, Argonne, Illinois 60441, USA}
\newcommand*{\ANLindex}{3}
\affiliation{\ANL}
\newcommand*{\ASU}{Arizona State University, Tempe, Arizona 85287, USA}
\newcommand*{\ASUindex}{4}
\affiliation{\ASU}
\newcommand*{\CSUDH}{California State University, Dominguez Hills, Carson, California 90747, USA}
\newcommand*{\CSUDHindex}{5}
\affiliation{\CSUDH}
\newcommand*{\CANISIUS}{Canisius College, Buffalo, New York 14208, USA}
\newcommand*{\CANISIUSindex}{6}
\affiliation{\CANISIUS}
\newcommand*{\CUA}{Catholic University of America, Washington, D.C. 20064, USA}
\newcommand*{\CUAindex}{7}
\affiliation{\CUA}
\newcommand*{\SACLAY}{CEA, Centre de Saclay, Irfu/Service de Physique Nucl\'aeaire, 91191 Gif-sur-Yvette, France}
\newcommand*{\SACLAYindex}{8}
\affiliation{\SACLAY}
\newcommand*{\CNU}{Christopher Newport University, Newport News, Virginia 23606, USA}
\newcommand*{\CNUindex}{9}
\affiliation{\CNU}
\newcommand*{\UCONN}{University of Connecticut, Storrs, Connecticut 06269, USA}
\newcommand*{\UCONNindex}{10}
\affiliation{\UCONN}
\newcommand*{\EDINBURGH}{Edinburgh University, Edinburgh EH9 3JZ, United Kingdom}
\newcommand*{\EDINBURGHindex}{11}
\affiliation{\EDINBURGH}
\newcommand*{\FU}{Fairfield University, Fairfield Connecticut 06824, USA}
\newcommand*{\FUindex}{12}
\affiliation{\FU}
\newcommand*{\FIU}{Florida International University, Miami, Florida 33199, USA}
\newcommand*{\FIUindex}{13}
\affiliation{\FIU}
\newcommand*{\FSU}{Florida State University, Tallahassee, Florida 32306, USA}
\newcommand*{\FSUindex}{14}
\affiliation{\FSU}
\newcommand*{\Genova}{Universit$\grave{a}$ di Genova, 16146 Genova, Italy}
\newcommand*{\Genovaindex}{101}
\affiliation{\Genova}
\newcommand*{\GWU}{The George Washington University, Washington, DC 20052, USA}
\newcommand*{\GWUindex}{15}
\affiliation{\GWU}
\newcommand*{\ISU}{Idaho State University, Pocatello, Idaho 83209, USA}
\newcommand*{\ISUindex}{16}
\affiliation{\ISU}
\newcommand*{\INFNFE}{INFN, Sezione di Ferrara, 44100 Ferrara, Italy}
\newcommand*{\INFNFEindex}{97}
\affiliation{\INFNFE}
\newcommand*{\INFNFR}{INFN, Laboratori Nazionali di Frascati, 00044 Frascati, Italy}
\newcommand*{\INFNFRindex}{17}
\affiliation{\INFNFR}
\newcommand*{\INFNGE}{INFN, Sezione di Genova, 16146 Genova, Italy}
\newcommand*{\INFNGEindex}{18}
\affiliation{\INFNGE}
\newcommand*{\INFNRO}{INFN, Sezione di Roma Tor Vergata, 00133 Rome, Italy}
\newcommand*{\INFNROindex}{19}
\affiliation{\INFNRO}
\newcommand*{\ORSAY}{Institut de Physique Nucl\'eaire ORSAY, Orsay, France}
\newcommand*{\ORSAYindex}{20}
\affiliation{\ORSAY}
\newcommand*{\ITEP}{Institute of Theoretical and Experimental Physics, Moscow, 117259, Russia}
\newcommand*{\ITEPindex}{21}
\affiliation{\ITEP}
\newcommand*{\JMU}{James Madison University, Harrisonburg, Virginia 22807, USA}
\newcommand*{\JMUindex}{22}
\affiliation{\JMU}
\newcommand*{\KNU}{Kyungpook National University, Daegu 702-701, Republic of Korea}
\newcommand*{\KNUindex}{23}
\affiliation{\KNU}
\newcommand*{\LPSC}{LPSC, Universite Joseph Fourier, CNRS/IN2P3, INPG, Grenoble, France}
\newcommand*{\LPSCindex}{24}
\affiliation{\LPSC}
\newcommand*{\UNH}{University of New Hampshire, Durham, New Hampshire 03824, USA}
\newcommand*{\UNHindex}{25}
\affiliation{\UNH}
\newcommand*{\NSU}{Norfolk State University, Norfolk, Virginia 23504, USA}
\newcommand*{\NSUindex}{26}
\affiliation{\NSU}
\newcommand*{\OHIOU}{Ohio University, Athens, Ohio  45701, USA}
\newcommand*{\OHIOUindex}{27}
\affiliation{\OHIOU}
\newcommand*{\ODU}{Old Dominion University, Norfolk, Virginia 23529, USA}
\newcommand*{\ODUindex}{28}
\affiliation{\ODU}
\newcommand*{\RPI}{Rensselaer Polytechnic Institute, Troy, New York 12180, USA}
\newcommand*{\RPIindex}{29}
\affiliation{\RPI}
\newcommand*{\URICH}{University of Richmond, Richmond, Virginia 23173, USA}
\newcommand*{\URICHindex}{30}
\affiliation{\URICH}
\newcommand*{\ROMAII}{Universita' di Roma Tor Vergata, 00133 Rome Italy}
\newcommand*{\ROMAIIindex}{31}
\affiliation{\ROMAII}
\newcommand*{\MSU}{Skobeltsyn Nuclear Physics Institute, Skobeltsyn Nuclear Physics Institute, 119899 Moscow, Russia}
\newcommand*{\MSUindex}{32}
\affiliation{\MSU}
\newcommand*{\SCAROLINA}{University of South Carolina, Columbia, South Carolina 29208, USA}
\newcommand*{\SCAROLINAindex}{33}
\affiliation{\SCAROLINA}
\newcommand*{\JLAB}{Thomas Jefferson National Accelerator Facility, Newport News, Virginia 23606, USA}
\newcommand*{\JLABindex}{34}
\affiliation{\JLAB}
\newcommand*{\UNIONC}{Union College, Schenectady, New York 12308, USA}
\newcommand*{\UNIONCindex}{35}
\affiliation{\UNIONC}
\newcommand*{\UTFSM}{Universidad T\'{e}cnica Federico Santa Mar\'{i}a, Casilla 110-V Valpara\'{i}so, Chile}
\newcommand*{\UTFSMindex}{36}
\affiliation{\UTFSM}
\newcommand*{\GLASGOW}{University of Glasgow, Glasgow G12 8QQ, United Kingdom}
\newcommand*{\GLASGOWindex}{37}
\affiliation{\GLASGOW}
\newcommand*{\WM}{College of William and Mary, Williamsburg, Virginia 23187, USA}
\newcommand*{\WMindex}{38}
\affiliation{\WM}
\newcommand*{\YEREVAN}{Yerevan Physics Institute, 375036 Yerevan, Armenia}
\newcommand*{\YEREVANindex}{39}
\affiliation{\YEREVAN}
\newcommand*{\VIRGINIA}{University of Virginia, Charlottesville, Virginia 22901, USA}
\newcommand*{\VIRGINIAindex}{25}
\affiliation{\VIRGINIA}
\newcommand*{\VT}{Virginia Polytechnic Institute and State University, Blacksburg, Virginia 24061, USA}
\newcommand*{\VTindex}{29}
\affiliation{\VT}

\newcommand*{\NOWSLAC}{SLAC National Accelerator Laboratory, Stanford, California 94309 USA}
\newcommand*{\NOWIMPERIAL}{Imperial College London, London SW7 2AZ, United Kingdom}
\newcommand*{\NOWMIT}{Massachusetts Institute of Technology, Cambridge, Massachusetts 02139, USA}
\newcommand*{\NOWINDIANA}{Indiana University, Bloomington, Indiana 47405, USA}
\newcommand*{\NOWSIENA}{Siena College, Loudonville, NY 12211, USA}
\newcommand*{\NOWGWU}{The George Washington University, Washington, DC 20052, USA}
\newcommand*{\NOWJLAB}{Thomas Jefferson National Accelerator Facility, Newport News, Virginia 23606, USA}
\newcommand*{\NOWLANL}{Los Alamos National Laborotory, New Mexico, USA}
\newcommand*{\NOWWM}{College of William and Mary, Williamsburg, Virginia 23187, USA}
%
%
\author {B.~Dey} 
\altaffiliation[Current address: ]{\NOWSLAC}
\affiliation{\CMU}
\email{biplabd@slac.stanford.edu}
\author {C.~A.~Meyer} 
\affiliation{\CMU}
\author {M.~Bellis} 
\altaffiliation[Current address: ]{\NOWSIENA}
\affiliation{\CMU}
\author{M.~Williams}
\altaffiliation[Current address: ]{\NOWMIT}
\affiliation{\CMU}
%
%
\author {K.~P.~Adhikari} 
\affiliation{\ODU}
\author {D.~Adikaram} 
\affiliation{\ODU}
\author {M.~Aghasyan} 
\affiliation{\INFNFR}
\author {M.~J.~Amaryan} 
\affiliation{\ODU}
\author {M.~D.~Anderson} 
\affiliation{\GLASGOW}
\author {S.~Anefalos~Pereira} 
\affiliation{\INFNFR}
% ...

\collaboration{The CLAS Collaboration}
\noaffiliation

\date{\today}

\title{Data analysis techniques, differential cross sections, and spin density matrix elements for the reaction $\myreac$}
%
%
\begin{abstract}

High-statistics measurements of differential cross sections and spin density matrix elements for the reaction $\gamma p \to \phi p$ have been made using the CLAS detector at Jefferson Lab. We cover center-of-mass energies ($\sqrts$) from 1.97 to 2.84~GeV, with an extensive coverage in the $\phi$ production angle. The high statistics of the data sample made it necessary to carefully account for the interplay between the $\phi$ natural lineshape and effects of the detector resolution, that are found to be comparable in magnitude. We study both the charged- ($\phi \to K^+ K^-$) and neutral- ($\phi \to K^0_S K^0_L$) $\kkb$ decay modes of the $\phi$. Further, for the charged mode, we differentiate between the cases where the final $K^-$ track is directly detected or its momentum reconstructed as the total missing momentum in the event. The two charged-mode topologies and the neutral-mode have different resolutions and are calibrated against each other. Extensive usage is made of kinematic fitting to improve the reconstructed $\phi$ mass resolution. Our final results are reported in 10- and mostly 30-MeV-wide $\sqrts$ bins for the charged- and the neutral-mode, respectively. Possible effects from $K^+ \Lambda^\ast$ channels with $p \kkb$ final-states are discussed. These present results constitute the most precise and extensive $\phi$ photoproduction measurements to date and in conjunction with the $\omega$ photoproduction results recently published by CLAS, will greatly improve our understanding of low energy vector meson photoproduction. 
\end{abstract} 

\maketitle


\section{\label{sec:intro}Introduction and Motivation}

Vector meson electro- and photoproduction have played an important role in our understanding of photon-hadron interactions in QCD. Sakurai~\cite{sakurai} first proposed that during interactions with hadrons, the photon (either real or virtual) behaves like an on-shell vector meson $V = \{\rho$, $\omega$, $\phi \}$. This is possible because the photon and the vector mesons share the same set of quantum numbers (see Ref.~\cite{bauer} for a review on the hadronic properties of the photon). Following Feynman~\cite{feynman}, if the real or virtual photon momentum is $q$, we can write the photon-hadron interaction amplitude in terms of a current $J_\mu(q^2)$ as $\langle \mbox{hadrons}|J_\mu(q^2)|0\rangle$, and one sees ``resonances'' at the values $q^2 = m^2_V$.

In the so-called vector meson dominance (VMD) model for photoproduction, a real photon can fluctuate into a virtual vector meson, which subsequently scatters off the target proton. Therefore, the amplitude $\mathcal{A}_{\gamma p \to V' p'}$ is related to $\mathcal{A}_{V  p \to V' p'}$ as
\begin{equation}
\mathcal{A}_{\gamma p \to V' p'} = \displaystyle\sum_{V_T} \frac{e}{\gamma_V} \mathcal{A}_{V p \to V' p'},
\label{eqn:vmd_amp}
\end{equation}
where $V_T$ indicates that the summation is only over the transverse polarization states of the vector meson (a real photon has no longitudinal polarization), $\gamma_V$ is the $V$-$\gamma$ coupling constant and $e$ is the electric charge. Traditionally, the $\phi$ vector meson has played a special role in our understanding of this VMD picture. Since the SU(6)-based quark model assigns an almost pure $|s\bar{s}\rangle$ configuration to the $\phi$, assuming the strangeness component of the proton to be small, the OZI rule~\cite{ozi} suppresses direct exchanges of quarks between the $\phi$ and the proton. Therefore, $\phi$ photoproduction is predicted to proceed by the exchange of color singlet gluonic objects.  From quite early on, several authors~\cite{freund_barger_cline} gave very general arguments that $\phi p$ scattering should proceed by the exchange of the universal Pomeron ($\Pom$) trajectory which has the same quantum numbers as the vacuum and the maximal Regge intercept of $\alpha_0 \approx 1$ (the Froissart bound~\cite{donnachie_book}). In terms of Regge theory, for $pp$ or $\pi N$ scattering at low energies, $t$-channel $\rho$ and $\omega$ exchanges can occur and the forward-angle differential cross sections exhibit ``shrinkage''. However, at very high energies, both the total and the differential cross section show only a slow (logarithmic) variation with the total energy and almost no shrinkage. This is also known as ``diffractive'' scattering at $s \to \infty$ and $t \to 0$, where $s$ and $t$ are the squares of the total center-of-mass (c.m.) energy and the exchanged momentum, respectively. The shape of the differential cross section ($d \sigma /dt$ plotted against $t$) resembles the intensity distribution in ordinary diffraction of light around a small object. For diffractive scattering with Pomeron exchange, the total cross section stays almost constant with energy and the width of the forward peak decreases only logarithmically with energy. That is, there is only a very slow shrinkage.

In the case of $\phi p$ scattering, since meson exchanges are suppressed, diffractive Pomeron exchange dominates even at low energies. The early (1972) SLAC beam-asymmetry measurements for $\phi p$ photoproduction by Halpern {\em et al.}~\cite{halpern} confirmed the dominance of natural-parity (Pomeron) exchange over unnatural-parity ($\pi$) exchange. It is this decoupling from the light quark ($u$ and $d$) sector that makes the $\phi$  a very ``clean'' system to study gluonic interactions, the gluonic structure of the Pomeron, for example. There is also speculation that at near threshold and forward angles, the $\phi$ channel will give access to the $0^{++}$ glueball $f_0(1710)$~\cite{nakano_toki,titov_lee_toki}.

Experimentally, most of the world data exist in the high energy diffractive region. These include results from DESY (Erbe 1968~\cite{abbhhm}; Alvensleben 1972~\cite{alvensleben}; Behrend 1978~\cite{behrend}), Cornell (McClellan 1971~\cite{mcclellan}; Berger 1972~\cite{berger}), SLAC (Anderson 1970~\cite{anderson_prd}; Anderson 1973~\cite{anderson_prl}; Ballam 1973~\cite{ballam}), Fermilab (Egloff 1979~\cite{egloff}; Busenitz 1989~\cite{busenitz}), Daresbury (Barber 1978~\cite{barber_78}; Barber 1982~\cite{barber_82}) and HERA (Derrick 1996~\cite{derrick}; Breitweg 2000~\cite{breitweg}). Due to the inherently small $\phi$ cross sections, these data are generally sparse with wide energy bin-widths and limited statistical precision. Also, since $d\sigma/dt$ drops exponentially with increasing $|t|$ and variations with the total c.m. energy ($\sqrt{s}$) are logarithmic, the ``natural'' scale for comparison purposes at these high energies seems to be logarithmic and not linear.

The earliest low energy near-threshold measurement (the ABBHHM results~\cite{abbhhm} had some low energy data as well) was at Bonn (Besch 1978~\cite{besch}), with more recent results coming from SAPHIR (Barth 2003~\cite{barth}) and LEPS (Mibe 2005~\cite{mibe}). Although both the Bonn and SAPHIR results covered the $E_\gamma=2.0$~GeV (or $\sqrts \approx 2.15$~GeV) energy region, it was the LEPS 2005 paper that first took note of a localized ``bump'' around $E_\gamma=2.0$~GeV, when a simple Pomeron exchange model predicts a smooth rise from threshold. To explain this non-monotonic behavior, theoretical groups have put forward two different explanations. First, the works by Ozaki {\em et al.}~\cite{ozaki} and Ryu {\em et al.}~\cite{ryu} relate this to a coupling between the $\phi p$ and $K^+ \Lambda(1520)$ channels. In the kinematic regime ${2~\mbox{GeV} \leq \sqrt{s} \leq 2.2~\mbox{GeV}}$, the $\phi p \to K^+ K^-p$ charged-mode and the $K^+ \Lambda(1520) \to K^+ p K^-$ decay-mode have the same final states. Therefore, rescattering effects can occur between the two channels. Our neutral-mode ($\phi \to K^0_S K^0_L$) results can play a critical role in this situation, since the $\phi$ neutral-mode does not share the same final states with $K^+ \Lambda(1520)$. Second, Kiswandhi {\em et al.}~\cite{kiswandhi,kiswandhi_prc} have attempted to explain this as a $J= \frac{3}{2}$ resonance of mass around $\sqrts ~\sim 2080$~MeV. Ultimately, the resolution could come from a combination of both approaches and would require a coupled-channel analysis of the $\phi$, $\Lambda(1520)$ and $\omega$ channels.

The first CLAS measurements at $E_\gamma = 3.6$~GeV (Anciant 2000~\cite{anciant}, McCormick 2004~\cite{mccormick}) also noted an interesting feature, a slight rise in the cross section at the backward-angles from nucleon exchanges via the $u$-channel. As mentioned earlier, the strangeness content of ordinary nucleons is usually assumed to be very small. However, $u$-channel nucleon exchanges points directly towards an appreciable strangeness content (possibly sea quarks) of the nucleon, or a violation of the OZI rule. We also note that a related CLAS analysis~\cite{odu} has presented differential cross sections for the purely neutral mode of $\phi$ photoproduction that are consistent with the more extensive results presented here.

In this work, we report on a precision study of $\phi$ photoproduction using a high statistics dataset. The $\phi$ mass ($\approx 1.0195$~GeV) is close to the the $\kkb$ production threshold. Therefore, the natural line-shape deviates from a relativistic-Breit-Wigner (rBW). We approximate the lineshape as a mass-dependent rBW with a $P$-wave barrier factor. Further, the $\phi$ natural width ($\Gamma_0 \approx 4.26$~MeV) is comparable to the CLAS resolution (1-2~MeV, see Fig.~\ref{fig:detector_res}). In general, the resolution can be a complicated mass-dependent function by itself (see Ref.~\cite{zach} for example). We approximate the resolution function as a single Gaussian and convolve it with the natural rBW lineshape to give the measured signal-lineshape. We incorporate three different $\kkb$ topologies that have different resolutions, and calibrate them against each other, to reduce systematic uncertainties on the $\phi$ lineshape. We also dwell on the issues pertaining to $K^+ \Lambda^\ast$ backgrounds.

For the spin-density matrix elements (SDME), it has long been observed that diffractive vector meson photoproduction roughly follows $s$-channel helicity conservation (SCHC), while $t$-channel helicity conservation (TCHC) is broken. Helicity non-conservation in any frame refers to the deviation of the $\rho^0_{00}$ SDME from 0 in that frame (see Sec.~\ref{sec:disc_tchc_schc}). SCHC indicates that in one particular reference frame (the Helicity frame), the outgoing vector meson has the same helicity as the incident photon. Gilman {\em et al.}~\cite{gilman} observed that even though in the Regge picture, the Pomeron acts like a spin-1 trajectory in the diffractive limit ($t \to 0$), its coupling is not well known. Naively, one ascribes the quantum numbers of the vacuum, $J^{PC} = 0^{++}$, to the Pomeron. This means that the coupling for the Pomeron in the $t$-channel is the same as the exchange of a spin-0 (scalar) particle. If so, one expects TCHC (no helicity flip in the $t$-channel), while experiments show violation of TCHC, and instead, support for SCHC. Phenomenologically, Gilman {\em et al.} showed that (in a sufficiently high energy limit) the spin-flip and non-flip components of the Pomeron in the $t$-channel exchange have to be related in a special manner to conserve helicity in the $s$-channel. However, there is no fundamental principle that predicts SCHC. With a fine binning in both $\sqrt{s}$ and $\cos \theta_{\mbox{\scriptsize c.m.}}^\phi$, where $\theta_{\mbox{\scriptsize c.m.}}^\phi$ is the meson production angle in the c.m. frame, our new precision results show that TCHC and SCHC are both violated.

With an unpolarized beam, one has access to only the $\rho^0$ elements, while $\rho^1$ and $\rho^2$ require a linear polarized beam, and $\rho^3$, a circularly polarized beam. Recently, the LEPS Collaboration~\cite{chang_leps_sdme} has published near-threshold measurements for the $\rho^1$ and $\rho^2$ SDME's. From their $\rho^1_{1-1}$ measurements in the Gottfried-Jackson frame, they have estimated a non-zero contribution from unnatural-parity ($\pi$, $\eta$) exchanges in the $t$-channel, at these low energies. With the beam and target both polarized, the FROST experiment~\cite{frost} at Jefferson Lab will give access to several of the double polarization observables~\cite{pichowsky} as well.


\section{\label{sec:exper}Experimental Setup}

The data that we use in this analysis were obtained using real photons produced via bremsstrahlung from a 4.023-GeV electron beam produced by the Continuous Electron Beam Accelerator Facility (CEBAF) at Jefferson Lab. The photons were energy-tagged by  measuring the momenta of the recoiling electrons with a dipole magnet and scintillator hodoscope system~\cite{sober}. A separate set of scintillators was used to make accurate timing measurements. The photon energy resolution was about $0.1 \%$ of the incident beam energy and the timing resolution was 120~ps. These tagged photons were then directed toward a $40$-cm-long cylindrical liquid-hydrogen cryotarget inside the CEBAF Large Angle Spectrometer (CLAS) detector system. Immediately surrounding the target cell was a ``start counter'' scintillator, used in the event trigger.

Both the start counter and the CLAS detector were segmented into sectors with a six-fold azimuthal symmetry about the beam line. A non-uniform toroidal magnetic field with a peak strength of 1.8~T was used to bend the trajectories of charged particles and a series of drift chambers was used for charged particle tracking. In this manner, CLAS could detect charged particles and reconstruct their momenta over a large fraction of the $4 \pi$ solid angle. The overall momentum resolution of the detector was $\sim 0.5 \%$. A system of $342$ scintillators placed outside the magnetic field and drift chamber regions provided timing information by measuring the time-of-flight (TOF) for each charged particle trajectory. A fast triggering and fast data-acquisition system (capable of running at $\sim 5$k~events/s) allowed for operating at a photon flux of a few times $10^7$~photons/s. Further details of CLAS can be found in Ref.~\cite{mecking}.


\section{\label{sec:data}Data}

The specific dataset that we analyze in this work was collected in the summer of 2004, during the CLAS ``$g11a$'' run period. Roughly 20 billion triggers were recorded during this time, out of which only a small fraction corresponded to $p \phi$ events. Each event trigger required a coincidence between the photon tagger Master OR (MOR) and the CLAS Level 1 trigger. The MOR consisted of a logical OR of the signals from the 40 tagger counters corresponding to the high-energy part of the tagged photon beam. The Level 1 trigger required that two tracks be present in the CLAS detector. A single track was defined on a sector-wise basis and required a coincidence between the start counter and the TOF counters within the given sector. The two-track trigger required that at least two sectors in CLAS satisfied the single-track trigger within a 150~ns window of each other. In addition to the MOR and Level 1 trigger, we required the tagger MOR and an OR of all the start counter hits occur within 15 ns of each other. This constraint was used to reduce out-of-time contamination. The beam structure contained one or a few photons within bunches separated by 2.0~ns.

During offline processing, before any physics analysis began, the CLAS detector sub-systems had to be calibrated. This included determining the relative offsets between the photon tagger, start counter and TOF counter times, as well as calibration of the drift times in the drift chambers and the pulse heights of the TOF scintillators. Energy and momentum corrections were made for individual particles to account for their energy and momentum losses during passage through several layers of the detector sub-systems. Corrections were also made to the incident photon energy ($E_\gamma$) to account for mechanical sagging in the tagger hodoscope. A discussion of the collection and calibration of this data set can be found in Ref.~\cite{my-thesis}. Several meson photoproduction channels ($\omega p$~\cite{omega_prc, omega_pwa}, $\eta^{(')} p $~\cite{eta_prc}, $K^+ \Lambda$~\cite{klam_prc}, $K^+ \Sigma^0$~\cite{ksig_prc}, $KY^\ast$~\cite{kei_prc1,kei_prc2}) have already been analyzed using this dataset. In the vector meson sector, the present work extends our recently published $\omega$ results~\cite{omega_prc} to the $\phi$ channel.


\section{\label{sect:event_sel} Reaction topologies and Event Selection}

In the reaction $\gamma p \to \phi p$, the $\phi$ subsequently decays most of the time into two kaons: $K^+ K^-$ (charged-mode) and $K^0_S K^0_L$ (neutral-mode). For the neutral-mode, the daughter $K^0_S$ further decays into $\pi^+ \pi^-$ ($60.2\%$) and $\pi^0 \pi^0$ ($39.7\%$). Since CLAS is optimized for detecting charged particles, we only employed the $K^0_S \to \pi^+ \pi^-$ decay in this analysis.

The ``charged-two-track'' topology was then defined as $\gamma p \to K^+ (K^-) p$, where only the proton and the $K^+$ were detected and the undetected $K^-$ was reconstructed as the missing 4-momentum using a 1-$C$ kinematic fit to a total missing mass of $m_{K^-} = 0.493$~GeV. The polarity setting of the drift chamber magnets inside CLAS bent negatively charged particles like $K^-$ inwards towards the beam line, where the detector acceptance was the lowest. Therefore, not detecting the $K^-$ led to a substantial increase in the overall statistics, allowing for a fine energy binning (10-MeV-wide $\sqrt{s}$ bins) and wide kinematic coverage for this topology. The main bulk of our results derives from this mode.

The ``charged-three-track'' topology was a subset of the charged-two-track data sample where all three final-state charged tracks were required to be detected. A 4-$C$ fit to zero missing 4-momentum and subsequent confidence level (CL) cut was used to remove background. Due to the 4-$C$ fit, this topology had a very high purity of $p K^+ K^-$ final states and the highest resolution in the reconstructed $\phi$ lineshape, but with very low acceptance. It is mainly used for understanding some of the systematics and our final set of results do not include this topology.

As mentioned in the introduction, it is also important to examine the neutral decay mode of the $\phi$, since this is relatively immune (see discussion in Sec.~\ref{sec:disc}B) to effects from the $K^+ \Lambda^\ast$ final states. We defined the ``neutral-mode'' topology as $\gamma p \to \pi^+ \pi^- (K^0_L) p$, where only the $\pi^+$, the $\pi^-$ and proton were detected, and the undetected $K^0_L$ was reconstructed as the missing 4-momentum using a 2-$C$ kinematic fit to $m_{K^0} = 0.497$~GeV total missing mass and $M(\pi^+ \pi^-) = m_{K^0}$. Since this topology required detection of a negatively charged $\pi^-$, the acceptance was substantially lower compared to the charged-two-track mode, especially at high energies and backward-angles. To bolster statistics, therefore, we employed wider 30-MeV-wide energy bins in $\sqrt{s}$ for this case.

\begin{table}
  \centering
  \begin{tabular}{lcc} \hline\hline
    \multirow{2}{*}{Cut} & \multicolumn{2}{c}{Topology} \\
    &  $ K^+ (K^-) p$, $ K^+ K^- p$  \;\;\;\; &  \;\;\;\;  $\pi^+ \pi^- (K^0_L) p \;\;\; $ \\ \hline
    CL & \checkmark & \checkmark \\ \hline
    Timing & \checkmark & \checkmark \\ \hline
    $K^+\Lambda^\ast$  & \checkmark & -- \\ \hline
    Fiducial cuts\;\;\; & \checkmark & \checkmark \\ \hline
    $M(\kkb)$ cut & \checkmark & \checkmark \\ \hline \hline
    \end{tabular}
  \caption[]{\label{table:pid_cuts} List of cuts applied to the two topologies in this analysis. The CL, timing and fiducial volume cuts applied to both topologies. The charged-mode analysis had an extra hard cut to remove $K^+\Lambda^\ast$ events. Lastly, both topologies had a ${1.0~\mbox{GeV} < M(\kkb) <1.06}$~GeV cut placed at the very end, after the completion of the signal-background separation process.}
\end{table}

Each event trigger recorded by CLAS consisted of one or more tagged photons. To begin the event selection process, at least two positively charged particle tracks were required to have been detected. These were hypothesized as a proton and a $K^+$ for the charged-modes, and as a proton and a $\pi^+$ for the neutral-mode. The charged-two-track (neutral-) mode topology required an extra negatively charged particle track that was assumed to be a $K^-$ ($\pi^-$). To minimize bias, all possible photon-particle combinations allowed by the given topology were taken to be a candidate signal event. Events with incorrectly assigned photons or particle hypotheses were removed by subsequent cuts. In the following sub-sections, we describe each of these event selection cuts, referring the interested reader to Ref.~\cite{my-thesis} for further details. Since the two topologies followed significantly different analysis chains, to avoid confusion, we list the various cuts as applicable to each of the charged- and neutral-mode topologies in Table~\ref{table:pid_cuts}.

\begin{figure}
\centering
\includegraphics[width=2.1in,angle=90]{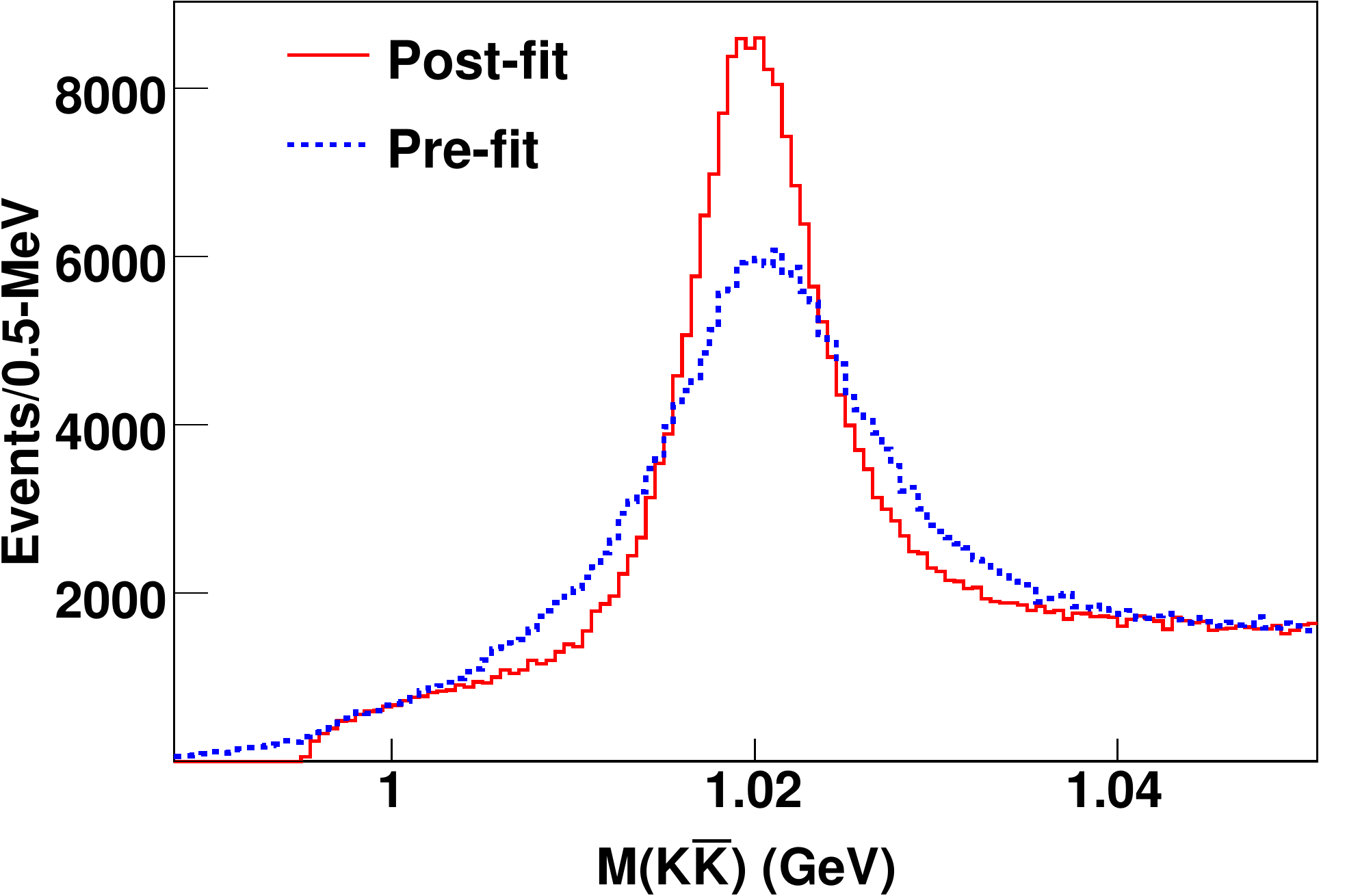}
\caption[]{(Color online) The dotted blue and the continuous red histograms represent the reconstructed $\phi$ mass prior to and after kinematic fitting, respectively, for the neutral-mode. The resolution is markedly improved due to the kinematic fit.}
\label{fig:neutral_kinfit_effect}
\end{figure}

\begin{figure*}
\begin{center}
\subfigure[]{
{\includegraphics[width=2.2in]{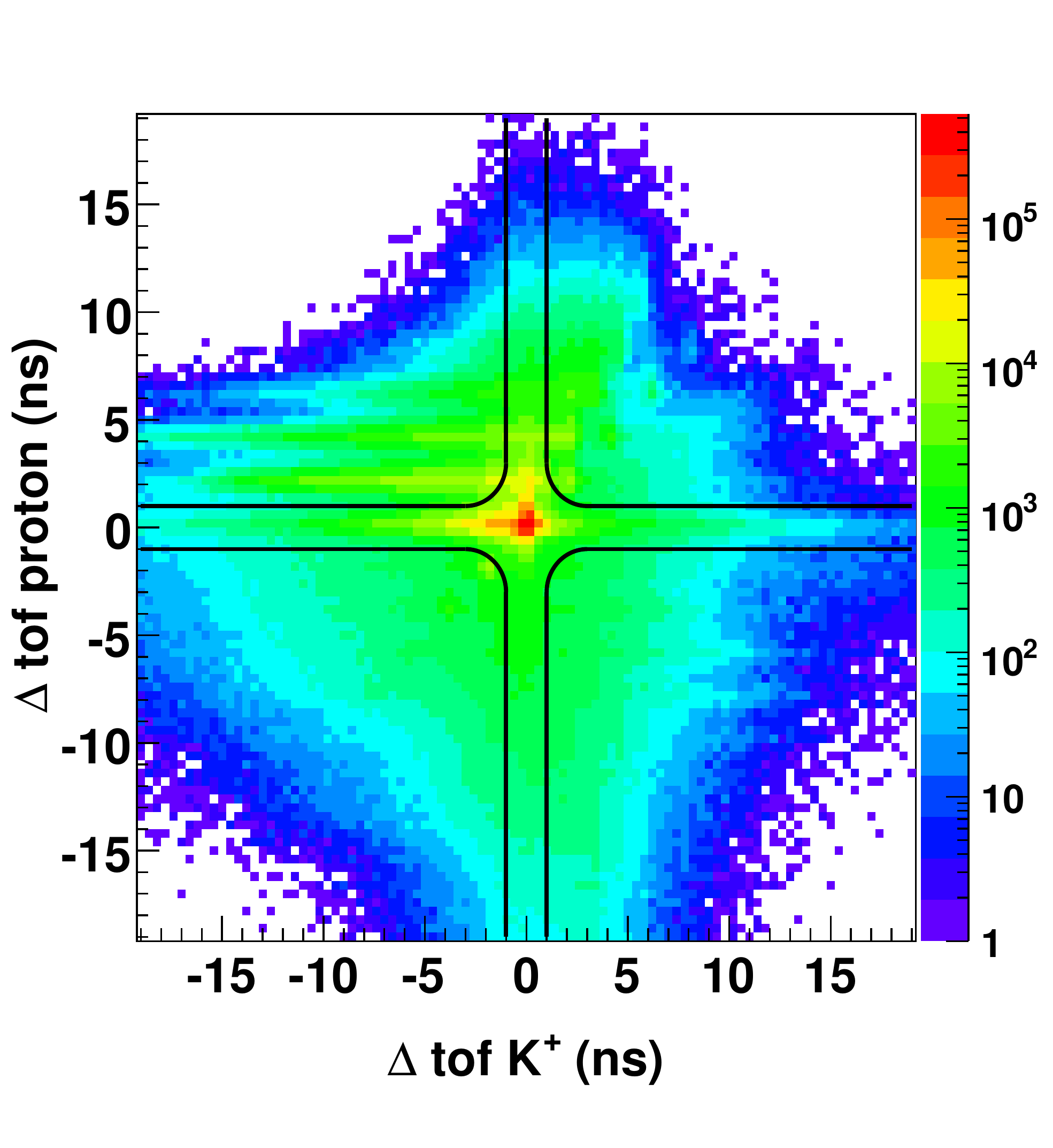}} 
}
\subfigure[]{
{\includegraphics[width=2.2in]{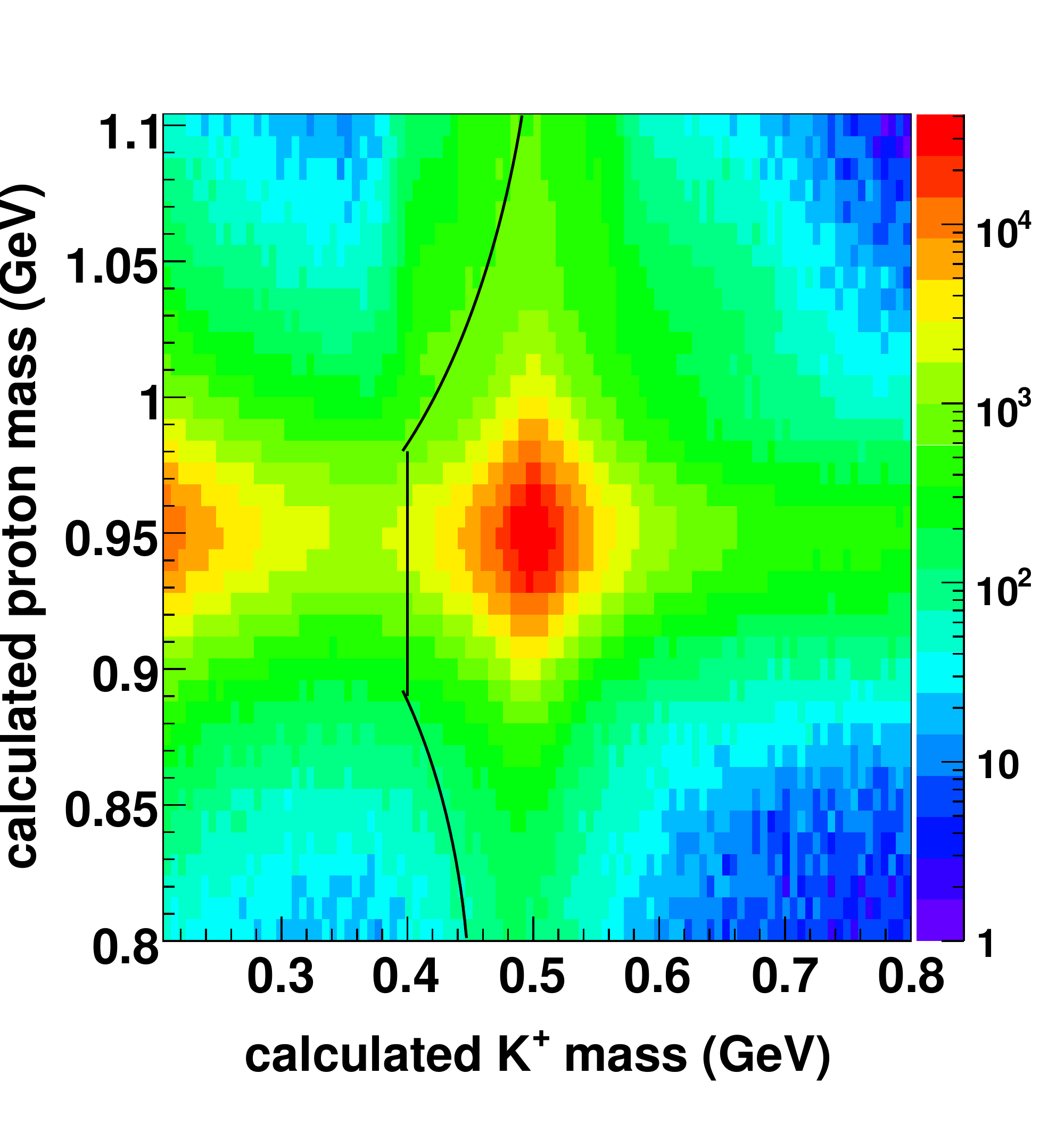}} 
}
\subfigure[]{
{\includegraphics[width=2.2in]{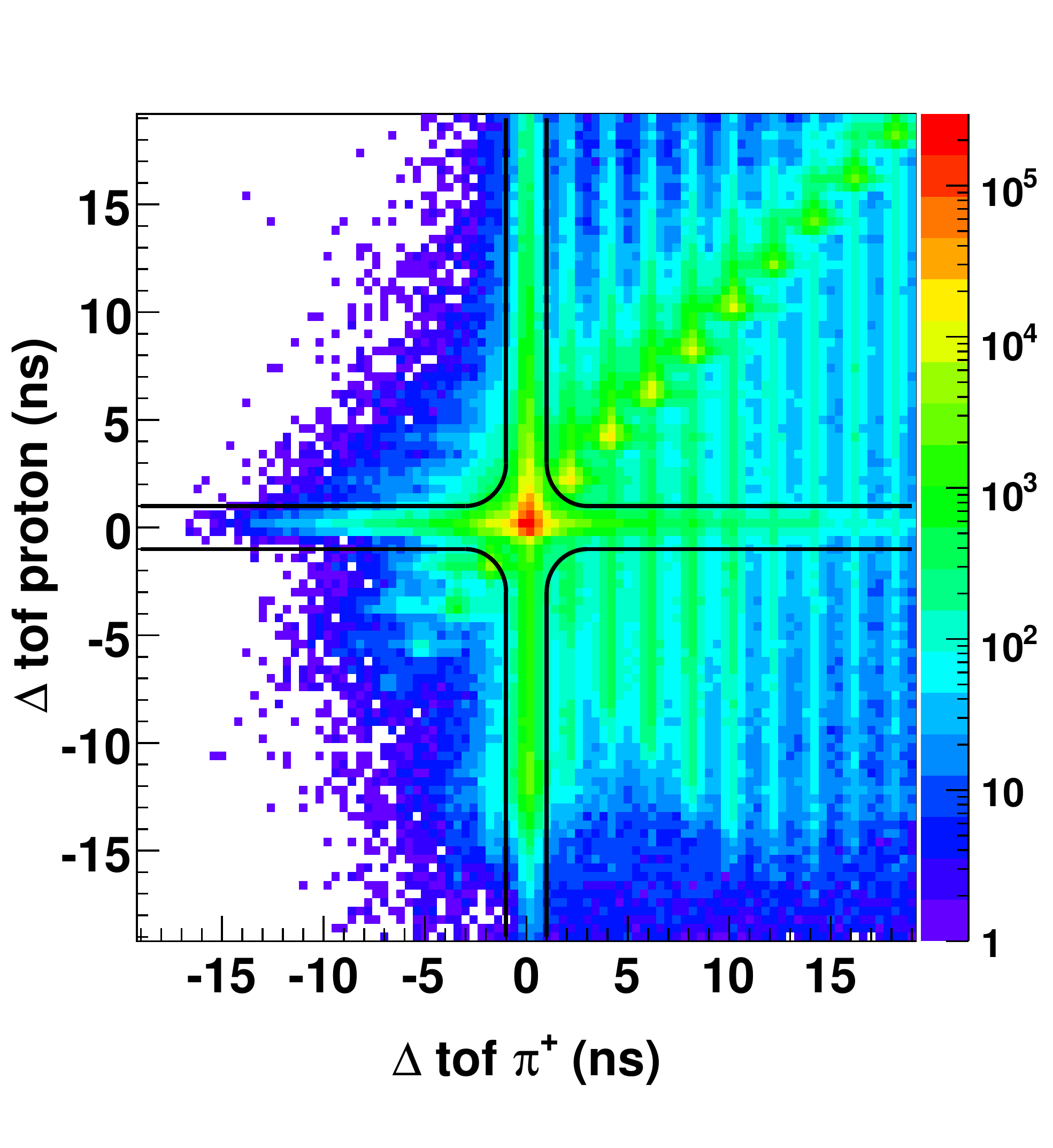}} 
}
\caption[]{(Color online) Timing cuts for background removal: (a) and (b) charged-two-track topology, (c) neutral-mode topology. In subfigures (a) and (c), events lying outside the ``iron-cross'' shaped quadruplet of black curves were removed from further analysis. For the charged-two-track mode, an additional cut was placed for further pion removal by rejecting events on the left of the curve in subfigure (b).}
\label{fig:timing_cuts}
\end{center}
\end{figure*}

\subsection{Confidence level cut}
Each event candidate in the charged-two-track data set was kinematically fit~\cite{my-thesis} to an overall $m_{K^-}$ missing mass hypothesis for the undetected $K^-$. For every event recorded by CLAS, both ``$K^+:p$'' and ``$p:K^+$'' combinations were treated as independent event hypotheses, where the two-particle assignment (separated by the colon) corresponds to the two detected positively charged particles. Similarly, for the neutral-mode, every event was kinematically fitted to an overall $m_{K^0}$ missing mass hypothesis for the undetected $K^0_L$. Both ``$\pi^+:p:\pi^-$'' and ``$p:\pi^+:\pi^-$'' combinations, corresponding to different particle assignments to the two positively charged tracks, were taken as independent event hypotheses. For the charged-three-track mode, both ``$K^+:p:K^-$'' and ``$p:K^+:K^-$'' combinations were allowed as event candidates. Incorrect particle hypothesis assignments were removed by subsequent timing cuts. The kinematic fitter adjusted the momentum of each individual detected particle, while constraining the total missing mass to be either $m_{K^-}$ (charged-two-track mode) or $m_{K^0}$ (neutral-mode), and the total missing 4-momentum to be zero for the charged-three-track mode. The shifts in the momenta, combined with the known detector resolution within the current experiment, gave a CL for the event to be the desired reaction. For the charged-two-track and neutral-mode topologies, only events with a CL greater than $5\%$ were retained for further analysis. For the charged-three-track mode, we required the CL to be greater than $1\%$.

Fig.~\ref{fig:neutral_kinfit_effect} demonstrates the effect of kinematic fitting for the neutral-mode topology. After kinematic fitting, both $K^0$'s were mass constrained, so that the threshold value of $M(\kkb)_{min} = 2 m_{K^0}$ was enforced. The reconstructed $\phi$ mass lineshapes were markedly different between before (blue dashed histogram) and after (red histogram) kinematic fitting. In topologies with the initial 4-momenta known and a single missing particle in the final states, kinematic fitting amounts to measurement of the missing particle's 4-momentum. In particular, the present analysis is very sensitive to the reconstructed $\phi$ mass resolution relative to the $\phi$ natural linewidth $\Gamma_0 \approx 4.26$~MeV (see also Sec.~\ref{sec:phi_resolution_discussion} for a discussion). Therefore kinematic fitting plays a critical role here.

\subsection{Timing cuts}

Track reconstruction through the different CLAS detector segments yielded both the momentum $\vec{p}$ and the path length $l$ from the reaction vertex to the TOF scintillator wall. The expected time-of-flight for a track hypothesized to be a particle of mass $m$ was then given by
\begin{equation}
  t_{exp} = \frac{l}{c}\sqrt{1 + \left(\frac{m}{|\vec{p}|}\right)^2}.
\end{equation}
CLAS also measured the time-of-flight $t_{meas}$ as the difference between the tagged photon's projected arrival time at the reaction vertex for the given event and the time the given particle track hits the TOF scintillator wall. The difference between these two time-of-flight calculations gave $\Delta tof = t_{meas} - t_{exp}$. For each track there was also a calculated mass $m_c$, given by
\begin{equation}
m_c =  \sqrt{\frac{|\vec{p}|^2(1-\beta^2)}{\beta^2 c^2}},
\end{equation}
where $\beta = l/(ct_{meas})$.

Timing information in the form of $\Delta tof$ or $m_c$ was used to place particle identification cuts on the proton and $K^+$ tracks for the charged-mode, and the proton and $\pi^+$ tracks for the neutral-mode. As mentioned earlier, for each pair of positively charged particle tracks, all possible combinations of particle assignments were considered and treated as independent event hypotheses. Figs.~\ref{fig:timing_cuts}a and ~\ref{fig:timing_cuts}c show our $\Delta tof$ cuts placed on proton-$K^+$ (charged-two-track mode) and proton-$\pi^+$ (neutral-mode) tracks, respectively, where events outside the ``iron-cross'' shaped quadruplet of black curves were rejected. The clusters along the diagonal in the $\Delta tof$ plots, more prominent in Fig.~\ref{fig:timing_cuts}c, were due to accidental coincidences with events in different beam bursts corresponding to the 2~ns radio-frequency pulses used by the CEBAF electron accelerator.

For the charged-mode, even after the application of the $\Delta tof$ iron-cross cut, a remnant pion leakage was still visible, as seen in the left hand side ``blob'' in Fig.~\ref{fig:timing_cuts}b. To further remove this pion background, we placed an additional cut on the calculated mass of the proton and $K^+$ tracks by rejecting events that lie on the left of the curve in Fig.~\ref{fig:timing_cuts}b.

\subsection{Overlap with $K^+\Lambda^\ast$}
\label{sec:phi_lambda1520_interference}

\begin{figure}
\begin{center}
\includegraphics[width=2.7in]{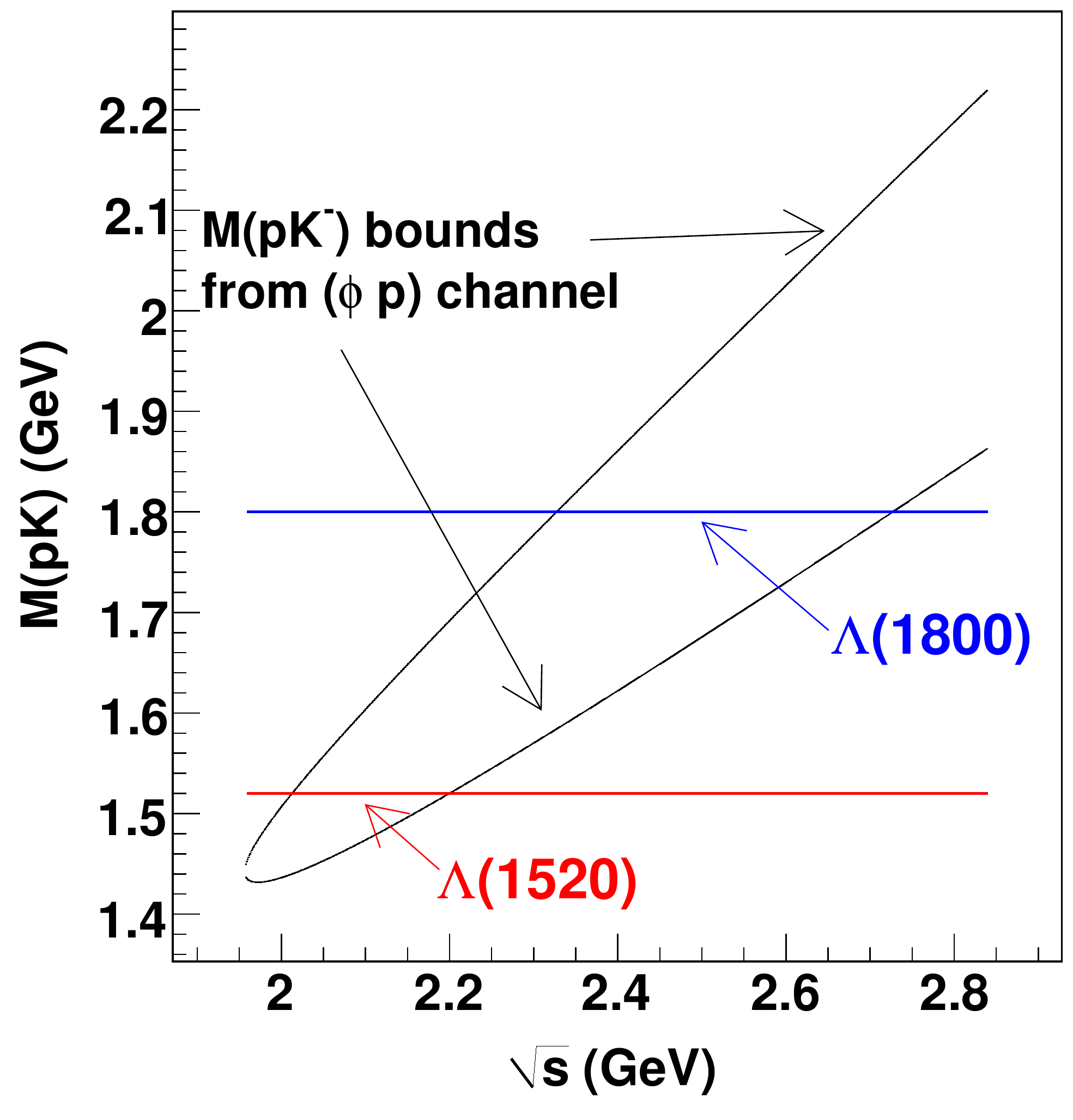}  
\caption[]{(Color online) $\phi$-$\Lambda^\ast$ overlap for the charged-mode topology: if the $M(K^+K^-)$ invariant mass is fixed at $m_\phi = 1.020$~GeV, $M(pK^-)$ is bounded between an upper and lower limit depending on $\sqrt{s}$. Phase-space overlaps with different $K^+ \Lambda^\ast$ production channels occur in different $\sqrt{s}$ ranges.}
\label{fig:lstar_bound}
\end{center}
\end{figure}

Consider the process $\gamma p \to M(\sqrt{s}) \to K^+ K^- p$ from the perspective of a 3-body decay, where $M(\sqrt{s})$ denotes a general $s$-channel state with total invariant mass $\sqrt{s}$. If $M(K^+ K^-)$ is fixed at $m_\phi = 1.02$~GeV, $M(p K^-)$ is bound between a minimum ($M_{pK,min}$) and a maximum ($M_{pK,max}$). The specific values of $M_{pK,min}$ and $M_{pK,max}$ depend on $\sqrt{s}$, and the masses of the $\phi$, the proton and the kaon. Fig.~\ref{fig:lstar_bound} shows the variation of $M_{pK,min}$ and $M_{pK,max}$ with $\sqrt{s}$.

\begin{figure}
\begin{center}
\hspace{0.05cm}
\subfigure[]{
{\includegraphics[width=2.7in]{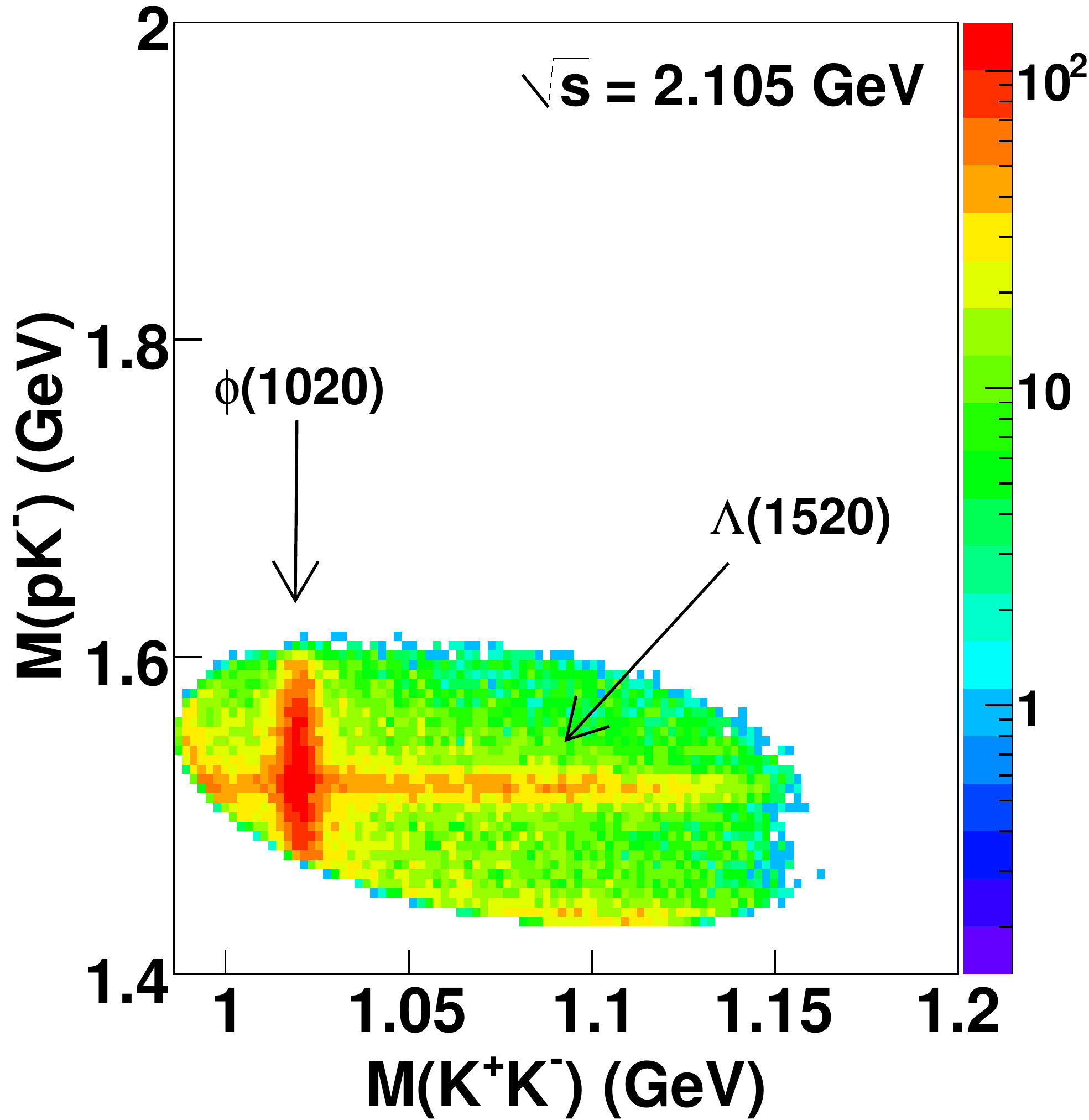}} 
}
\hspace{0.05cm}
\subfigure[]{
{\includegraphics[width=2.7in]{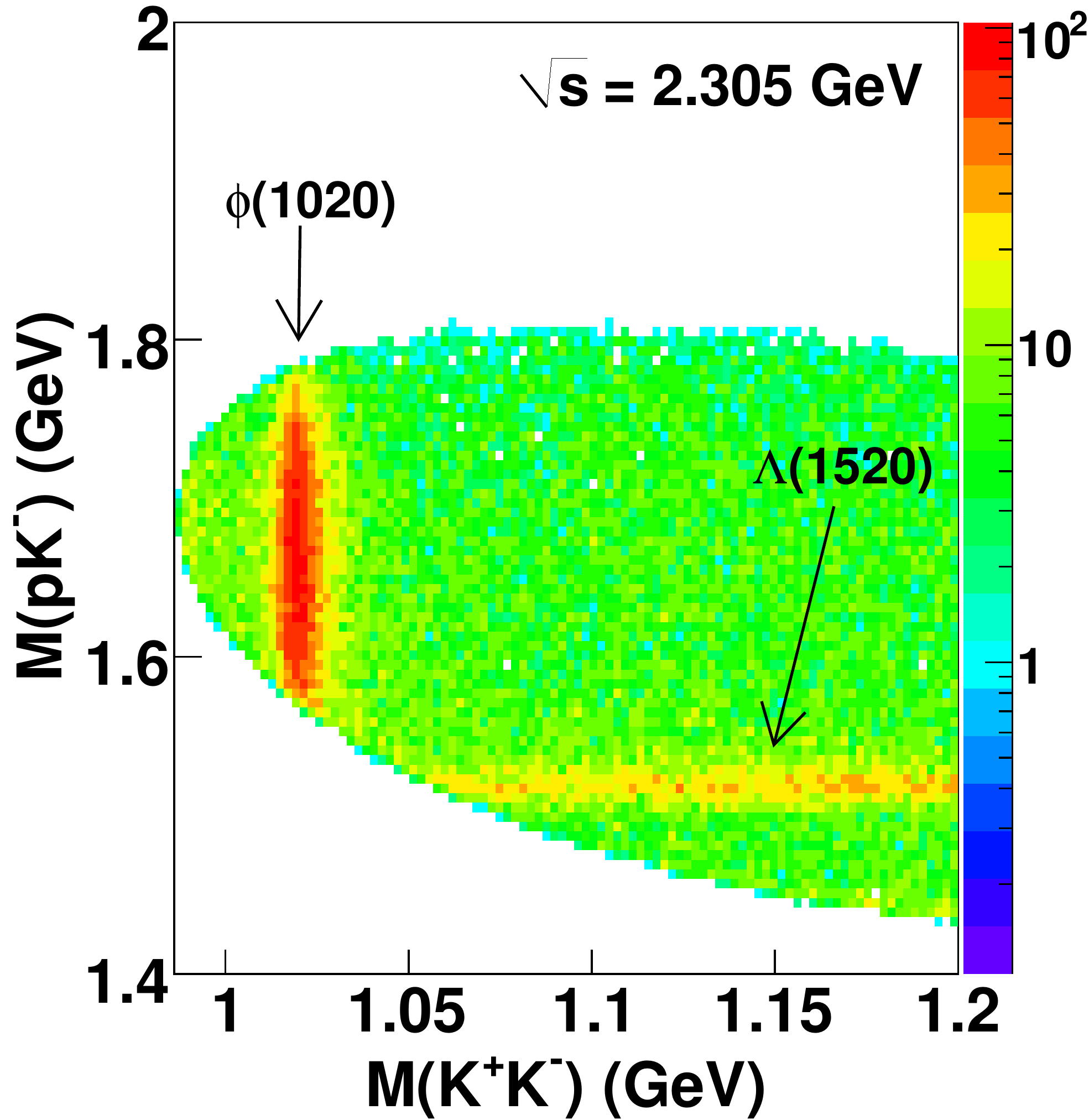}} 
}
\caption[]{(Color online) $\phi$-$\Lambda(1520)$ overlap for the charged-two-track mode topology: between $\sqrt{s} =2$ and 2.2~GeV, the $\Lambda(1520)$ mass lies within the allowed $M(p K^-)$ bounds and the $\phi p$ and $K^+ \Lambda(1520)$ channels overlap in phase-space. A plot of $M(pK^-)$ vs. $M(K^+K^-)$ is shown in (a) and (b) for two different energies in the charged-two-track mode. (a) Shows the overlap for the $\sqrt{s} = 2.105$~GeV bin, while (b) shows that above $\sqrt{s} \approx 2.2$~GeV, the $\phi$ and $\Lambda(1520)$ ``bands'' are separated.}
\label{fig:lambda1520_bound}
\end{center}
\end{figure}

In the region 2~GeV$\leq\sqrt{s}\leq 2.2$~GeV, $M(p K^-) = 1.52$~GeV lies within these bounds. This implies that in this energy regime, the kinematics for production of $\phi p$ and $K^+ \Lambda(1520)$ overlap in phase-space. Figs.~\ref{fig:lambda1520_bound}a~and~\ref{fig:lambda1520_bound}b show plots of $M(pK^-)$ vs. $M(K^+K^-)$ at two different energies for the charged-two-track mode. The $\phi$ and $\Lambda(1520)$ ``bands'' are clearly visible in both figures. Above $\sqrt{s} \approx 2.2$~GeV there is no overlap between the $\phi$ and the $\Lambda(1520)$. To reduce the $\Lambda(1520)$ contamination in the charged modes, we keep the additional option of placing a hard cut around the $\Lambda(1520)$ mass as
\begin{equation}
|M(p K^-) - 1.52| \geq 0.015~\text{GeV}.
\label{eqn:lam1520_hard_cut}
\end{equation}

\begin{figure}
\begin{center}
\hspace{0.05cm}
\subfigure[]{
{\includegraphics[width=2.7in]{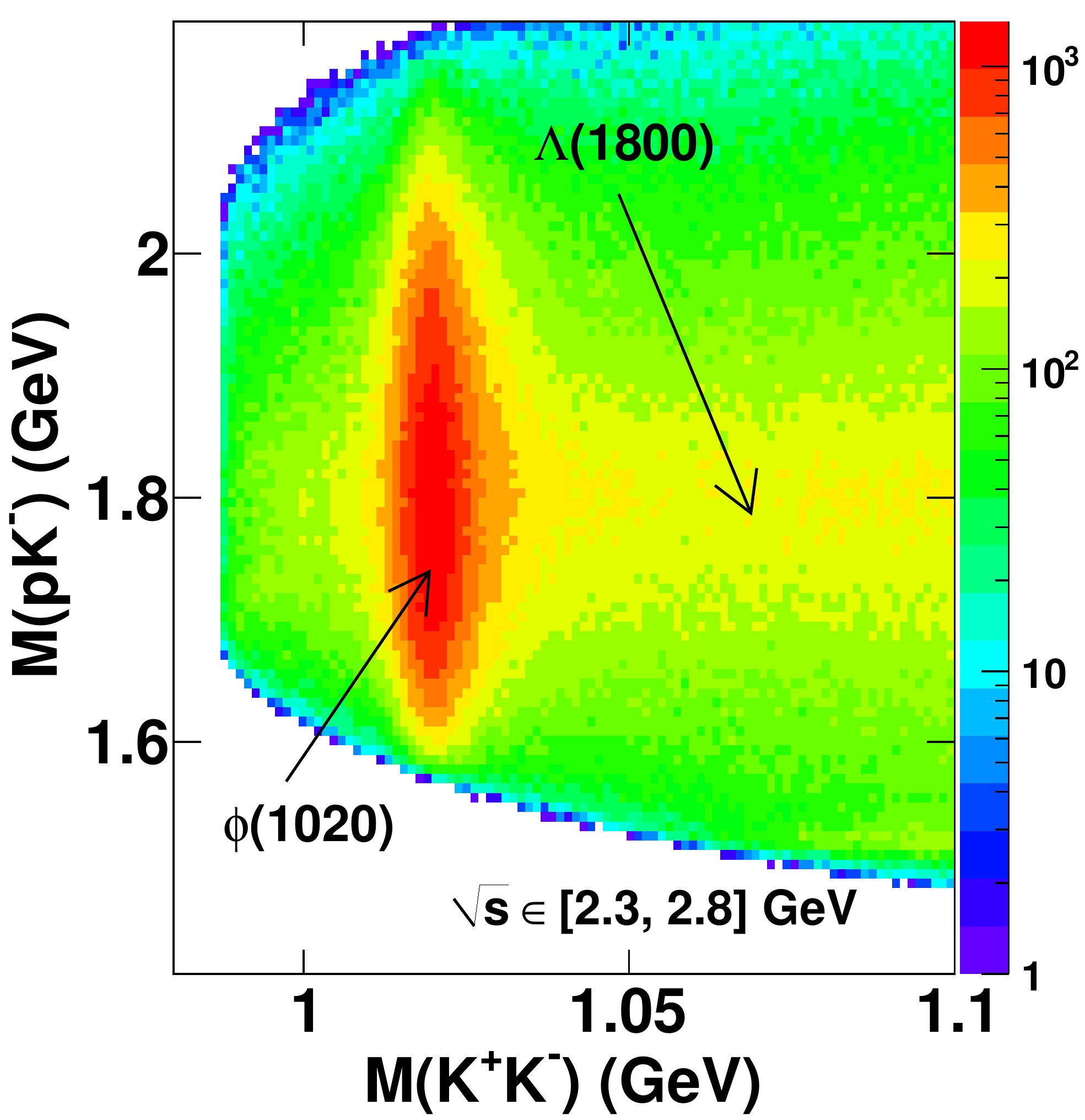}} 
}
\hspace{0.05cm}
\subfigure[]{
{\includegraphics[width=2.7in]{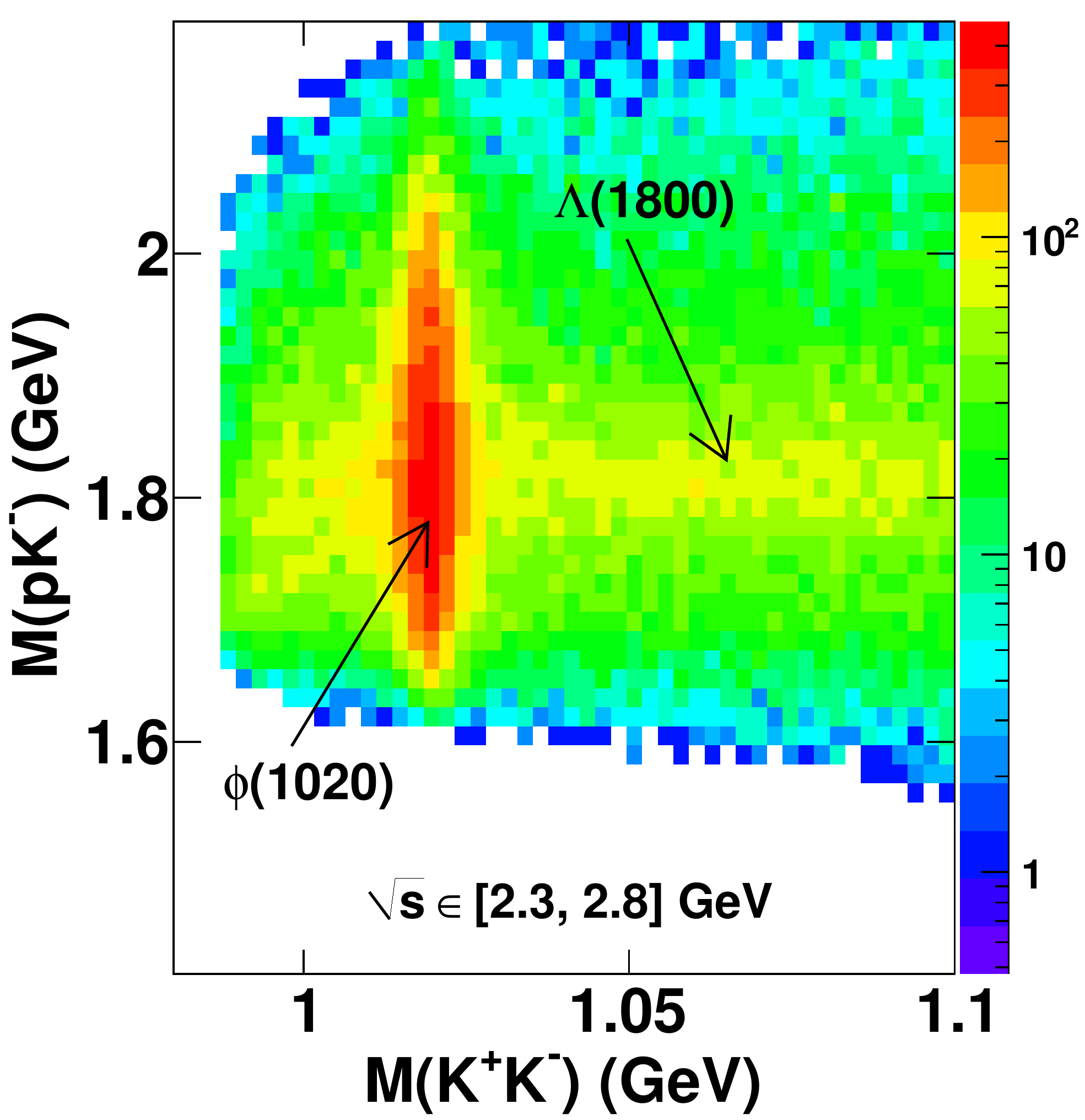}} 
}
\caption[]{(Color online) $\phi$-$\Lambda(1800)$ overlap for the charged-track mode topology: between $\sqrt{s} =2.3$ and 2.8~GeV, several broad $\Lambda^\ast$ states with masses around 1800~MeV within the allowed $M(p K^-)$ bounds. A plot of $M(pK^-)$ {\em vs.} $M(K^+K^-)$ is shown for the charged (a) two-track and (b) three-track mode.}
\label{fig:lambda1800_bound}
\end{center}
\end{figure}

Similarly, there are several broad $\Lambda^\ast$ states around 1800~MeV that overlap with the $\phi$ in the region 2.3~GeV$\leq\sqrt{s}\leq 2.8$~GeV. Figs.~\ref{fig:lambda1800_bound}a and \ref{fig:lambda1800_bound}b show the $M(pK^-)$ {\em vs.} $M(K^+K^-)$ distributions in this $\sqrt{s}$ region for the charged two- and three-track topologies, respectively. Given that these higher $\Lambda^\ast$ states have large widths, it is difficult to place hard cuts without significant signal loss in the $\phi$. Moreover, from Fig.~\ref{fig:lambda1800_bound}, the poorer resolution in the two-track compared to the three-track means that the $M(pK^-)$ cuts have to be tighter, for the former case. We consider the cuts
\begin{equation}
|M(p K^-) - 1.8| \geq \delta_{1800},
\label{eqn:lam1800_hard_cut}
\end{equation}
where $\delta_{1800} = 0.04$~GeV for charged-two-track and 0.02~GeV for the charged-three-track mode.

\subsection{Effectiveness of cuts}

The effectiveness of these cuts can be gauged by the percentage of signal events lost due to them. The $M(\kkb)$ distributions were fit to a signal lineshape (see Sec.~\ref{sec:lineshape}) plus a quartic background function before and after placing the cuts. From this study, the losses in signal yields due to the cuts were estimated to be $\sim4.5\%$ and $\sim5\%$ for the charged- and neutral-mode topologies, respectively. We quote these as the upper limits of the systematic uncertainties in our particle identification/event selection for this analysis.

\subsection{Fiducial cuts}

In addition to the above particle identification cuts, fiducial volume cuts were required to remove events belonging to regions where our understanding of the detector performance was relatively poor. These cuts were motivated by differences in an empirical efficiency calculation between the actual data and Monte Carlo which indicated discrepancies in the forward-angle region and at the boundaries of the six sectors of the CLAS detector due to edge effects in the drift chambers. Therefore, events with any particle trajectory falling near the sector boundary regions were removed. A $\phi_{lab}$-dependent cut on $\cos \theta_{lab}$ along with a hard cut at $\cos \theta_{lab} \geq 0.985$ removed extremely forward-going particles that coincided with the beam-dump direction. Localized inefficiencies within the CLAS detector volume such as inside the drift chambers were accounted for by placing trigger efficiency cuts on the Monte Carlo data as functions of $\phi_{lab}$, $\theta_{lab}$ and $|\vec{p}|$ for each particle track. Additional cuts were placed on backward-going tracks ($\cos \theta_{lab} \leq -0.5$). A minimum proton momentum cut at 375~MeV removed slow moving protons, whose energy losses were difficult to model in the detector simulation. Events with particles corresponding to poorly performing TOF scintillator counters were removed as well.


\section{\label{sec:lineshape}The $\phi$ lineshape}

We first define our notation for the 2-body breakup momentum of a parent particle of mass $m$ decaying into two daughters of masses $m_1$ and $m_2$ as
\begin{align}
p_{m\to m_1 m_2} =\frac{\sqrt{(m^2 - (m_1 +m_2)^2)(m^2 - (m_1-m_2)^2)}}{2m}.
\end{align}
The phase-space element contribution from each such decay is
\begin{align}
\frac{dN}{dm^2} \propto \left( \frac{p_{m\to m_1 m_2}}{m} \right).
\end{align}

The $\phi$ has three main decay modes, viz., $\kp\km$, $\ks\kl$ and $\pip \pim \piz$, with the approximate branching fractions being 0.489, 0.342 and 0.153, respectively. The neutral and charged kaons have different masses and the ratio between the neutral and charged $\kkb$ branching fractions is consistent with a $p^3_{\phi\to\kkb}$ dependence on the breakup momentum, as expected from a $P$-wave decay. The mass-dependent rBW amplitude for $\phi\to \kkb$ is given by
\begin{align}
\mathcal{A}_{\phi \to \kkb} \sim \left(\frac{p}{p_0}\right)^L \frac{B^L(p,R)}{B^L(p_0,R)} \frac{1}{m^2_0 - m^2 - im_0\Gamma_{\mbox{\scriptsize total}}}.
\end{align}
Here $m \equiv M(\kkb)$ is the $\phi$ mass, $p \equiv p_{\phi \to \kkb}$ is the mass-dependent breakup momentum and, $p_0$ is $p$ evaluated at the $\phi$ mass $m = m_0$. $B^L(p,R)$ is the phenomenological Blatt-Weisskopf barrier factor with $R \approx 5$~GeV$^{-1}$, corresponding to a meson radius of 1~fm. For a $P$-wave decay, $L=1$ and this is given by
\begin{align} 
B^{L=1}(p,R) = \frac{1}{\sqrt{1+p^2R^2}}.
\end{align}

The total mass-dependent width, $\Gamma_{\mbox{\scriptsize total}}$, comprises four parts
\begin{align}
\Gamma_{\mbox{\scriptsize total}} = \Gamma_{\kp\km} + \Gamma_{\ks\kl} + \Gamma_{3\pi} + \Gamma_4,
\end{align}
corresponding to the different decay modes of the $\phi$ ($\Gamma_4$ accounts for any mode other than $K\overline{K}$ and $3\pi$). For each of the $\kkb$ modes, the contribution is
\begin{align}
\Gamma_{\kkb} = \Gamma_0 \;\mathcal{B}_{\kkb}\; \frac{p^3}{p^3_0}\;  \frac{m_0}{m}\; \frac{1 + p^2_0R^2}{1 + p^2R^2},
\end{align}
where $\mathcal{B}_X$ denotes the branching fraction of $\phi$ into a particular final state $X$. The dynamics of the $\phi \to 3 \pi$ decay is more complicated and beyond the scope of this work; we assume the corresponding amplitude to be approximately constant in the small range of $M(\kkb)$ that we are interested in. To incorporate the mass-dependence from the 3-body phase-space factor, we consider $m_{12}$ as the mass of the $(\pip,\pim)$ system. The 2-body phase-space element is then
\begin{align}
\frac{dN}{dm^2 dm^2_{12}} \sim \left(\frac{p_{m\to m_{12} m_{\piz}}}{m} \right) \times \left( \frac{p_{m_{12} \to m_{\pi} m_{\pi}} }{m_{12}}\right),
\end{align}
which gives
\begin{align}
\frac{dN}{dm} \sim \displaystyle \int_{2 m_\pi}^{m - m_{\scriptsize \pi^0 }}\left(p_{m\to m_{12} m_{\piz}}\right) \left( p_{m_{12} \to m_{\pi} m_{\pi}} \right) dm_{12}.
\end{align}
This integration is performed numerically. The $\phi \to 3\pi$ contribution to $\Gamma_{\mbox{\scriptsize total}}$ is then
\begin{align}
\Gamma_{3\pi} = \Gamma_0 \times \mathcal{B}_{3\pi} \times \frac{dN/dm}{\left[dN/dm\right]_0},
\end{align}
where $\left[dN/dm\right]_0$ corresponds to computation at $m = m_0$. The fourth component, we simply take as a constant:
\begin{align}
\Gamma_4 = \Gamma_0 (1- \mathcal{B}_{\kp \km} - \mathcal{B}_{\ks\kl} - \mathcal{B}_{3\pi}).
\end{align}

The mass-dependent natural lineshape in each $\sqrts$ bin is then
\begin{align}
\tilde{\mathcal{S}}(m) & \sim \frac{p_{\sqrts\to m\;m_p}}{\sqrts} |\mathcal{A}|^2\times \left( \frac{p}{m} \right) \nonumber \\
&\sim p_{\sqrts\to m\;m_p} \frac{p^3}{m} \frac{1}{1 + p^2R^2 }\frac{1}{(m^2_0-m^2)^2 + (m_0\Gamma_{\mbox{\scriptsize total}})^2},
\end{align}
where $p_{\sqrts\to m\;m_p}/\sqrts$ is the phase-space factor for the c.m. mass pseudoparticle decaying into a $\phi$ and a proton. Near production threshold, this factor suppresses events with large values of $m$.

We next convolve the natural lineshape with a Gaussian function to obtain our ansatz for the measured signal lineshape
\begin{align}
\mathcal{S}(m) = \displaystyle \int_{m-5\sigma}^{m+5\sigma}\tilde{\mathcal{S}}(m') \frac{e^{-\frac{(m-m')^2}{2\sigma^2}}}{\sqrt{2\pi} \sigma}dm',
\label{eqn:phi_lieshape_res}
\end{align}
where $\sigma$ represents the smearing due to the detector resolution. Ideally, $\sigma$ should incorporate any possible $m$ dependence as well. However, since our $\phi$ mass window is narrow enough, we neglect this effect, and treat $\sigma$ as a constant for each fit. However, the resolution can depend on kinematics and this is appropriately accounted for during our signal background separation procedure, by performing independent fits in small phase-space regions (see Sec.~\ref{sec:sig_bkgd}).


\section{\label{sec:sig_bkgd}Signal Background Separation}

The event selection cuts were very effective in cleaning the data sample for both topologies. Further removal of background, non-$\phi p$ events is done by an event-based technique that preserves correlations between all independent kinematic variables~\cite{my-thesis,jinst_williams}. The motivation behind this approach, as opposed to a more conventional sideband-subtraction method, is as follows.

For a reaction with multiple decays, such as in the present case, there are several independent kinematic variables (decay angles, for instance). To perform a background subtraction, one typically bins the data in a particular variable, such as the production angle $\cmangle$. This is because both the background shape and scale can vary widely within the range of the kinematic variable chosen. However, this binning in a single variable generally does not preserve correlations present in the other independent kinematic variables of interest. Therefore, one needs to bin the data in multiple kinematic variables, such that in any particular bin, the background level (both shape and scale) remains roughly the same. Finally, the event-based fits using partial wave amplitudes and Monte Carlo simulation, to calculate the detector acceptance (see Sec.~\ref{sec:acc}), are specifically intended to reproduce the correlations present in the data. Thus, the need for a more sophisticated background separation approach that preserves multi-dimensional correlations in the signal component of the data.

To execute this technique for a given event, first, an $N_c$ number of ``closest neighbor'' events are chosen in the phase space of all independent kinematic variables. $N_c$ is typically of the order of a hundred. These $N_c +1$ events are then fitted to a signal function $s(m)$ plus a background function $b(m)$ using an event-based, unbinned maximum likelihood method (the fit variable $m$ being $M(\kkb)$). Once the functions $s_i(m)$ and $b_i(m)$ have been obtained from this fit for the $i^{th}$ event, the event is assigned a signal quality factor $Q_i$ given by:
\begin{equation}
Q_i = s_i(m_i)/\left( s_i(m_i) + b_i(m_i)\right). 
\label{eqn:qval_defn}
\end{equation}
The $Q$-factor is then used to weigh the event's contribution for all subsequent calculations. In particular, the signal yield in a kinematic bin with $N$ events is obtained as
\begin{equation}
\mathcal{Y} = \sum\limits_{i}^{N} Q_i.
\end{equation}

\subsection{Specific issues for the $\phi$ channel}
\label{sec:phi_resolution_discussion}

The above-mentioned procedure has already been employed for several other photoproduction channels~\cite{omega_prc, omega_pwa, eta_prc, klam_prc, ksig_prc} from the same dataset, with excellent results. For the $\phi$, however, there are some unique issues which need to be carefully dealt with. The main source of the problem is that the $\phi$ mean mass ($m_0 \approx 1.019$~GeV) is very close to the $\kkb$ threshold ($\approx 0.99$~GeV). Therefore, conventional side-band subtraction is difficult to perform, since there are fewer events on the low mass side. Furthermore, the natural width of the $\phi$ ($\Gamma_0 \approx 4.26$~MeV) is non-negligible, leading to two complications. First, there are theoretical difficulties associated with determining the exact lineshape when multiple channels (in this case, the $\phi(1020)$ and $f_0(980)$) open up close to each other near threshold~\cite{flatte}. Second, $\Gamma_0$ is of the same order as the detector resolution ($\mathcal{O}(1~\text{MeV})$). The convolution of the resolution function with an asymmetric natural lineshape can result in a complicated measured lineshape.

Previous experiments often used a Gaussian function to represent the $\phi$ signal. In this analysis, we use an asymmetric relativistic Breit-Wigner appropriately convolved with a Gaussian (as described in Sec.~\ref{sec:lineshape}) to better characterize the measured $\phi$ lineshape. The next issue involves the background function shape and scale. The general scheme found in previous $\phi$ analyses~\cite{anciant, barth, mibe} has been to assume various ``template'' shapes arising from an underlying $\kkb$ phase-space, $a_0/f_0$ and the $K^+\Lambda^\ast$ channels (for the charged-mode). Several problems arise from this approach. First, all
three of the above physics backgrounds correspond to a $p \kkb$ final-state topology, which in turn implies that the particle identification procedure has cleanly separated any pion leakage, a highly improbable prospect. Second, the use of the $a_0/f_0$ or $K^+\Lambda^\ast$ template shapes require a good understanding of the production mechanism of these channels themselves, which we do not have as yet. This is especially true of the $S$-wave $a_0/f_0$, which remains poorly understood. Third, we have found that the background underneath the $\phi$ is dependent on the phase-space region one is examining, and fits to cumulative yields are almost certainly bound to be incorrect. The only proper way of performing signal-background separation for the $\phi$ is to bin the data in all independent kinematic variables $\{\sqrt{s}, \cmangle, \cos \zeta, \varphi\}$ ($\zeta$ and $\varphi$ are the polar and azimuthal decay angles for $\phi \to \kkb$) and perform independent fits in each phase-space volume.

\begin{figure}
\begin{center}
\includegraphics[width=3.4in]{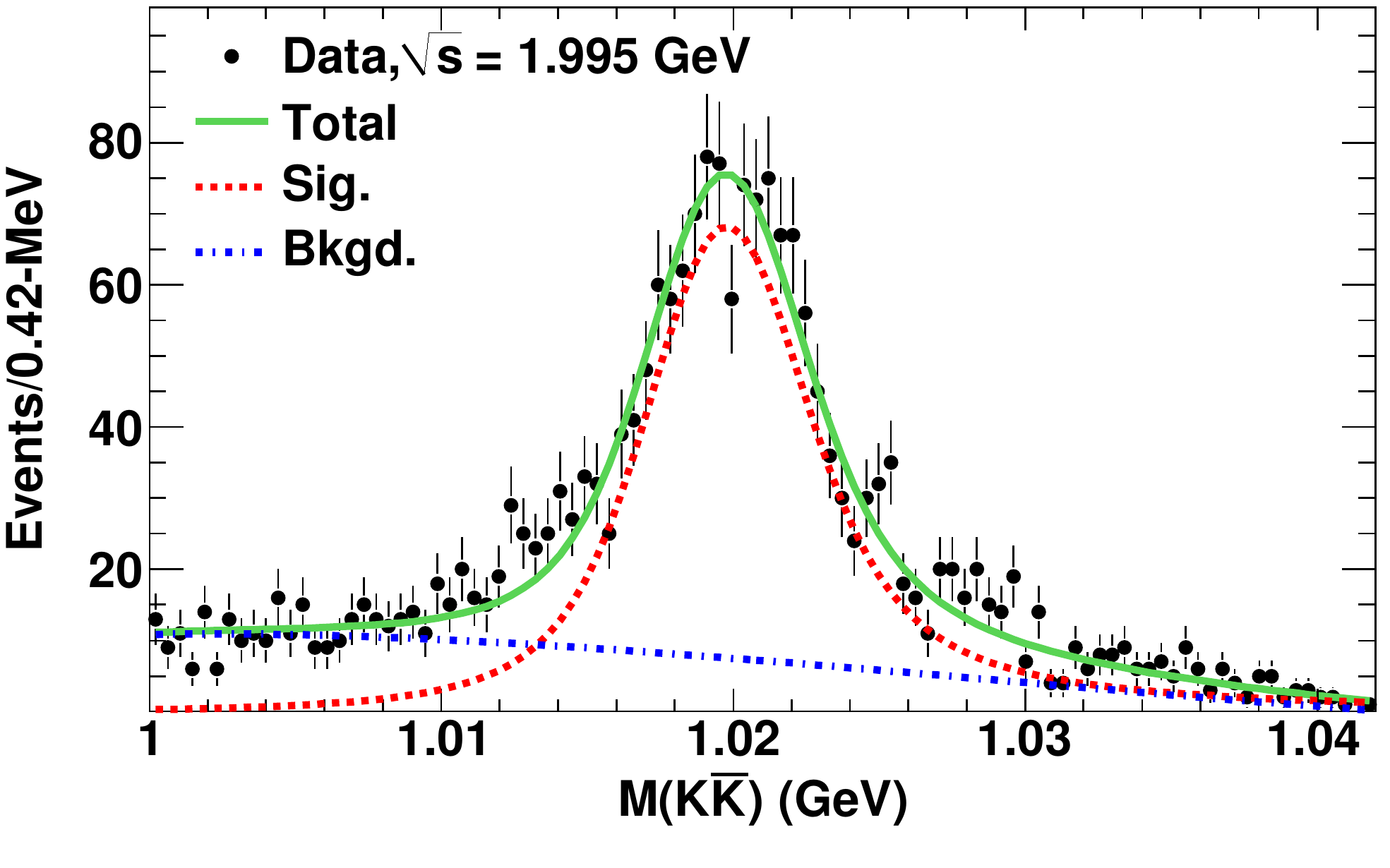} 
\caption[]{(Color online) Sample global fit for in the charged-two-track dataset in the near-threshold $\sqrt{s} = 1.995$~GeV bin. The phase-space suppression in the high-mass side is clearly visible. \label{fig:sample_global_fit_1995_ch2tr}}
\end{center}
\end{figure}

\begin{figure}
\begin{center}
\includegraphics[width=3.3in]{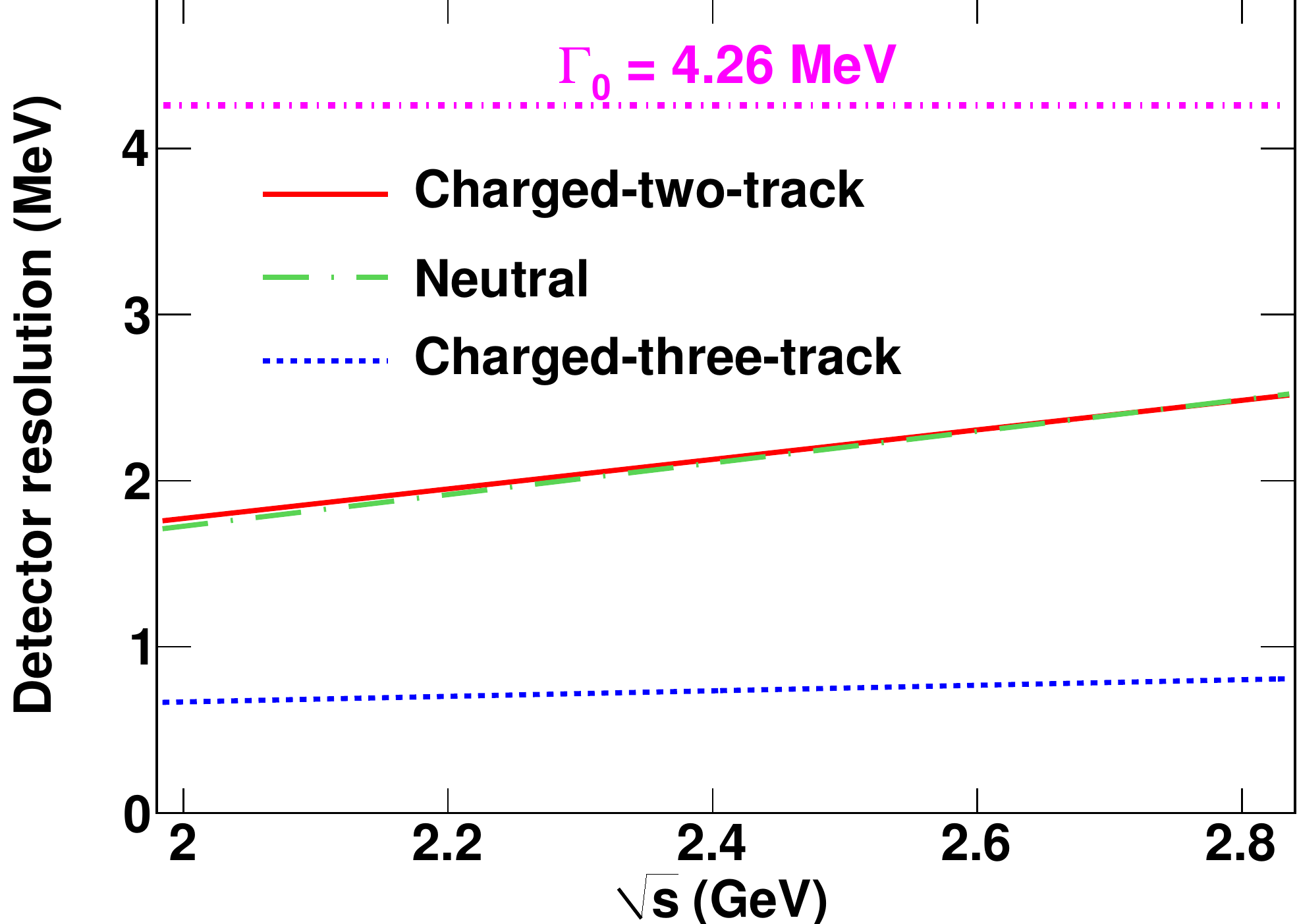} 
\caption[]{(Color online) Energy-dependence of the detector resolution $\sigma$ in Eq.~\ref{eqn:phi_lieshape_res} extracted from global fits in $\sqrt{s}$-bins for the three topologies. These should be compared to the $\phi$ natural line-width $\Gamma_0$.}
\label{fig:detector_res}
\end{center}
\end{figure}

In this work, therefore, we have specifically avoided the use of such templates for the background. Instead, we perform independent fits for each event in a suitably small volume of phase-space denoted by the $N_c$ number of closest-neighbor points. We assume a core linear background function (corresponding to a flat production amplitude), but include the phase-space factors corresponding to the two-body decays $M(\sqrt{s}) \to p \phi$ and $\phi \to \kkb$, as described in Sec.~\ref{sec:lineshape}. The signal lineshape is given by Eq.~\ref{eqn:phi_lieshape_res}.

Fig.~\ref{fig:sample_global_fit_1995_ch2tr} shows a global fit in the near-threshold bin $\sqrt{s} = 1.995$~GeV for the charged-two-track dataset. The dwindling phase-space at higher $M(\kkb)$, applicable to both the signal and background lineshapes is clearly visible here. Fig.~\ref{fig:detector_res} shows the energy-dependence of the detector resolution term $\sigma$ in Eq.~\ref{eqn:phi_lieshape_res}, extracted via such global fits (integrated over all angles) in each $\sqrt{s}$ bin. The resolutions worsen with increasing track momenta at higher $\sqrt{s}$. The charged-two-track and neutral topologies have almost the same resolution in the entire $\sqrt{s}$ range, while the charged-three-track has a markedly better resolution due to all final-state tracks being detected directly, which enables the use of the 4-$C$ kinematic fit. Since the detector resolution and the $\phi$ natural linewidth are of same order of magnitude, the measured signal lineshape is highly sensitive to the resolution. By improving the reconstructed $\phi$ mass resolution, kinematic fitting plays an important role in this analysis.

We note that since fits are done for individual events in small phase-space bins, the assumed linearity of the background is local. The start values of the signal and background lineshapes are taken from the global fits, but, event-by-event, both the signal and the background functions in Eq.~\ref{eqn:qval_defn} can be different. Fits with several different values of $N_c$ were tried out and were seen to give stable results. Our final results are present with $N_c = 100$ and $\phi$ mass range $M(\kkb) \in [1.0,1.06]$~GeV.

\subsection{Results}

\begin{figure}
\hspace{-0.33cm}\includegraphics[width=3.5in]{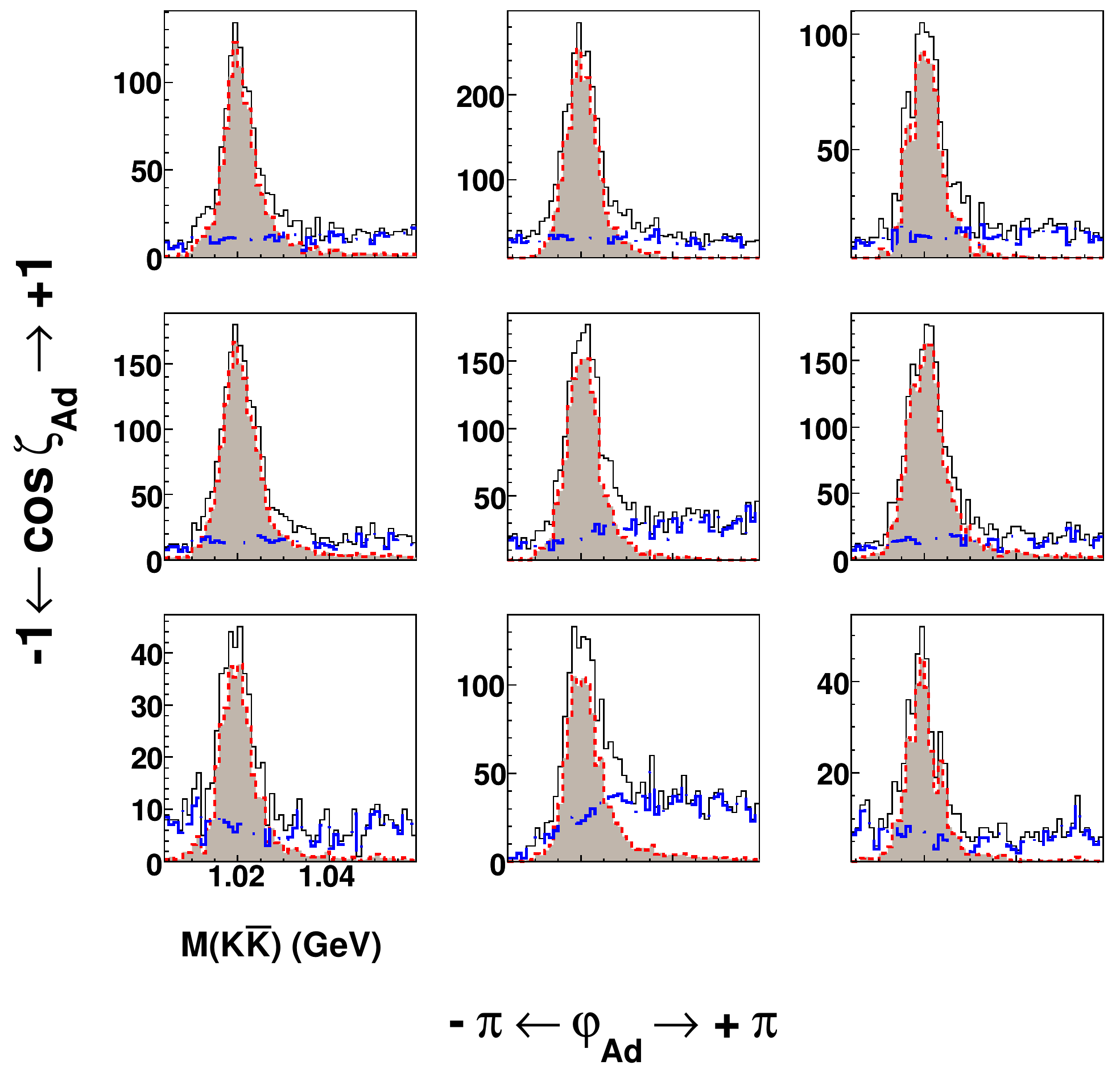} 
\caption[]{(Color online) Signal-background separation quality checks for the charged-two-track topology in the $\sqrt{s} \in [2.12,2.15]$~GeV and $\cmangle > 0.33$ kinematic regime, further broken up into nine phase-space volumes in the $\phi \to \kkb$ decay angles in the Adair (Ad) frame. The red dashed and blue dot-dashed histograms represent the extracted signal and background components, respectively. The $y$-axis units are number of events per 1-MeV-wide $M(\kkb)$  bin.}
\label{fig:W2120-2150_phase_space_quality_checks}
\end{figure}

Fig.~\ref{fig:W2120-2150_phase_space_quality_checks} shows the signal-background separation quality checks for the charged-two-track topology in the $\sqrt{s} \in [2.12,2.15]$~GeV and $\cmangle > 0.33$ kinematic regime. Events are further categorized into nine phase-space volumes in the $\phi \to \kkb$ decay angles. The red dashed and blue dot-dashed histograms represent the extracted signal and background components, respectively. The $y$-axis units are number of events per 10-MeV-wide $\sqrt{s}$ bin. The dependence of the background (both in shape and size) on phase-space is clearly borne out. Fig.~\ref{fig:W2120-2150_phase_space_quality_checks} corresponds to the forward-angle regime in $\cmangle$. The background levels were found to be different in the mid- and backward-angles. This strong phase-space dependance arises because the composition and dynamics of the background components vary in phase-space. Similar checks were performed in other kinematic regions and topologies as well.

\subsection{Further discussion}

Seen in one way, there is some difference in philosophy between our approach and that in some of the previous analyses -- instead of ``subtracting away the background'', we are ``pulling out the signal''. Furthermore, by performing independent fits in very small regions of phase-space for each event (where the background shape is assumed to be roughly constant), we are not making any a priori guesses about the global features of the background. Even if there is an $f_0/a_0$ or $\Lambda^\ast$ underneath the $\phi$, unless this background interferes strongly with the $\phi$, our method should work properly and provide a much better handle on the systematics of the signal-background separation process than in the older methods. While there is certainly an $S$-wave underneath the $\phi$, the extent of the $S$-wave contribution is usually estimated to be small, the presently accepted value being at the percent level~\cite{sp-wave_bibrzycki,sp-wave_fries}. We leave the $S$-$P$-wave interference issue as beyond the scope of this work.


\section{\label{sec:acc}Detector Acceptance}

Detector efficiency was modeled using GSIM, a GEANT-based simulation package of the CLAS detector. A Monte Carlo sample containing $10^8$ $\gamma p \rightarrow \phi p$ events were pseudo-randomly generated according to phase-space distributions and allowed to propagate through the simulation. The simulator also handled the decay of the $\phi$ into the $\kkb$ charged- and neutral-modes according to the corresponding branching fractions. An additional momentum smearing algorithm was applied to better match the resolution of the Monte Carlo with the real data. After processing, the ``raw'' ({\em i.e.}, original phase space generated) events yielded a set of ``accepted'' Monte Carlo events. The ``accepted'' Monte Carlo data then underwent the exact same series of event reconstruction, analysis cuts and energy-momentum correction steps as applied to the real data events.

To account for the characteristics of the event trigger used in this experiment, an additional correction for the Monte Carlo came from a trigger efficiency study using the $\gamma p \to p \pi^+ \pi^-$ channel (see Refs.~\cite{my-thesis,bellis_trigger_clas_note} for details). This study computed the probability that an individual particle trajectory did not fire the trigger, when the reaction kinematics strongly demanded (via total missing mass) that the particle should have been there. The average effect of this correction was found to be 5-6$\%$.

To form a more accurate characterization of the detector acceptance pertaining to the kinematics of the reaction of interest, one should use a Monte Carlo event generator based on a physics model, instead of a simple phase-space generator. Typically, this is achieved in an iterative fashion; one starts with phase-space generated Monte Carlo events, extracts the differential cross sections, fits these cross sections to a model and uses the model to generate new Monte Carlo events for the next iteration. After several such iterations, the accepted Monte Carlo and data distributions are expected to resemble each other to a fair degree.

However, the above procedure assumes that one has a very good handle on the signal-background separation. For a complicated reaction with multiple decay angles, the detector acceptance can depend on several kinematic variables and it becomes more difficult to disentangle the effect of the detector acceptance on signal events from that on the background. Our signal-background separation procedure, as described in the previous section, specifically addresses this issue. By weighting every event by its $Q$-value, we are able to produce distributions of any particular kinematic variable that include only signal events.

In the next step, we expand the scattering amplitude $\mathcal{M}$ for the complete reaction $\gamma p \to \phi p$ in a basis of $s$-channel production amplitudes $\mathcal{A}^{J^{P}}$:
\begin{equation}
\mathcal{M}_{\vec{m}}(\vec{x},\vec{\alpha}) \approx \linebreak 
\displaystyle \sum_{J=\frac{1}{2}}^{\frac{11}{2}} \sum_{P=\pm} \mathcal{A}^{J^{P}}_{\vec{m}}(\vec{x},\vec{\alpha}),
\label{eqn:scat_amp_exp}
\end{equation}
where $\vec{m}=\{m_{\gamma}, m_i, m_\phi, m_f\}$ denotes spin projections quantized along the beam direction for the incident photon, target proton, intermediate $\phi$ and final-state proton, respectively. The vector $\vec{x}$ represents the set of kinematic variables that completely describes the reaction, while the vector $\vec{\alpha}$ denotes a set of 56 fit parameters, estimated by a fit to the data distribution using the method of extended unbinned maximum likelihoods. The only assumption made here is that any distribution can be expanded in terms of partial waves (denoted by the spin-parity combination $J^P$). Ideally, one needs to use a ``complete'' basis of such $J^P$ waves, but we found that a ``large-enough'' ($J^P = \frac{1}{2}^{\pm}, \frac{3}{2}^{\pm}, \ldots ,  \frac{11}{2}^{\pm}$) set of waves was sufficient to fit the data very well. The $s$-channel $J^P$ waves were constructed using the relativistic Rarita-Schwinger formalism~\cite{rarita} and numerically evaluated using the {\tt qft++} software package~\cite{qft_pack}. A full description of the amplitude construction and fitting procedure can be found in Refs.~\cite{my-thesis,omega_prc}, but a salient feature was a $Q$-value weighted contribution of each data event to the total negative log likelihood
\begin{eqnarray}
-\ln{\cal L} & = & \sum_{i}^{N_{data}} Q_{i}\,\ln I_{i} + \mathcal{N},
\end{eqnarray}
where the intensity $I_i$ corresponds to the differential rate calculated using the $\mathcal{M}$ amplitudes and the normalization integral $\mathcal{N}$ that ensures that all probabilities are normalized to unity, is numerically calculated using the Monte Carlo. This function is then minimized with respect to the parameters $\vec{\alpha}$ to obtain the fit results.

\begin{figure}
\includegraphics[width=6cm,angle=90]{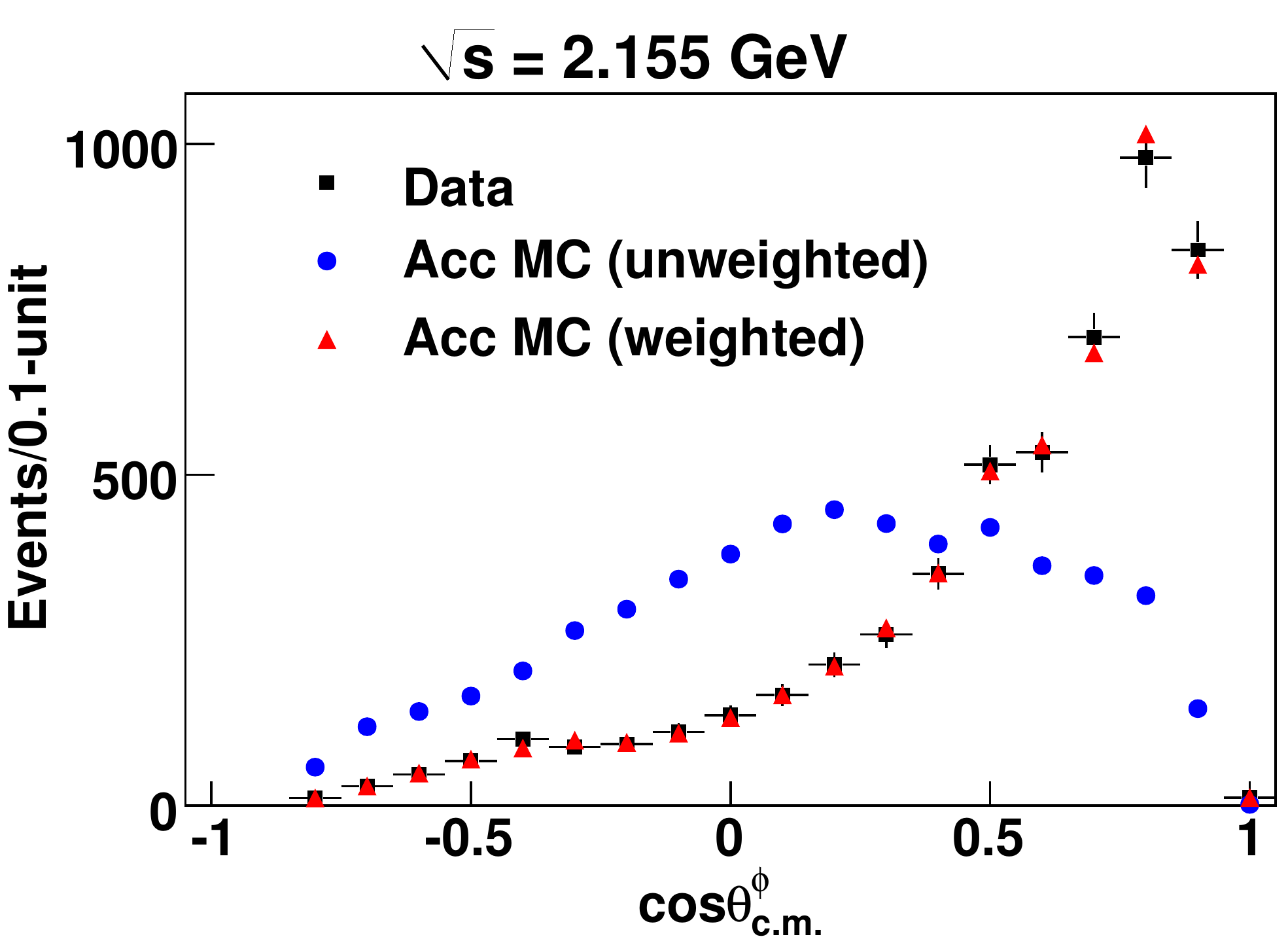}
\caption[]{\label{fig:weighted_acc_1d}
  (Color online)
  Shown are the $\cmangle$ distributions for the data, accepted Monte Carlo and accepted Monte Carlo weighted by the PWA fit in the $\sqrt{s} = 2.155$~GeV bin for the charged-two-track topology. Weighting by the fit results brings the weighted Monte Carlo distribution into excellent agreement with the data.
}
\end{figure}

Based on these fit results, each accepted Monte Carlo event was assigned a weight $I_{i}$ given by,
\begin{equation}
  I_{i} = \displaystyle \sum_{m_{\gamma},m_i,m_f}|\sum_{m_\phi} \mathcal{M}_{\vec{m}}(\vec{x}_{i},\vec{\alpha}) \times \mathcal{M}_{\phi \to \kkb}|^{2} \Delta \Phi_i,
\end{equation}
where we have coherently summed over the intermediate $\phi$ spins and $\Delta \Phi_i$ is the phase-space element. The accepted Monte Carlo, weighted by the fits, matched the data in all physically significant distributions and correlations. Fig.~\ref{fig:weighted_acc_1d} shows comparisons between the data and the accepted MC weighted by the fit results in the production angle for the charged-mode $\sqrt{s} = 2.155$~GeV bin. The detector acceptance as a function of the kinematic variables $\vec{x}$ was then calculated as
\begin{equation}
  \eta_{\,wtd}(\vec{x}) = \left( \displaystyle \sum_{i}^{N_{acc}} I_{i} \right) / \left( \displaystyle \sum_{j}^{N_{raw}}I_{j}\right),
\label{eqn_wtd_acceptance}
\end{equation}
where $N_{raw}$ and $N_{acc}$ denote the number of events in the given kinematic bin for the raw and the accepted Monte Carlo data sets, respectively.

\section{\label{sec:norm}Normalization}

To calculate differential cross sections, the data yields were normalized by the photon flux and the target factors as
\begin{eqnarray}
  \frac{d\sigma}{d \cmangle}(\sqrt{s},\cmangle) = \left( \frac{A_{t}}{\mathcal{F}(\sqrt{s})\rho_{t}\ell_{t}N_{A}} \right) \times \nonumber \\
 \;\;\;\; \frac{\mathcal{Y}(\sqrt{s},\cmangle)}{(\Delta \cmangle)\eta(\sqrt{s},\cmangle)},
\end{eqnarray}
where $A_{t}$, $\rho_{t}$, and $\ell_{t}$ are the target atomic weight, density and length, respectively, $N_{A}$ is the Avogadro constant, $\mathcal{F}(\sqrt{s})$ is the photon flux incident on the target for the given $\sqrt{s}$ bin, $\Delta \cmangle$ is the angular binning width, and $\mathcal{Y}(\sqrt{s},\cmangle)$ and $\eta(\sqrt{s},\cmangle)$ are the number of data events and the acceptance for the given kinematic bin, respectively.

Photon flux normalization for this analysis was carried out by measuring the rate of out-of-time electrons at the photon tagger, that is, hits that did not coincide with any event recorded by CLAS~\cite{gflux}. Corrections were made to account for photon losses along the beam line and the detector dead-time.

A separate correction to the photon flux normalization was required to account for the fact that only the first two-thirds of the photon tagger counters (1-40) went into the trigger. ``Accidental'' events corresponding to tagger counters 41-61 could trigger if a simultaneous hit occurred in the lower (1-40) counters within the same time window. Such ``accidental'' events would be triggered as usual and recorded by CLAS just as any other ``normal'' event. However, the photon flux calculation would not incorporate the associated photon corresponding to an invalid tagger counter. By utilizing the trigger rates in counters 1-40 and assuming a Poisson distribution for the probability of occurrence of such ``accidental'' events, we were able to correct for this feature. Faulty tagger electronics prevented accurate electron rate measurements for photons in the energy bins $\sqrt{s}$~=~2.735~and~2.745~GeV~\cite{my-thesis}. Differential cross sections are therefore not reported at these two energies. However, polarization measurements do not depend on flux normalizations and are reported in these two bins.


\section{\label{sec:syst}Uncertainties}

The statistical uncertainties for the differential cross sections were comprised of the uncertainty in the data yield and the acceptance calculation. For the $i^{th}$ event, the covariance matrix from the signal-background fit described in Sec.\ref{sec:sig_bkgd} gave the uncertainty $\sigma_{Q_i}$ in our estimate of the signal quality factor $Q_i$. Summing up these uncertainties, assuming 100$\%$ correlation for events in a given $(\sqrt{s}, \cmangle)$ bin, the statistical uncertainty in the data yield was given by
\begin{equation}
  \sigma^2_{data} = \mathcal{Y}+ 
  \left(\sum\limits_i^{N_{data}} \sigma_{Qi}\right)^2.
\end{equation}
The relative statistical uncertainty in the acceptance calculation was computed using the expression for binomial errors
\begin{equation}
\delta \eta/\eta =  \sqrt{\frac{1/ \eta - 1}{N_{raw}}}.
\end{equation}

An overall detector acceptance uncertainty between 4~to~6$\%$, depending on $\sqrt{s}$ was estimated based on previous studies~\cite{omega_prc,ksig_prc} on the same dataset. A study of the cross sections for three different reactions ($\omega p$, $K^+ \Lambda$ and $\eta p$) using the same (present) data set in comparison with other experiments gave a flux normalization uncertainty of $7\%$. Data collection for the present experiment occurred in bunches of about 10~million event triggers (called ``runs''). Our estimated photon flux normalization uncertainty from a ``run''-wise comparison of flux-normalized yields was $3.2\%$~\cite{ksig_prc}. Combining these in quadrature with the contributions from photon transmission efficiency ($0.5\%$), a current-dependent normalization ($3\%$) and target density and length ($0.2\%$), we quote an overall normalization uncertainty of $8.3\%$. The other contributions come from the $\phi \to \kkb$ branching fractions ($0.5\%$ and $0.4\%$~\cite{pdg} for the charged- and neutral-modes, respectively) and $\phi$ full-width ($0.9\%$)~\cite{pdg}. A list of all the systematic uncertainties pertaining to $d\sigma /d \cmangle$ measurements for each of the two topologies is given in Table~\ref{table:systematics}.

\begin{table}
  \centering
  \begin{tabular}{lrr} \hline \hline
    \multirow{2}{*}{Source of Uncertainty\;\;\;} & \multicolumn{2}{c}{ Topology} \\
    &  $ K^+ (K^-) p$  &   $\pi^+ \pi^- (K^0_L) p$ \\ \hline
    Particle ID & 4.5\% & 5\% \\
    Kinematic Fitter & 3\% & 3\% \\
    Relative Acceptance & 4\%-6\% & 4\%-6\% \\
    Normalization & 8.3\% & 8.3\% \\
    $\phi \to \kkb$ BF & 0.5\%  & 0.4\%  \\
    $\phi$ Full Width & 0.9\%  & 0.9\%  \\ 
    Overall estimate & 10.7\%-11.6\% \;& \;10.9\%-11.8\%  \\
    \hline \hline
    \end{tabular}
  \caption[]{\label{table:systematics} List of systematic uncertainties for this analysis.}
\end{table}


\section{Spin density matrix elements}

\begin{figure}
 \begin{center}
\includegraphics[width=3.4in]{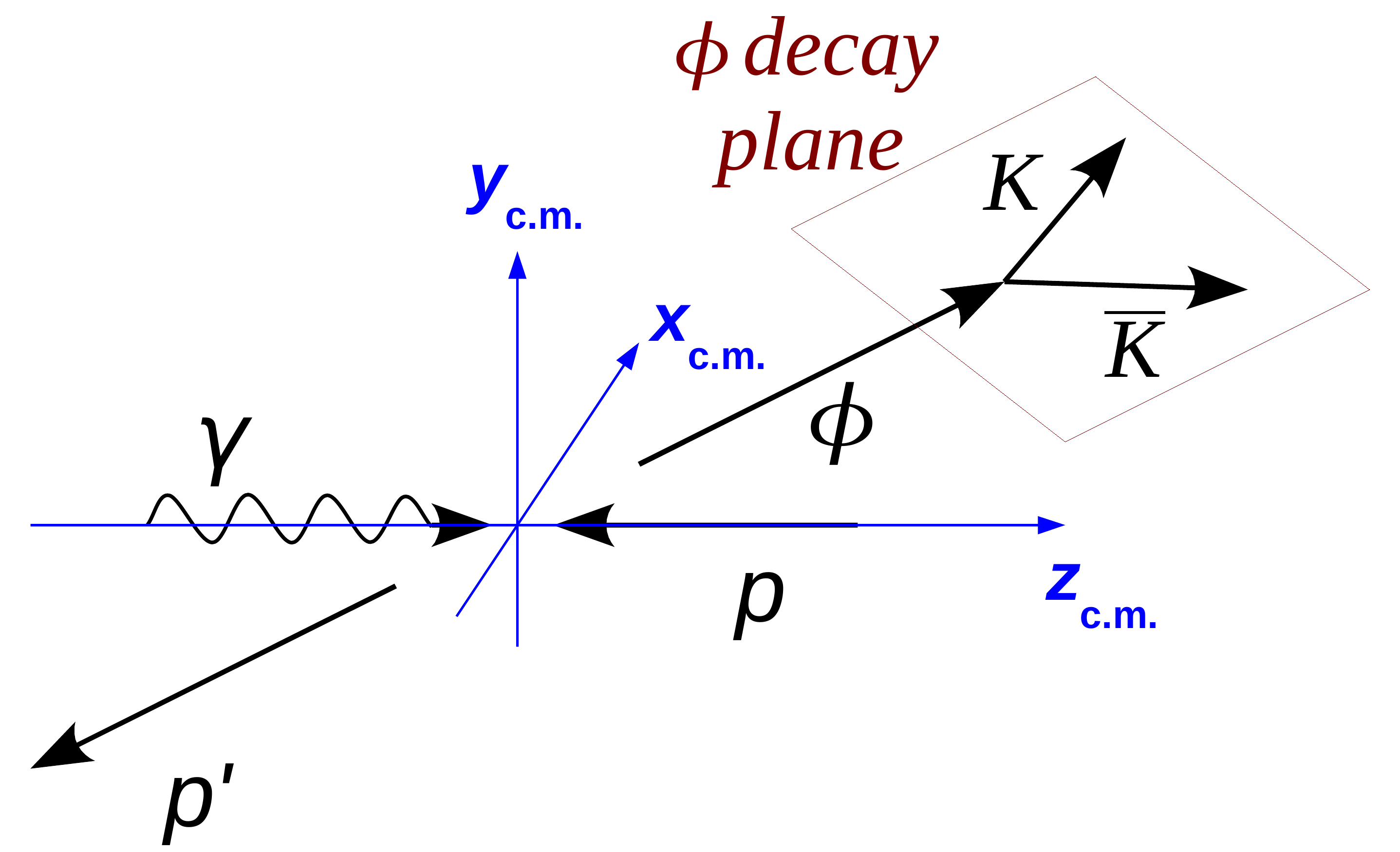} 
\caption[]{(Color online) A schematic diagram of the reaction chain $\gamma p \to \phi (\to \kkb)p'$ in the overall c.m. frame. The beam direction is taken as the positive $z$-axis, and the $y$-axis is normal to the $\phi$ production plane.}
\label{fig:phi_decay_angles}
\end{center}
\end{figure}

A massive vector boson like the $\phi$ meson has three physical spin components. As in Eq.~\ref{eqn:scat_amp_exp}, we take the beam direction as our spin quantization axis, as well as the positive $\hat{z}$ direction. The $\hat{y}$ direction is the normal to the production plane ({\em i.e}, along $\vec{p}_\gamma \times \vec{p}_\phi$), and $\hat{x} = \hat{y} \times \hat{z}$. This is schematically shown in Fig.~\ref{fig:phi_decay_angles}. The three spin operators are
\begin{subequations}
\begin{eqnarray}
   S_x &=& \frac{1}{\sqrt{2}} \left(\begin{array}{ccc} 0&1&0 \\ 1&0&1 \\ 0&1&0 \end{array}\right), \\
   S_y &=& \frac{1}{\sqrt{2}} \left(\begin{array}{ccc} 0&-i&0 \\ i&0&-i \\ 0&i&0 \end{array}\right),\;\; \mbox{and} \\
   S_z &=& \left(\begin{array}{ccc} 1&0&0 \\ 0&0&0 \\ 0&0&-1 \end{array}\right),
\end{eqnarray}
\end{subequations}
and a pure spin-state $|\alpha\rangle$ is an eigenstate of the full $\vec{S}$ operator. For a classical ensemble of states, the spin of the vector particle is described by the density matrix $\rho = \sum w_\alpha |\alpha\rangle \langle \alpha |$, where the sum is over a complete basis of states and $w_\alpha$ is the classical probability of finding the particle in the state $| \alpha \rangle$. For the general case, however, $\rho$ will not be diagonal and the different polarization states will be correlated. A general $3\times3$ complex matrix $\rho$ has $2\times 3^2$ real elements. Hermiticity constrains the diagonal elements of $\rho$ to be real (3 real elements) and the off-diagonal elements to be conjugate transpose of each other (3 complex elements or 6 real elements). The unit trace constraint further reduces the number of independent elements by 1. Therefore, in all, the most general $3\times3$ density matrix will have 8 real and independent elements. A convenient basis to expand the density matrix is given by three rank-1 tensors, $S_i$ ($i = x$, $y$, $z$) and five rank-2 tensors $\tau_{ij}$ given by
\begin{equation}
\tau_{ij} = \frac{3}{2}(S_iS_j + S_j S_i) - 2\delta_{ij}.
\end{equation}
Therefore, by construction, in the tensorial space indexed by the two rank-1 tensors $S_i$ and $S_j$, $\tau_{ij}$ is symmetric and traceless.

The above tensors were written in the Cartesian basis. Following Ref.~\cite{kloet_chiang_tabakin}, we switch to the helicity basis where the spin-1 operators are written as~\cite{pichowsky}
\begin{equation}
S_{1\pm1} = \mp \frac{S_x \pm iS_y}{\sqrt{2}}, \;\;\; S_{10} = S_z.
\end{equation}
Explicitly, they are
\begin{subequations}
\begin{eqnarray}
   S_{10} &=& \left(\begin{array}{ccc} 1&0&0 \\ 0&0&0 \\ 0&0&-1 \end{array}\right) \\ 
   S_{11} &=& -\left(\begin{array}{ccc} 0&1&0 \\ 0&0&1 \\ 0&0&0 \end{array}\right) \\ 
   S_{1-1} &=&  \left(\begin{array}{ccc} 0&0&0 \\ 1&0&0 \\ 0&1&0 \end{array}\right).
\end{eqnarray}
\end{subequations}
In the helicity basis, the rank-2 operators $\tau_{2\mu}$ are given by the tensor products $[S_1\otimes S_1]_\mu$, $\mu = \{0,\pm1,\pm2\}$. Substituting the appropriate Clebsch-Gordan coefficients, the tensor polarization operators are
\begin{subequations}
\begin{eqnarray}
\tau_{22} &=& S_{11}S_{11}\\
\tau_{2-2} &=& S_{1-1}S_{1-1}\\
\tau_{21} &=& \frac{1}{\sqrt{2}} (S_{11}S_{10} + S_{10}S_{11})\\
\tau_{2-1} &=& \frac{1}{\sqrt{2}} (S_{1-1}S_{10} + S_{10}S_{1-1}), \;\; \mbox{and}\\
\tau_{20} &=& \frac{1}{\sqrt{6}} (S_{11}S_{1-1} + 4S_{10}S_{10} + S_{1-1}S_{11}).
\end{eqnarray}
\end{subequations}

For the sake of completeness, we give the explicit form of these five matrices:
\begin{subequations}
\begin{eqnarray}
   \tau_{22} &=& \left(\begin{array}{ccc} 0&0&1 \\ 0&0&0 \\ 0&0&0 \end{array}\right) \\
   \tau_{2-2} &=& \left(\begin{array}{ccc} 0&0&0 \\ 0&0&0 \\ 1&0&0 \end{array}\right) \\ 
   \tau_{21} &=&  -\frac{1}{\sqrt{2}}\left(\begin{array}{ccc} 0&1&0 \\ 0&0&-1 \\ 0&0&0 \end{array}\right) \\
   \tau_{2-1} &=&  -\frac{1}{\sqrt{2}}\left(\begin{array}{ccc} 0&0&0 \\ -1&0&0 \\ 0&1&0 \end{array}\right) \\
   \tau_{20} &=&  \frac{1}{\sqrt{6}}\left(\begin{array}{ccc} 1&0&0 \\ 0&-2&0 \\ 0&0&1 \end{array}\right).
\end{eqnarray}
\end{subequations}
The full expression of the density matrix is then given as
\begin{equation}
\rho = \frac{1}{3} \left[I + \frac{3}{2} \vec{S}\cdot \vec{P} + \sqrt{3} \tau\cdot T \right],
\end{equation}
with the vector polarizations defined as
\begin{equation}
P_{1\pm1} = \mp \frac{P_x \pm iP_y}{\sqrt{2}}, \;\;\; P_{10} = P_z,
\end{equation}
and the sum over the tensor polarizations defined as
\begin{equation}
\tau\cdot T = \displaystyle\sum_{\mu = 0, \pm1, \pm2} (-1)^\mu \tau_{2 -\mu} T_{2\mu}.
\end{equation}
Therefore, the density matrix can be written as
\begin{widetext}
\begin{equation}
\label{eqn:density_matrix_all_pols_full_expression}
\rho_{\lambda \lambda'} =\;\; \left(\begin{array}{ccc} \rho_{-1-1}&\rho_{-10} &\rho_{-11} \\ \rho_{0-1}&\rho_{00} &\rho_{01} \\ \rho_{1-1}&\rho_{10} &\rho_{11} \end{array}\right) =  \frac{1}{3}\left(\begin{array}{ccc} 1 + \frac{3}{2}P_{10} + \sqrt{\frac{1}{2}} T_{20} & -\frac{3}{2}P_{11} + \sqrt{\frac{3}{2}}T_{2-1} &\sqrt{3}T_{2-2} \\ \frac{3}{2}P_{1-1} - \sqrt{\frac{3}{2}}T_{21} &1 - \sqrt{2} T_{20} & -\frac{3}{2}P_{11} - \sqrt{\frac{3}{2}}T_{2-1}  \\ \sqrt{3}T_{22}& + \frac{3}{2}P_{1-1} + \sqrt{\frac{3}{2}}T_{21} &  1 - \frac{3}{2}P_{10} + \sqrt{\frac{1}{2}} T_{20}  \end{array}\right).
\end{equation}
\end{widetext}
For unpolarized beam and target polarizations, parity conservation leads to the condition ${\rho_{\lambda \lambda'} = (-1)^{\lambda - \lambda'} \rho_{-\lambda -\lambda'}}$. Along with the hermiticity property $\rho_{\lambda \lambda'} = \rho^\ast_{\lambda' \lambda}$, Eq.~\ref{eqn:density_matrix_all_pols_full_expression} implies~\cite{kloet_chiang_tabakin} $P_x = P_z = 0$, $T_{20}$, $T_{21}$ and $T_{22}$ be real, $T_{2-1} = -T_{21}$, and $T_{2-2} = T_{22}$, so that the most general form of the density matrix for unpolarized photoproduction is given as (following the sign conventions in Ref.~\cite{schilling})
\begin{equation}
\rho^{0} = \left( \begin{array}{ccc} \frac{1-\rho^0_{00}}{2} & \rho^0_{10}  & \rho^0_{1-1} \\ 
                                 \rho^{0\ast}_{10}        & \rho^0_{00}  & -\rho^{0\ast}_{10} \\
                                 \rho^0_{1-1}          & -\rho^0_{10} & \frac{1-\rho^0_{00}}{2} \end{array} \right),
\label{eqn:schilling_rho}
\end{equation}
where $\rho^0_{00}$ and $\rho^0_{1-1}$ are purely real and only $\rho^0_{10}$ has both real and imaginary parts (the superscript denotes the unpolarized beam-target case). The physical interpretation of the SDME's are~\cite{kloet_chiang_tabakin} $P_y = -2\sqrt{2} Im \rho^0_{10}$, $T_{20} = \frac{1}{2} (1 - 2 \rho^0_{00})$, $T_{21} = - \sqrt{6} Re \rho^0_{10}$ and $T_{22} = \sqrt{3} \rho^0_{1-1}$.

Even though the density matrix given by Eq.~\ref{eqn:schilling_rho} consists of four real independent observables, for the vector meson decaying into pseudoscalar mesons ($\phi \to \kkb$, $\rho \to \pi \pi$ or $\omega \to \pi \pi \pi$), there are only three measurable quantities. For these decays, the intensity distribution is given by the Schilling's equation~\cite{schilling}
\begin{eqnarray}
\mathcal{I} (\sqrt{s},\cmangle) &\sim& \frac{1}{2}(1 - \rho^0_{00}) + \frac{1}{2} (3 \rho^0_{00} -1) \cos^2\zeta \nonumber \\
         && \;\;\;\;\;\;- \sqrt{2} Re \rho^0_{10} \sin 2\zeta \cos \varphi \nonumber \\ && \;\;\;\;\;\;- \rho^0_{1-1} \cos 2\varphi,
\label{eqn:schilling}
\end{eqnarray}
where $\zeta$ and $\varphi$ are the polar and azimuthal angles of the vector meson decay into pseudoscalar mesons (see Eq.~\ref{eqn:phi2kk_angles}), and we have explicitly retained the $(\sqrt{s},\cmangle)$ dependence. Since Eq.~\ref{eqn:schilling} does not include $Im \rho^0_{10}$, $P_y$ is not measurable. Since $P_x$ and $P_z$ are also constrained to be 0 for unpolarized beam-target configurations, the vector polarization $\vec{P}$ is not measurable at all. Kloet {\em et al.}~\cite{kloet_chiang_tabakin} have shown that the only way to measure the vector polarization is through leptonic decays of the vector mesons, with the additional requirement that one of the lepton spins also be measured.

\subsection{Helicity conservation and choice of reference frames}
\label{sec:wigner_rotations}

\begin{figure}
 \begin{center}
\includegraphics[width=3.4in]{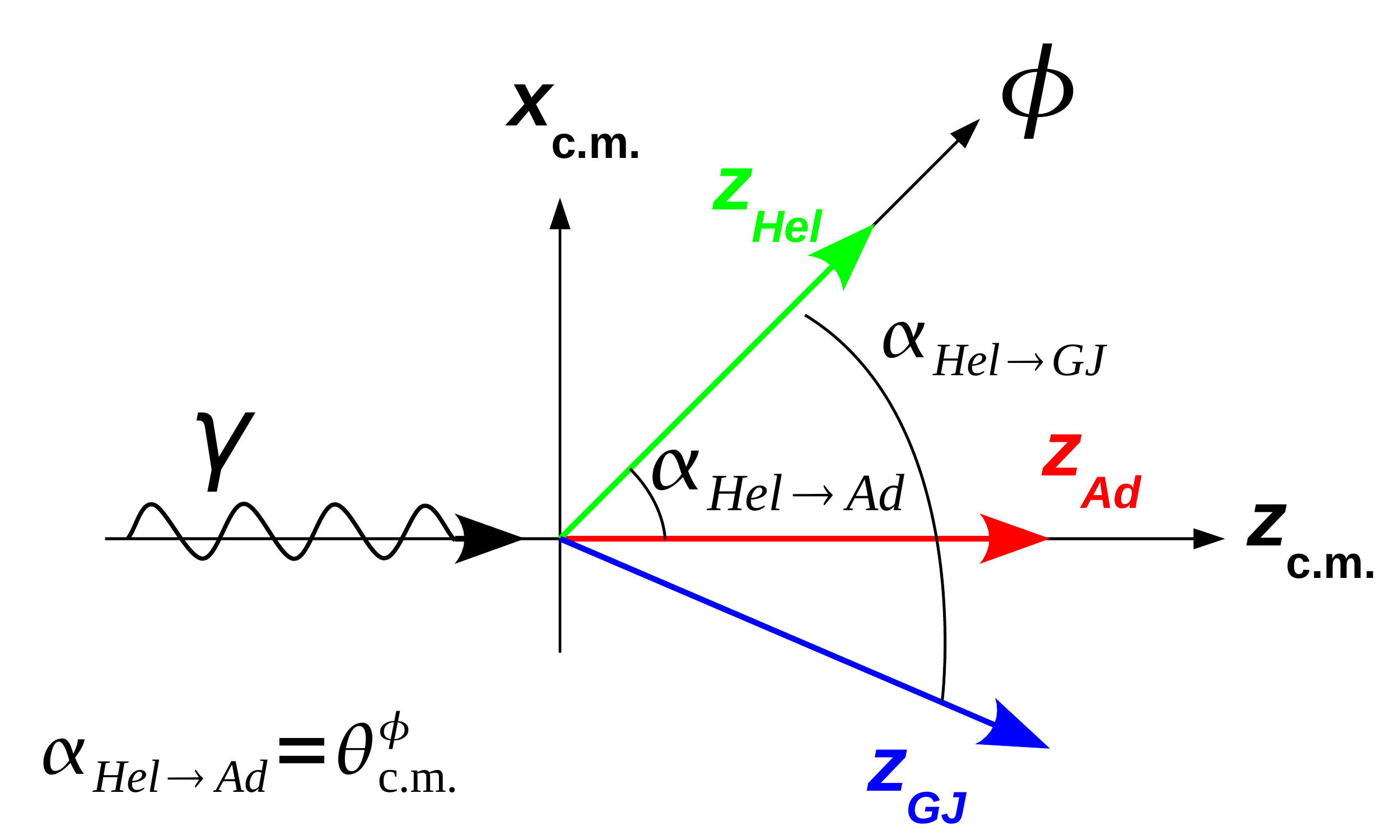} 
\caption[]{(Color online) The spin-quantization axes for the Helicity (Hel, in green), Adair (Ad, in red) and Gottfried-Jackson (GJ, in blue) frames, in relation to the overall c.m. frame. The $z$-axis for the c.m. frame points along the beam direction and coincides with $z_{Ad}$. Since $z_{Hel}$ points along the direction of the $\phi$ meson, the angle between the Helicity and Adair frames is $\phith$. The Gottfried-Jackson frame is defined as the direction of the incoming photon, as seen in the rest frame of the $\phi$ meson. The angle between the Helicity and Gottfried-Jackson frame is given by Eq,~\ref{eqn:phi_hel_ad_gj}b.}
\label{fig:phi_hel_ad_gj}
\end{center}
\end{figure}

The choice of the reference frame for the two decay angles in the intensity distribution of Eq.~\ref{eqn:schilling} can be made to emphasize various exchange mechanisms. The reaction is shown in the c.m. frame in Fig.~\ref{fig:phi_decay_angles}. However, since the SDME's are not Lorentz invariant quantities, an analyzing direction for the vector-meson must be chosen. Three common choices exist in the literature, the Adair frame, the Helicity frame, and the Gottfried-Jackson frame, as shown in Fig.~\ref{fig:phi_hel_ad_gj}. In the Adair (Ad) frame, the polarization axes for both the incoming and outgoing states are chosen as the $z$-axis (along the beam direction).

The Adair frame is convenient when the production mechanism conserves spin in the $s$-channel c.m. frame. For the Helicity (Hel) frame, the vector meson direction in the c.m. system defines the quantization axis. This is preferred for $s$-channel helicity conservation (SCHC). Under the assumptions of SCHC, $\rho^{Hel}_{00} = \rho^{Hel}_{10} = \rho^{Hel}_{1-1} = 0$~\cite{gilman}. For the Gottfried-Jackson (GJ) frame, one makes a further boost to the vector meson rest-frame from the overall c.m. frame. The quantization axis is along the direction of the incoming photon seen in the vector meson rest-frame. For a $t$-channel exchange of $X$, the momentum of the incoming photon and $X$ is collinear in the GJ frame. Therefore the $\rho$ elements measure the degree of helicity flip due to the $t$-channel exchange of $X$ in the GJ frame. For example, if the $t$-channel exchange particle is a $J^P = 0^+$ state, then no helicity flip will occur (TCHC) and the vector meson will have the same helicity as the incoming photon. For this case $\rho^{GJ}_{00} = \rho^{GJ}_{10} = \rho^{GJ}_{1-1} = 0$~\cite{gilman}. The quantization axes for these three frames are shown in Fig.~\ref{fig:phi_hel_ad_gj}.

It is clear that knowing the $\rho$ elements in one frame, one can immediately calculate them in any other frame by a Wigner rotation. The $y$-axis is always the normal to the vector meson production plane; $\hat{y} = \hat{k} \times \hat{q}/|\hat{k} \times \hat{q}|$, where $\hat{k}$ is the incoming photon direction and $\hat{q}$ is the outgoing vector meson ($\phi$) direction. The choice of the $z$-axis is frame dependent, as described above. For the Adair frame $\hat{z} = \hat{k}$, for the Helicity frame, $\hat{z} = \hat{q}$, and for the GJ frame, $\hat{z} = \hat{k'}$, where $\hat{k'}$ points along the incoming photon direction in the vector meson rest frame. Once the $y$- and the $z$-axis have been fixed, $\hat{x} = \hat{y} \times \hat{z}$. Let $\hat{\pi}$ be the direction of the daughter $K$ (for $\phi \to \kkb$) in the chosen reference frame. Then the angles $\zeta$ and $\varphi$ in Eq.~\ref{eqn:schilling} are given as~\cite{schilling}:
\begin{equation}
\label{eqn:phi2kk_angles}
\cos \zeta = \hat{\pi} \cdot \hat{z},\; \cos \varphi = \frac{\hat{y} \cdot \left( \hat{z} \times \hat{\pi} \right)}{|\hat{z} \times \hat{\pi}| }, \;\sin \varphi = -\frac{\hat{x} \cdot \left( \hat{z} \times \hat{\pi} \right)}{|\hat{z} \times \hat{\pi}| }.
\end{equation}
In the Rose convention of the signs (this is followed by Schilling in Ref.~\cite{schilling}), the Wigner rotation matrix for a spin-1 system by an angle $\alpha$ is
\begin{equation}
   d^1(\alpha) = \left(\begin{array}{ccc} \frac{1}{2} (1 + \cos \alpha) & -\frac{1}{\sqrt{2}} \sin \alpha  & \frac{1}{2} (1 - \cos \alpha)  \\ \frac{1}{\sqrt{2}} \sin \alpha &\cos \alpha &  -\frac{1}{\sqrt{2}} \sin \alpha\\  \frac{1}{2} (1 - \cos \alpha) &\frac{1}{\sqrt{2}} \sin \alpha  & \frac{1}{2} (1 + \cos \alpha) \end{array}\right).
\end{equation}
To rotate the density matrix from reference frame $A$ to $B$, the transformation is
\begin{equation}
\rho^B = d^1(-\alpha_{A\to B}) \rho^A d^1(\alpha_{A\to B}).
\end{equation}
The rotation angles (counter-clockwise is positive) are given by~\cite{schilling}
\begin{subequations}
\label{eqn:phi_hel_ad_gj}
\begin{eqnarray}
\alpha_{Ad \to Hel} &=& \phith\\
\alpha_{Hel \to GJ} &=& - \cos^{-1}\left(\frac{\beta - \cmangle}{ \beta  \cmangle -1} \right)\\
\alpha_{Ad \to GJ} &=&  \alpha_{Ad \to Hel} + \alpha_{Hel \to GJ} ,
\end{eqnarray}
\end{subequations}
where $\beta = |\vec{p}_K |/E_K$ is the velocity of the daughter kaon in the $\phi$ rest frame (for the $\phi \to \kkb$ decay).

\subsection{``PWA'' method and ``Schilling's'' method of SDME extraction in the Adair frame}

The expansion of the production amplitudes using partial wave analysis (PWA) techniques in Sec.~\ref{sec:acc} allows for an elegant way of extracting the SDME's. For this, we follow Schilling~\cite{schilling} and express the SDME's in terms of the production amplitudes as
\begin{equation}
\rho_{m_\phi m'_\phi} =  \frac{\displaystyle\sum_{m_f m_\gamma m_i} \mathcal{M}_{m_\phi m_f m_i m_\gamma} \mathcal{M}^\ast_{m'_\phi m_f m_i m_\gamma} }{ \displaystyle\sum_{m_\phi m_f m_\gamma m_i} |\mathcal{M}_{m_\phi m_f m_i m_\gamma}|^2},
\label{eqn:sdme_adair}
\end{equation}
where $\mathcal{M}_{m_\phi m_f m_i m_\gamma}$ are the same amplitudes as in the PWA fit in Eq.~\ref{eqn:scat_amp_exp}, and $m_\gamma$, $m_i$ and $m_f$ and $m_\phi$ are the spins of the incoming photon, target proton, outgoing proton and the $\phi$ vector meson, respectively. Note that the $\phi \to \kkb$ decay portion of the full amplitude in Eq.~\ref{eqn:scat_amp_exp} occurs as a constant factor that cancels between the numerator and the denominator in Eq.~\ref{eqn:sdme_adair}. The $\phi$ decay portion of the full amplitude in Eq.~\ref{eqn:scat_amp_exp} can therefore be suppressed for SDME extraction.

The above ``PWA'' method is completely equivalent to a direct application of the Schilling's expression for the intensity given by Eq.~\ref{eqn:schilling} (``Schilling's'' method). Since the spin-quantization axis for our PWA amplitudes was along the beam direction, the PWA method of extraction yields results in the Adair frame. The PWA expansion was specifically tuned to represent distributions in all kinematic variables, in particular, the intensity distribution given by Eq.~\ref{eqn:schilling}. The equivalence between the two methods were demonstrated previously~\cite{my-thesis,omega_prc}. The final results for the SDME's we present in this analysis use the PWA method.


\section{Results}
\label{sec:results}

\subsection{Differential cross section results}

\begin{figure*}
\includegraphics[width=6.5in]{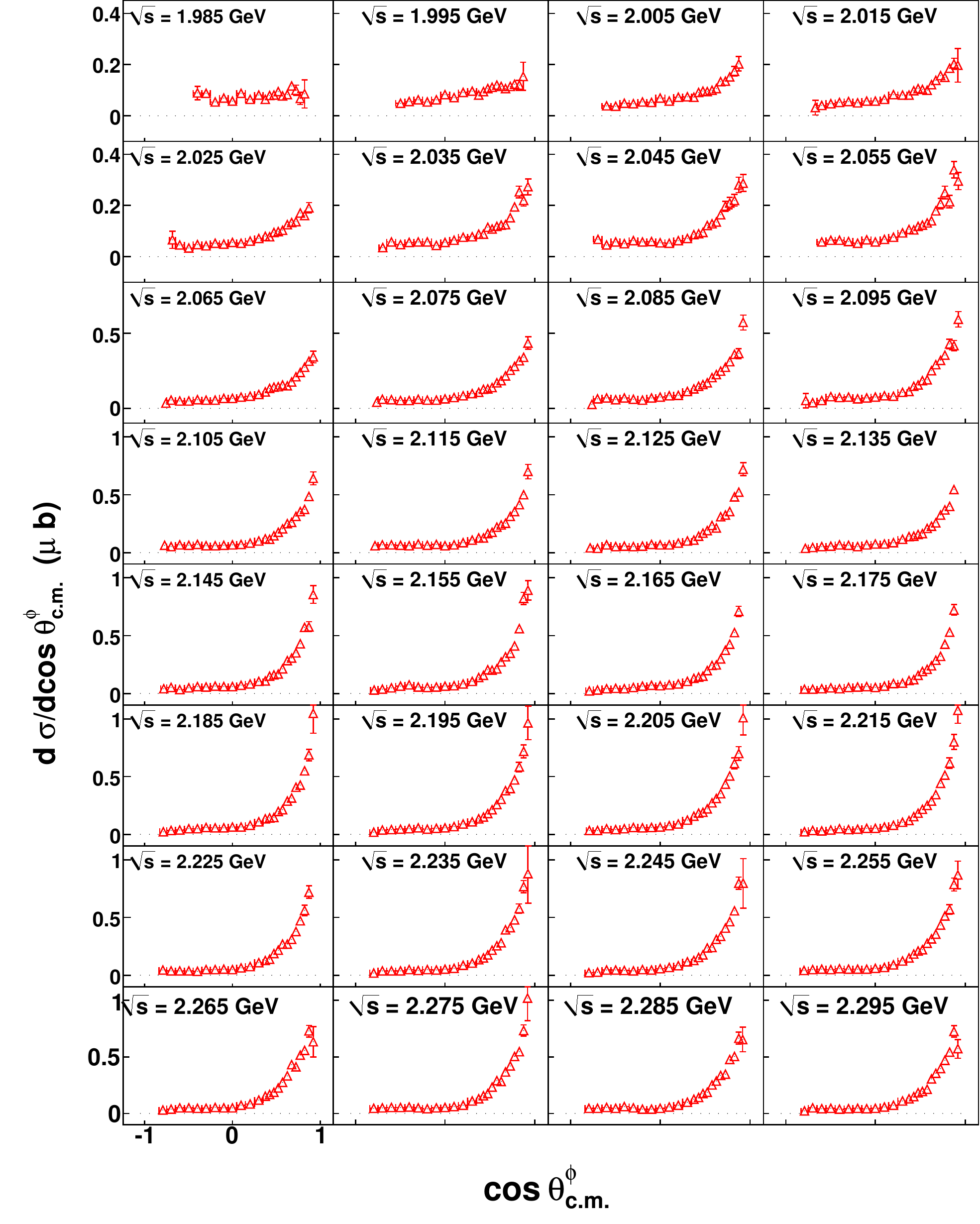} 
\caption[]{\label{fig:dsig_charged_1} (Color online) $\frac{d\sigma}{d\cmangle}$ ($\mu$b) {\em vs.} $\cmangle$: Differential cross section results for the charged-mode topology in the energy range 1.98~GeV~$\leq \sqrt{s} <$~2.3~GeV. The centroid of each 10-MeV-wide bin is printed on the plots. The $y$-axis range is constant over each horizontal row and is shown in the left-most column for every row. All error bars represent statistical uncertainties only.
}
\end{figure*}


\begin{figure*}
\includegraphics[width=6.5in]{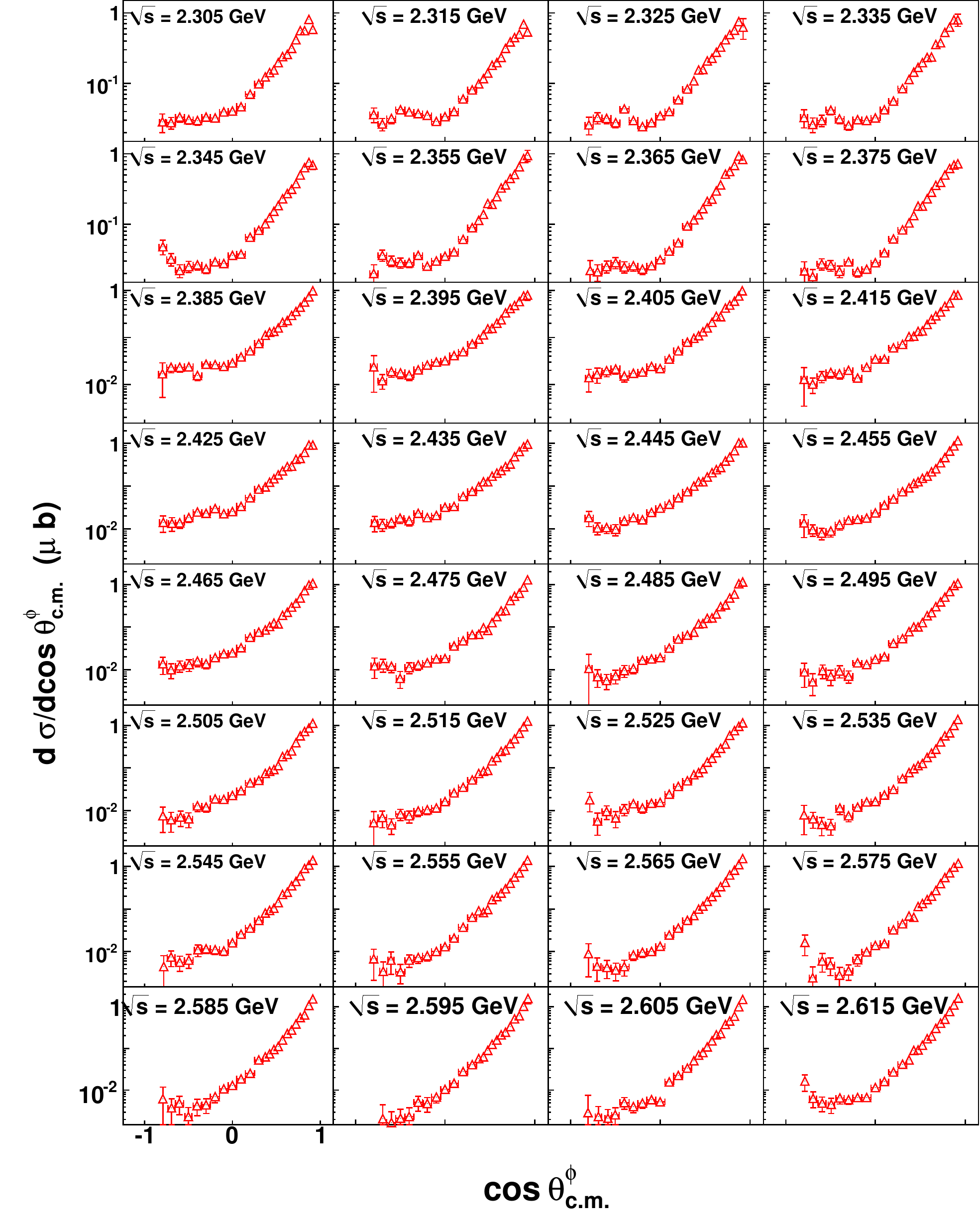} 
\caption[]{\label{fig:dsig_charged_2} (Color online) $\frac{d\sigma}{d\cmangle}$ ($\mu$b) {\em vs.} $\cmangle$: Differential cross section results for the charged-mode topology in the energy range 2.3~GeV~$\leq \sqrt{s} <$~2.62~GeV. The centroid of each 10-MeV-wide bin is printed on the plots. The $y$-axis range is constant over each horizontal row and is shown in the left-most column for every row. All error bars represent statistical uncertainties only.
}
\end{figure*}

\begin{figure*}
\includegraphics[width=6.5in]{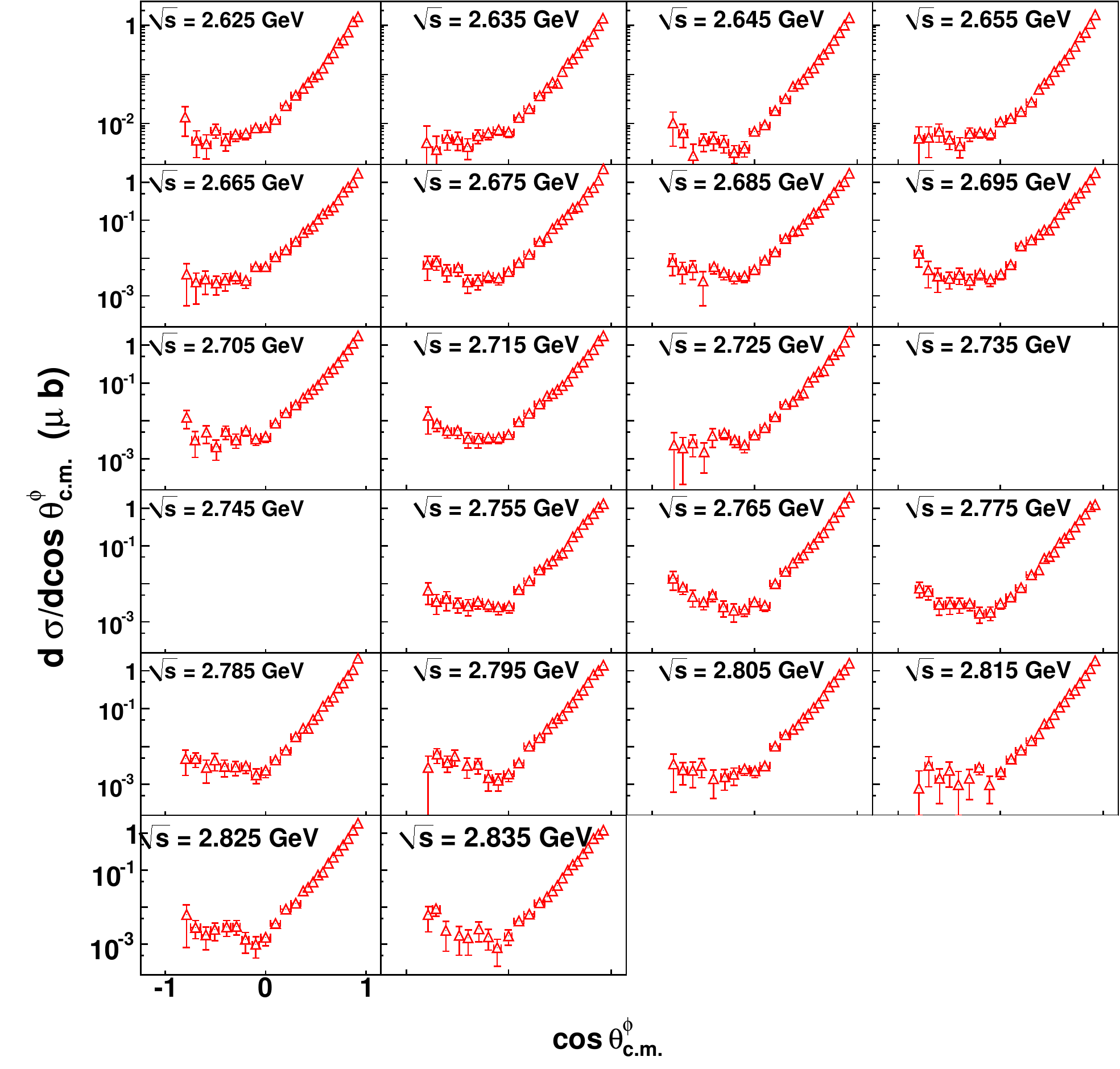} 
\caption[]{\label{fig:dsig_charged_3} (Color online) $\frac{d\sigma}{d\cmangle}$ ($\mu$b) {\em vs.} $\cmangle$: Differential cross section results for the charged-mode topology in the energy range 2.62~GeV~$\leq \sqrt{s} <$~2.84~GeV. The centroid of each 10-MeV-wide bin is printed on the plots. The $y$-axis range is constant over each horizontal row and is shown in the left-most column for every row. No results are presented for the bins $\sqrt{s} = 2.735$ and 2.745~GeV due to the normalization issues, as described in Sec~\ref{sec:norm}. All error bars represent statistical uncertainties only.
}
\end{figure*}

\begin{figure*}
\includegraphics[width=6.5in]{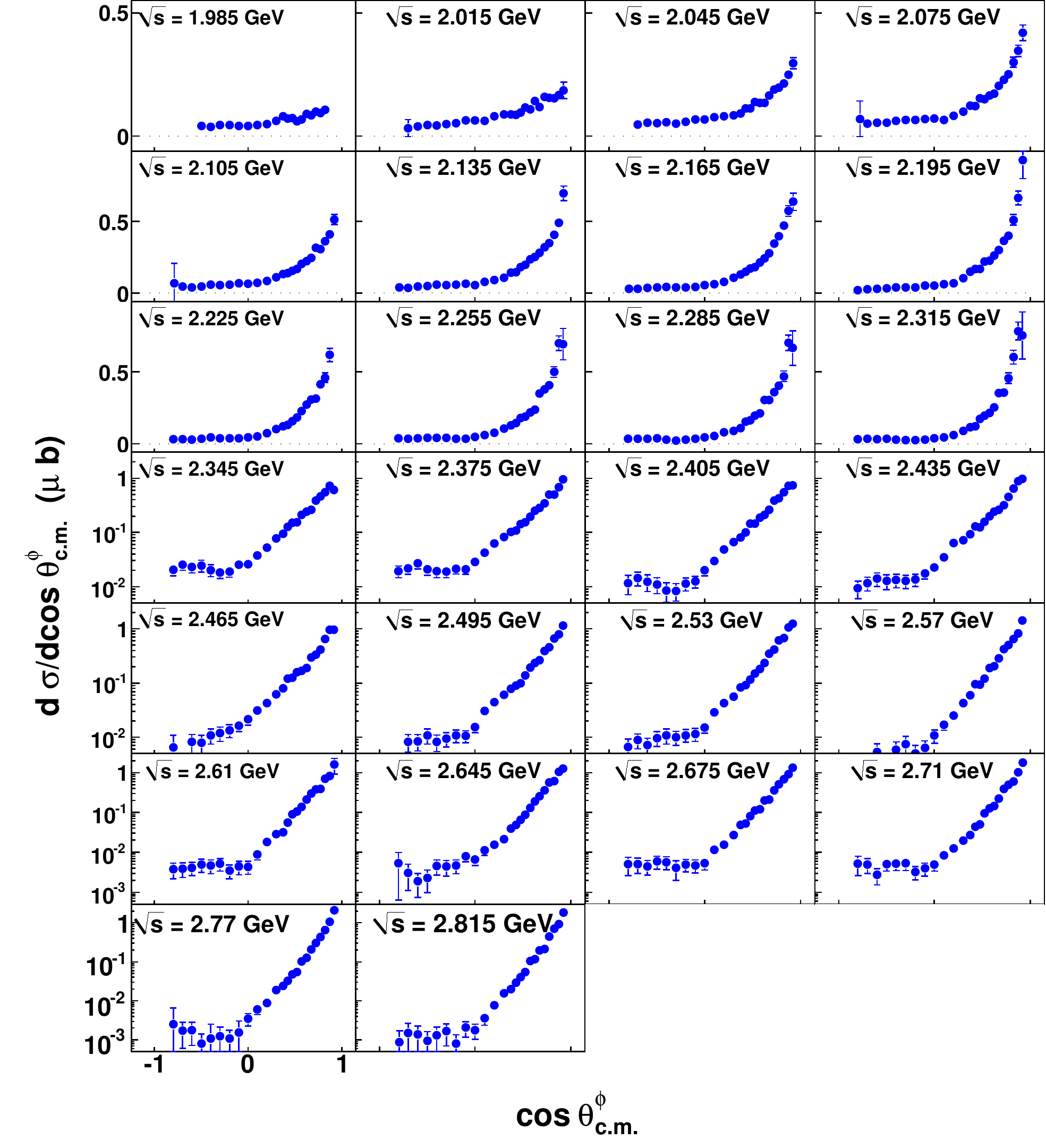} 
\caption[]{\label{fig:dsig_neutral}
  (Color online)
  $\frac{d\sigma}{d\cmangle}$ ($\mu$b) {\em vs.} $\cmangle$: Differential cross section results for the neutral-mode topology. The minimum(maximum) bin-width is 30-MeV(50-MeV) and the bin-centroid is printed on the plots. The $y$-axis range is constant over each horizontal row and is shown in the left-most column for every row. Note that no events are included from the $\sqrt{s} \in [2.73,2.75]$~GeV region. All error bars represent statistical uncertainties only.}
\end{figure*}

From here on, by charged-mode, we will denote only the charged-two-track topology; no final results are presented for the charged-three-track topology. Figs.~\ref{fig:dsig_charged_1}-\ref{fig:dsig_charged_3} show our differential cross section results in different energy bins for the charged-mode. Unless otherwise mentioned, for all plots, we nominally include the $\Lambda^\ast$ cuts as described in Sec.~\ref{sec:phi_lambda1520_interference}. Fig.~\ref{fig:dsig_neutral} shows the differential cross sections for the neutral-mode. The energy binning for the charged-mode is uniformly 10-MeV-wide, while the minimum bin-width for the neutral-mode is 30-MeV-wide (a few bins at high energies are 40- and 50-MeV wide). We do not report cross section results for the bins $\sqrt{s} = 2.735$ and 2.745~GeV, due to normalization issues, as described in Sec.~\ref{sec:norm}.

The diffractive nature of $\phi$ production means that most of the yields are concentrated in the forward-angle regime. Hence, we choose 0.1-unit-wide $\cmangle$ bins for $\cmangle \leq 0.35$ and finer 0.05-unit-wide $\cmangle$ bins in the $\cmangle > 0.35$ forward-angle regime, where more statistics are available.

\begin{figure*}
\includegraphics[width=7in]{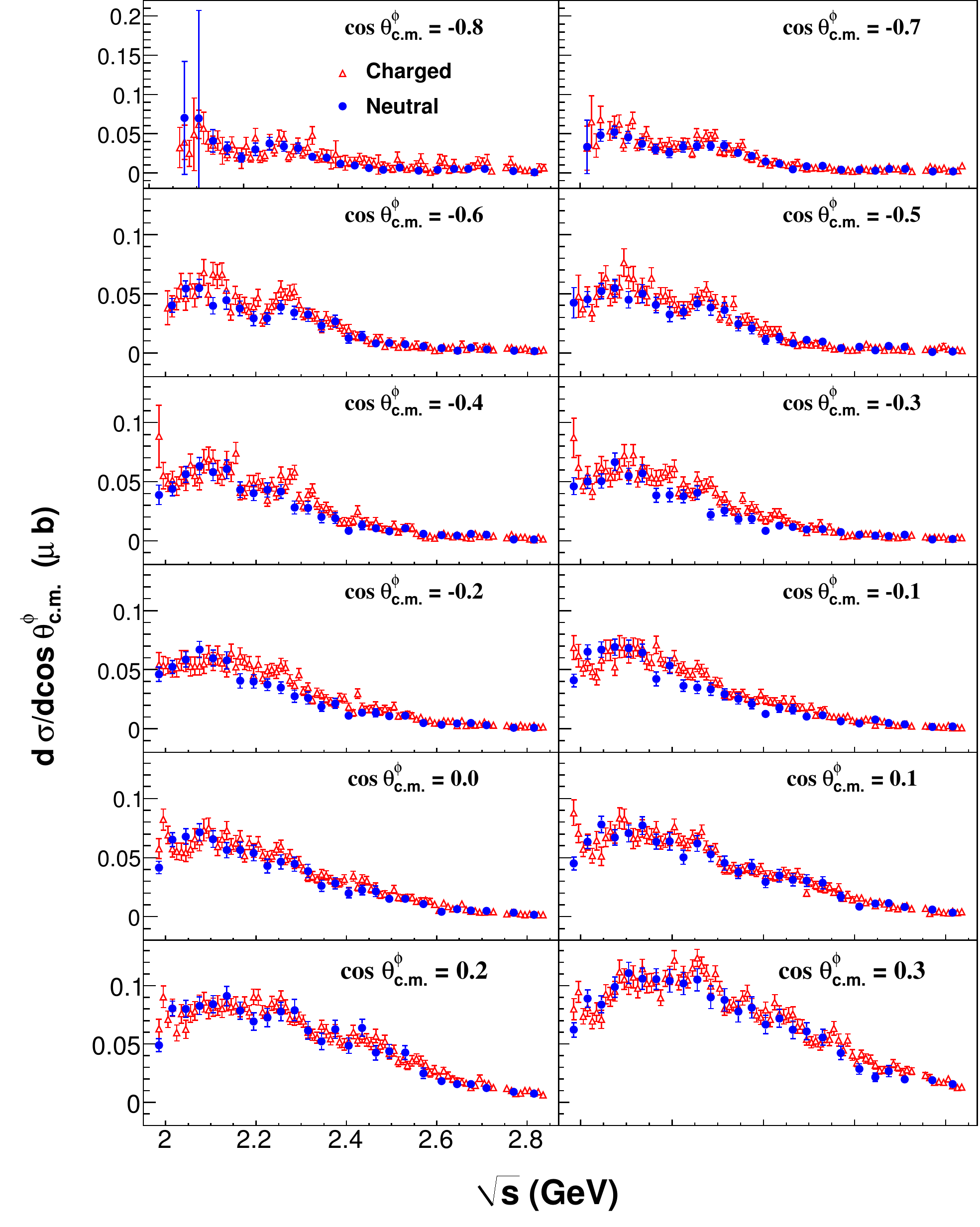} 
\caption[]{\label{fig:dsig_charged_neutral_0}
  (Color online)
  Comparison of the charged- and neutral-mode $d \sigma /d \cmangle$ results in 0.1-$\cmangle$ bins, at mid- and backward-angles. The charged-mode results comprise the two-track dataset with additional $\Lambda^\ast$ cuts as explained in the text. The $y$-axis range is constant over each horizontal row and is shown in the left-most column for every row. All error bars represent statistical uncertainties only.}
\end{figure*}

\begin{figure*}
\includegraphics[width=7in]{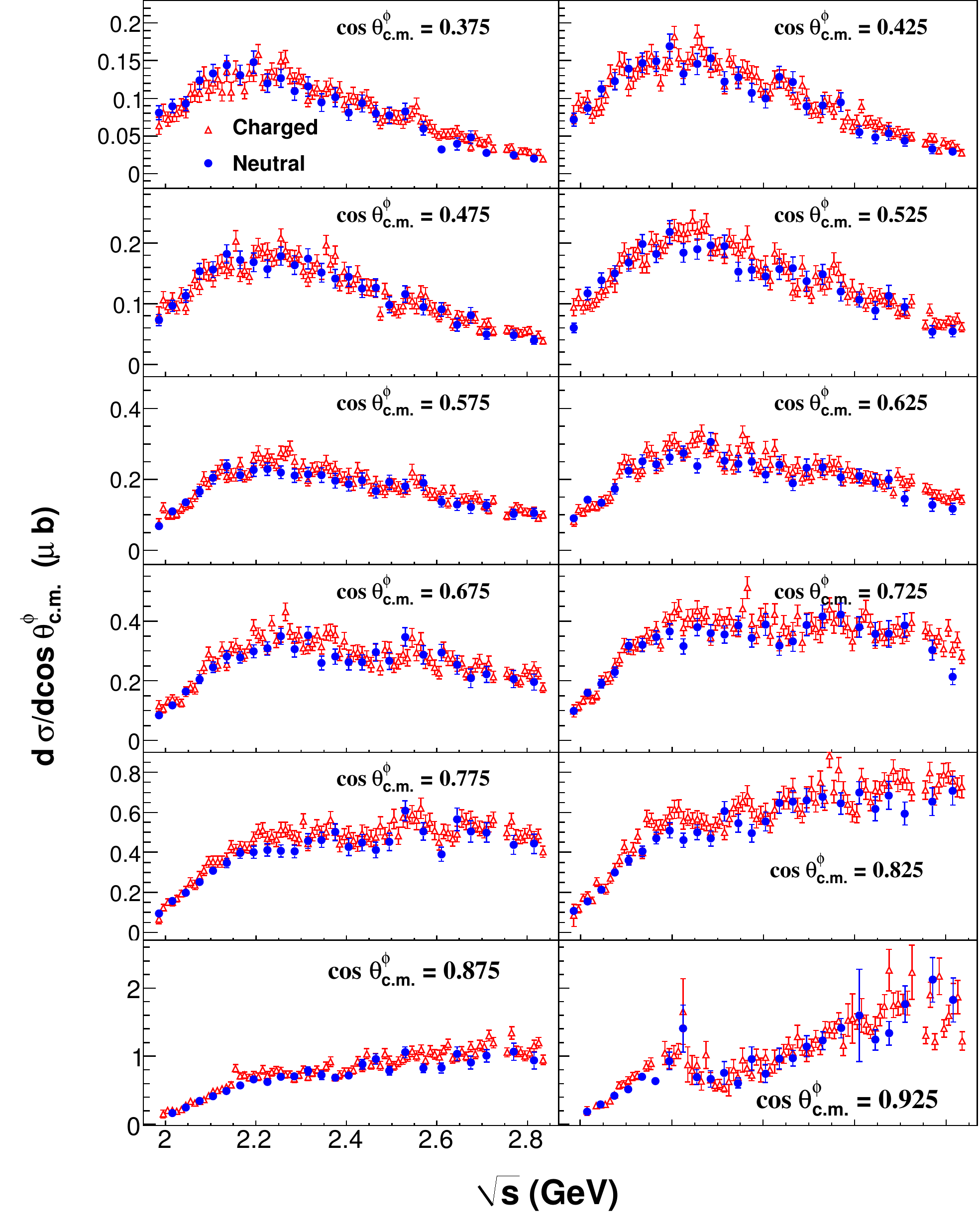} 
\caption[]{\label{fig:dsig_charged_neutral_1}
  (Color online)
  Comparison of the charged- and neutral-mode $d \sigma /d \cmangle$ results in 0.05-$\cmangle$ bins, at forward-angles. The charged-mode results comprise the two-track dataset with additional $\Lambda^\ast$ cuts as explained in the text. The $y$-axis range is constant over each horizontal row and is shown in the left-most column for every row. All error bars represent statistical uncertainties only.
}
\end{figure*}

Figs.~\ref{fig:dsig_charged_neutral_0} and \ref{fig:dsig_charged_neutral_1} show comparisons between the charged- and neutral-mode differential cross section results. Note that the angular bins are uniformly 0.1-unit- and 0.05-unit-wide in $\cmangle$ for Fig.~\ref{fig:dsig_charged_neutral_0} and Fig.~\ref{fig:dsig_charged_neutral_1}, respectively. The two sets of results should not be taken as independent measurements, since the topology-wise analyses were not performed blind to each other. For any future theory fits to these data, we suggest that the charged- and neutral-mode results be taken together as a single set of measurements involving some degree of correlation. Any remnant difference between the two modes should be taken as an additional systematic uncertainty. With this caveat in mind, Figs.~\ref{fig:dsig_charged_neutral_0} and \ref{fig:dsig_charged_neutral_1} show reasonable to good agreement between the two modes.

The forward-most angular bin shows a localized structure around $\sqrt{s} \approx 2.2$~GeV as already mentioned in Sec.~\ref{sec:intro}. This feature is discussed further in Sec.~\ref{sec:fwd_angle_bump}. We note that the structure is present in both the modes.

\subsection{Comparison with previous world data for differential cross sections}

\begin{figure}
\includegraphics[width=3.3in]{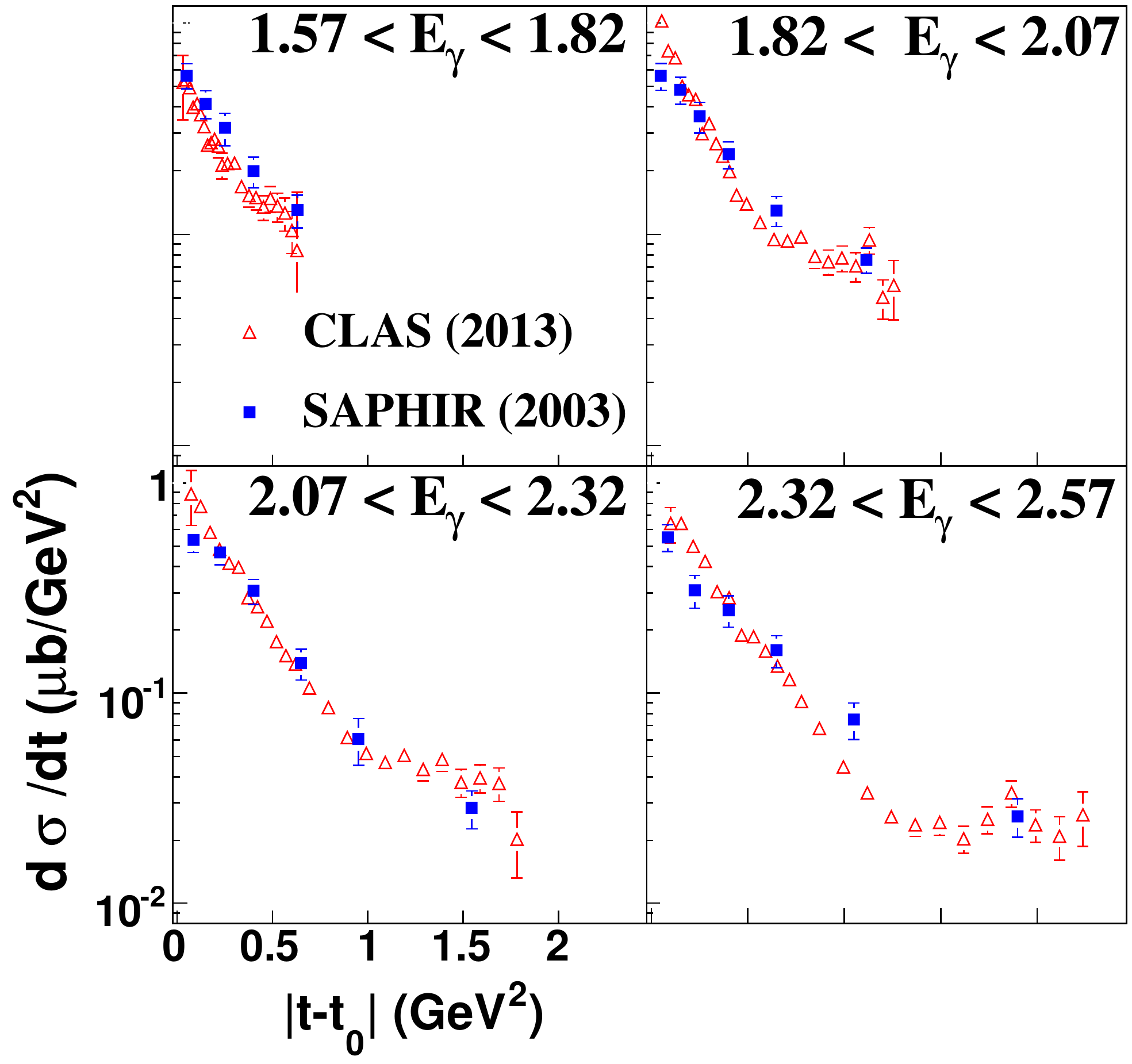} \\ 
\caption[]{\label{fig:saphir_clas_compare}
  (Color online) Comparison between the current CLAS (red triangles) and SAPHIR 2003~\cite{barth} (blue circles) results. The SAPHIR binning in $E_\gamma$ (GeV) is printed on each pad and the CLAS results are at the bin-centers for each SAPHIR energy bin. The CLAS results are taken from the charged-mode topology. All error bars represent statistical uncertainties only.}
\end{figure}

Previous world data for $\phi$ photoproduction cross sections are generally scarce and no world data exists for the neutral-mode topology at all. We therefore restrict our discussion in this section to the charged-mode topology only, and unless otherwise mentioned, inclusive of the hard $\Lambda^\ast$ cuts. Most of the earlier results have very low statistics, wide energy bins and forward-angle coverage only. The current analysis incorporates substantial improvements on all of these factors along with sophisticated data analysis techniques. Therefore, we suggest that caution be taken while interpreting these comparisons. For low energy and forward-angle kinematics, there are two previous results from the SAPHIR (2003, Barth {\em et al.}~\cite{barth}) and LEPS (2005, Mibe {\em et al}~\cite{mibe}) Collaborations. Both data sets have wide energy binnings, $E_\gamma \approx 200$-MeV-wide and 100-MeV-wide bins for SAPHIR and LEPS, respectively. However, the common feature in both results is that of a prominent enhancement around $E_\gamma \approx 2$~GeV (${\sqrt{s} \approx 2.2~\mbox{GeV}}$) in the forward-angle $d \sigma / dt$, in agreement with our current results.

\begin{figure}
\includegraphics[width=3.3in]{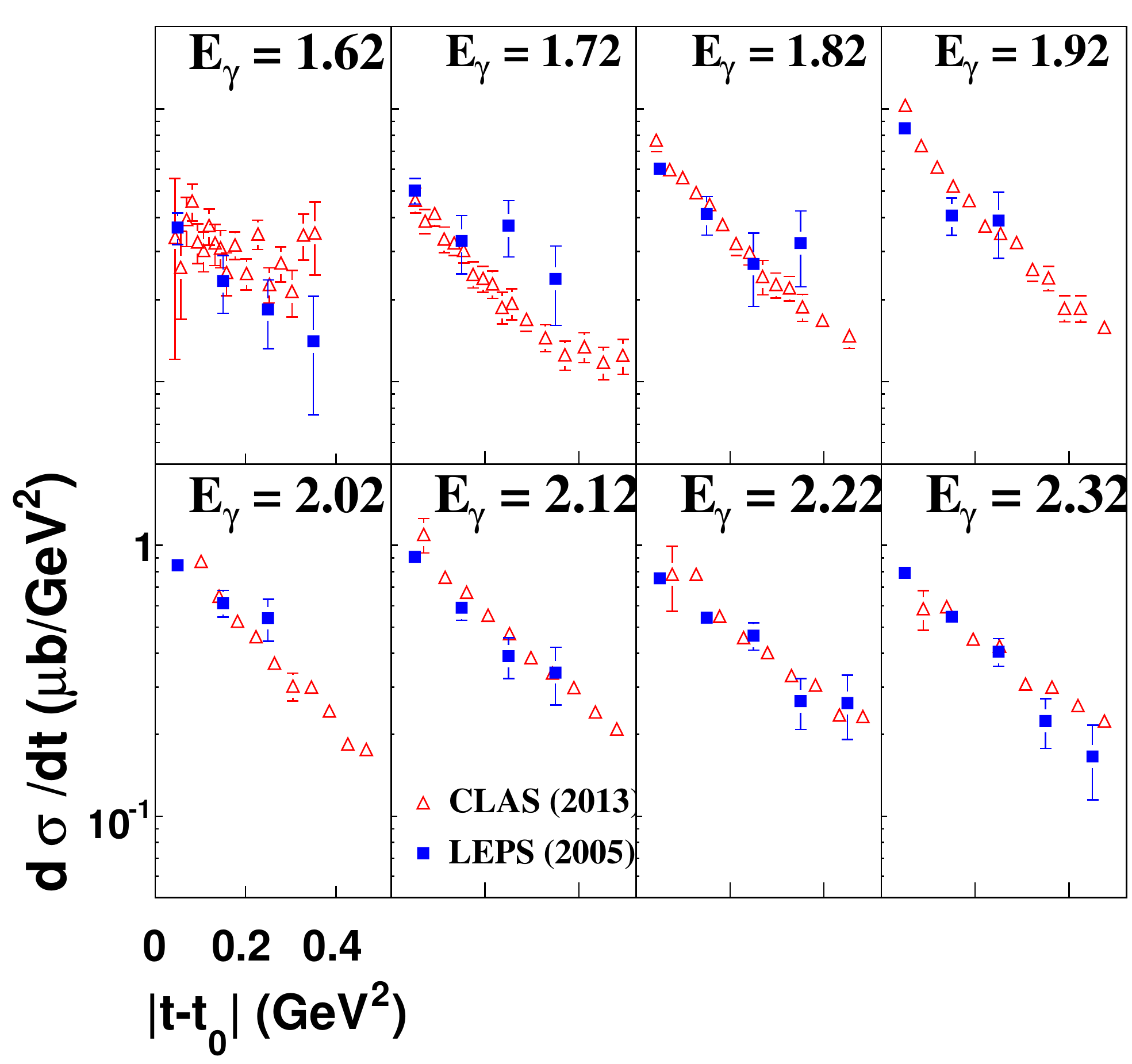}\\ 
\caption[]{\label{fig:leps_clas_compare}
  (Color online) Comparison between the current CLAS (red triangles) and LEPS 2005~\cite{mibe} (blue circles) results. The LEPS data had $E_\gamma = 200$~MeV wide bins (the bin-center is printed on each pad). The present CLAS results are taken from the charged-mode topology. All error bars represent statistical uncertainties only.}
\end{figure}

Since the SAPHIR and LEPS kinematics were mostly at forward-angles, these results were presented as $d\sigma/dt$ {\em vs.} $|t - t_0|$, where $t_0$ was the value of $t$ at $\cmangle = 0$. From the phenomenology of diffractive production, $d\sigma/dt$ was expected to show a simple exponential fall off with $|t - t_0|$. The conversion of $\cmangle$ to $t$ or $|t - t_0|$ depends on $\sqrt{s}$. With wide energy bins, it is not immediately clear which $\sqrt{s}$ should be chosen for this conversion. Therefore, we convert our results into the units chosen by SAPHIR and LEPS and make independent comparisons with both of them. Since our energy binning is much finer (10-MeV-wide in $\sqrt{s}$), we overlay our results at the energy bin-center of the SAPHIR or LEPS results. Figs.~\ref{fig:saphir_clas_compare} and~\ref{fig:leps_clas_compare} show the comparison between our results with SAPHIR and LEPS, respectively.

\begin{figure}
\includegraphics[width=3.3in,angle=90]{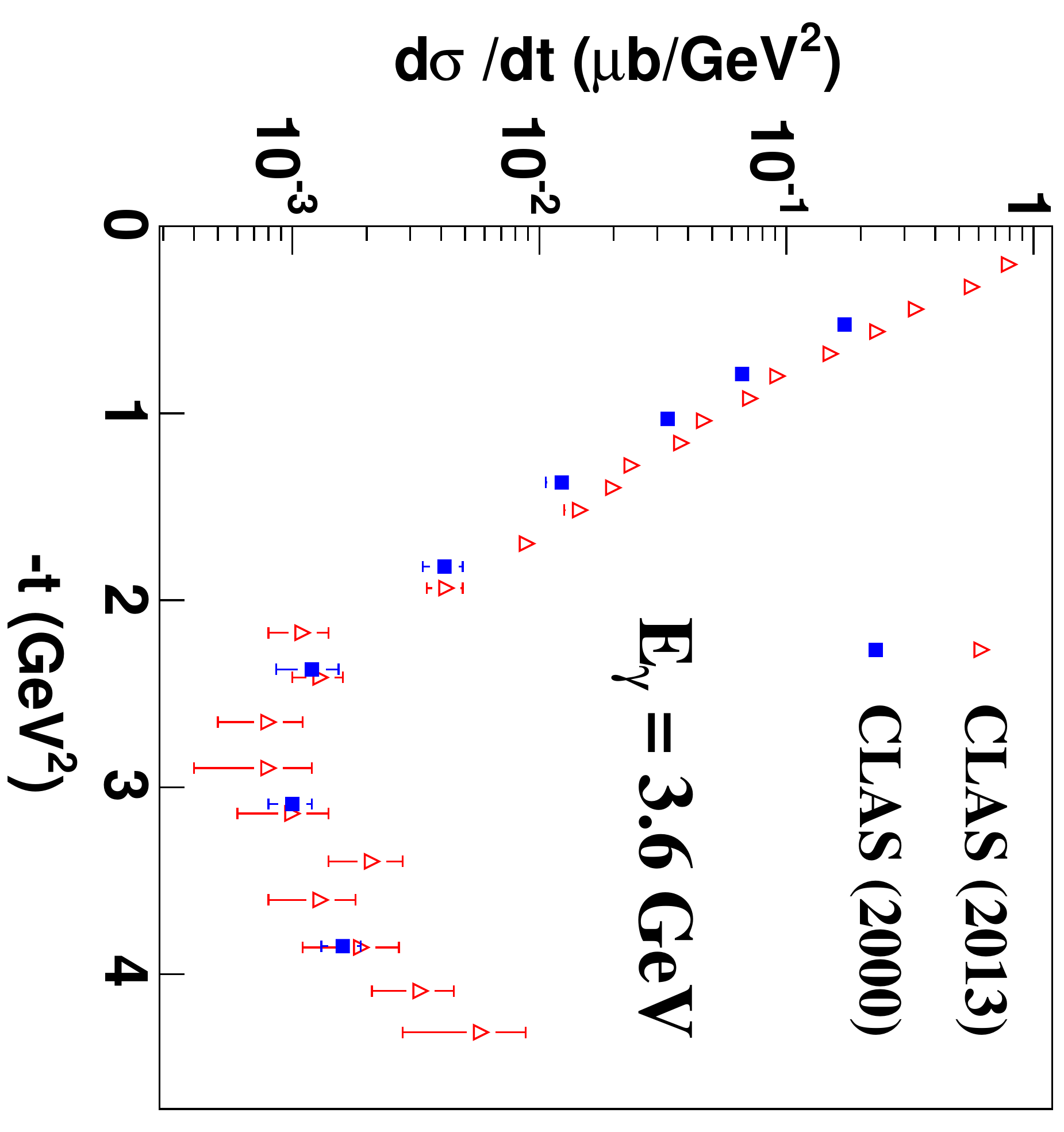}\\ 
\caption[]{\label{fig:anciant_clas_compare}
  (Color online) Comparison between the current CLAS (red triangles) and 2000 CLAS~\cite{anciant} (blue circles) results at the energy bin-center $E_\gamma = 3.6$~GeV. The CLAS 2000 energy binning was 600-MeV in $E_\gamma$, while the current CLAS binning is 10-MeV-wide in $\sqrt{s}$.}
\end{figure}

The only existing world data for large $|t|$ are the CLAS (2000, Anciant {\em et al.}~\cite{anciant}) results for a bin-center at $E_\gamma = 3.6$~GeV (tagged photon energy range 3.3 to 3.9~GeV). The chief motivation of the previous CLAS experiment was to investigate whether $u$-channel processes (at small $u$ or large $t$) contribute to the $\phi$ channel. Assuming that the $\phi$ is almost pure $|s \bar{s}\rangle$ and the strangeness content in ordinary nucleons is small, the coupling constant $g_{\phi NN}$ is expected to be small and therefore nucleon exchanges in the $u$-channel are supposed to the suppressed. However, as shown in Fig.~\ref{fig:anciant_clas_compare}, both the CLAS 2000 and the current CLAS results show a small but distinct rise in the backward-angles, suggestive of a non-negligible value for $g_{\phi NN}$.

\begin{figure}
\includegraphics[width=3.3in,angle=90]{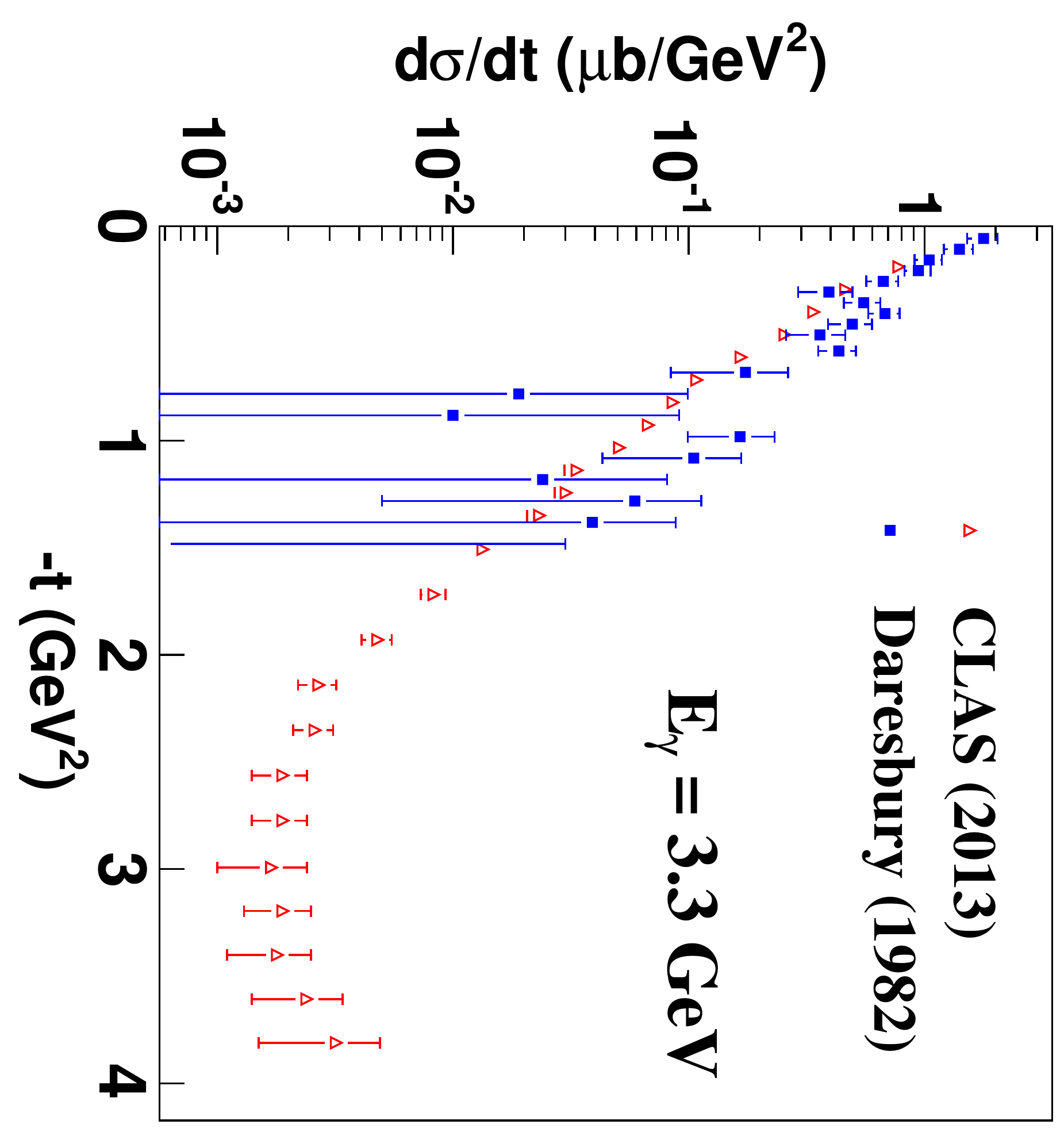}\\ 
\caption[]{\label{fig:daresbury_clas_compare}
  (Color online) Comparison between the current CLAS (red triangles) and 1982 Daresbury~\cite{barber_82} (blue circles) results at the bin-center $E_\gamma = 3.3$~GeV. The Daresbury energy bins were 1-GeV-wide in $E_\gamma$, while the CLAS binning is 10-MeV-wide in $\sqrt{s}$.}
\end{figure}

Last, Fig.~\ref{fig:daresbury_clas_compare} compares our results with the $E_\gamma = 3.3$~GeV bin-center results from Daresbury (1982, Barber {\em et al.}~\cite{barber_82}). The Daresbury binning was 1-GeV in $E_\gamma$, and away from the $t\to 0$ region, the error bars are large. Overall, within the limitations of statistical uncertainties, agreement between the two results is fair.

\subsection{Spin density matrix elements results}

\begin{figure*}
  \includegraphics[width=7in]{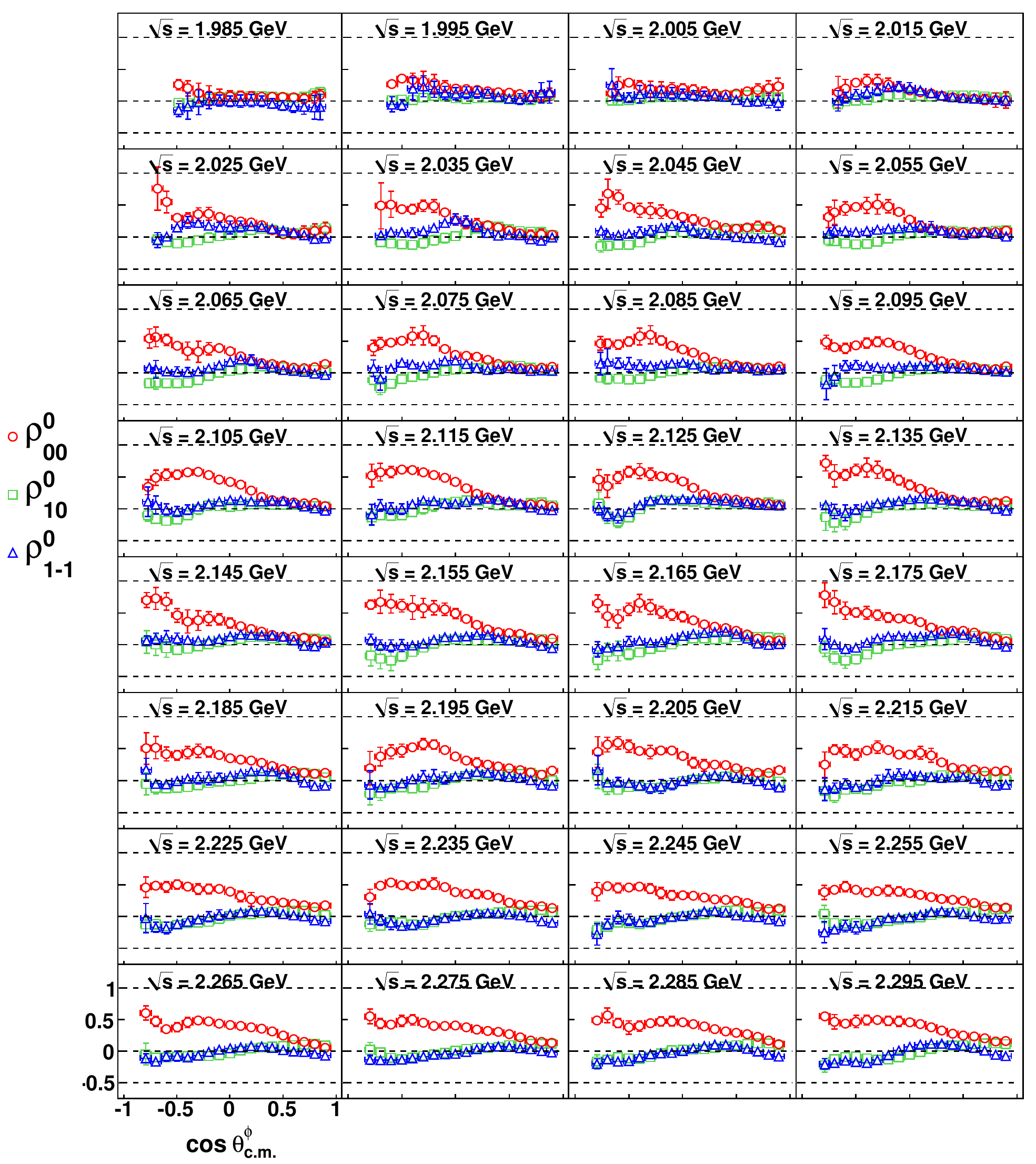} 
\caption[]{\label{fig:sdme_charged_15mev_0}
 (Color online) SDME {\em vs.} $\cmangle$: spin density matrix elements in the Adair frame for the charged-mode (with $\Lambda^\ast$ cuts) topology in the energy range 1.98~GeV~$\leq \sqrt{s} <$~2.28~GeV. The centroid of each 10-MeV-wide bin is printed on the plots. All error bars represent statistical uncertainties only.}
\end{figure*}

\begin{figure*}
    \includegraphics[width=7in]{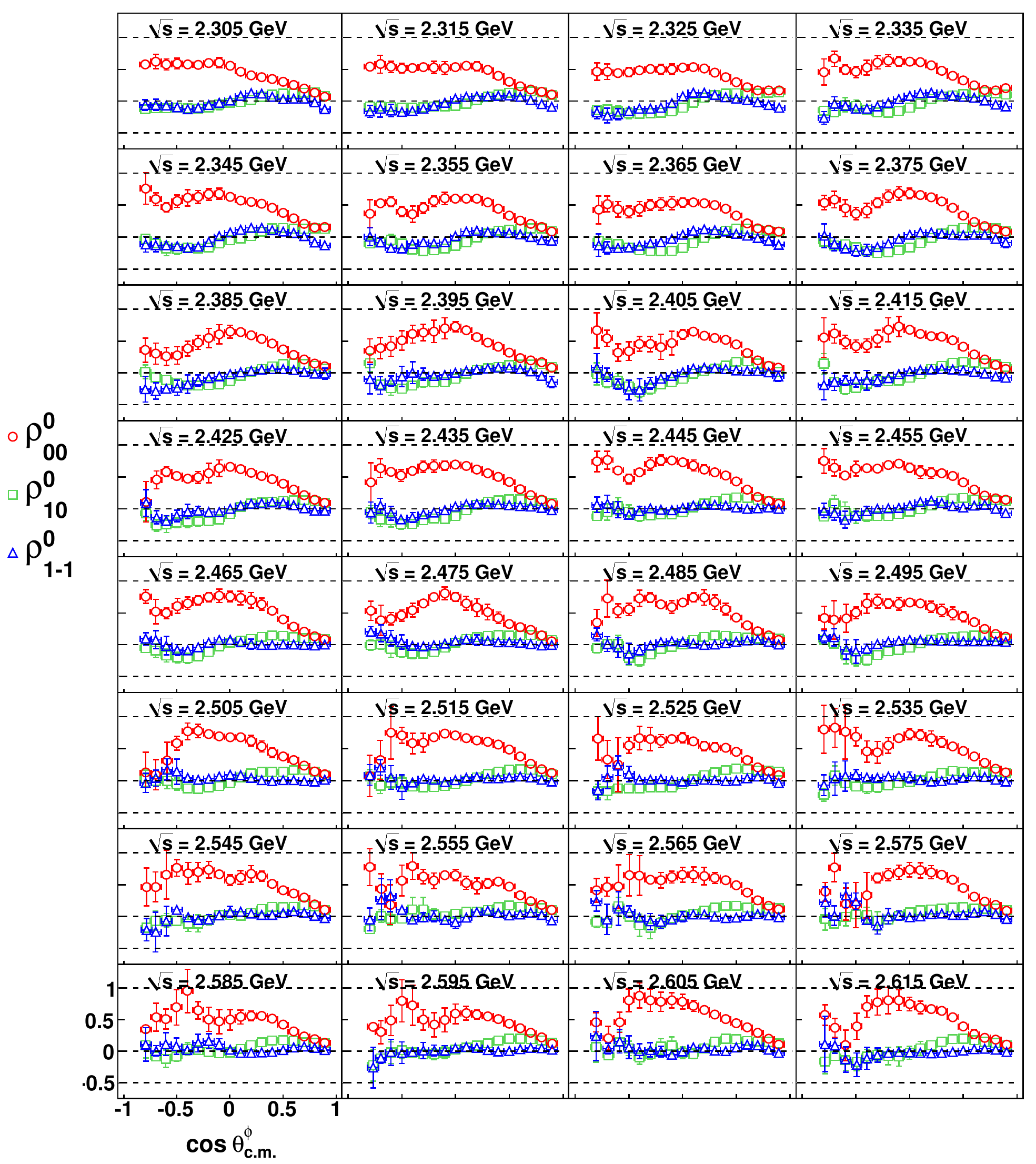} 
\caption[]{\label{fig:sdme_charged_15mev_1}
  (Color online) SDME {\em vs.} $\cmangle$: spin density matrix elements in the Adair frame for the charged-mode (with $\Lambda^\ast$ cuts) topology in the energy range 2.28~GeV~$\leq \sqrt{s} <$~2.62~GeV. The centroid of each 10-MeV-wide bin is printed on the plots. All error bars represent statistical uncertainties only.}
\end{figure*}

\begin{figure*}
    \includegraphics[width=7in]{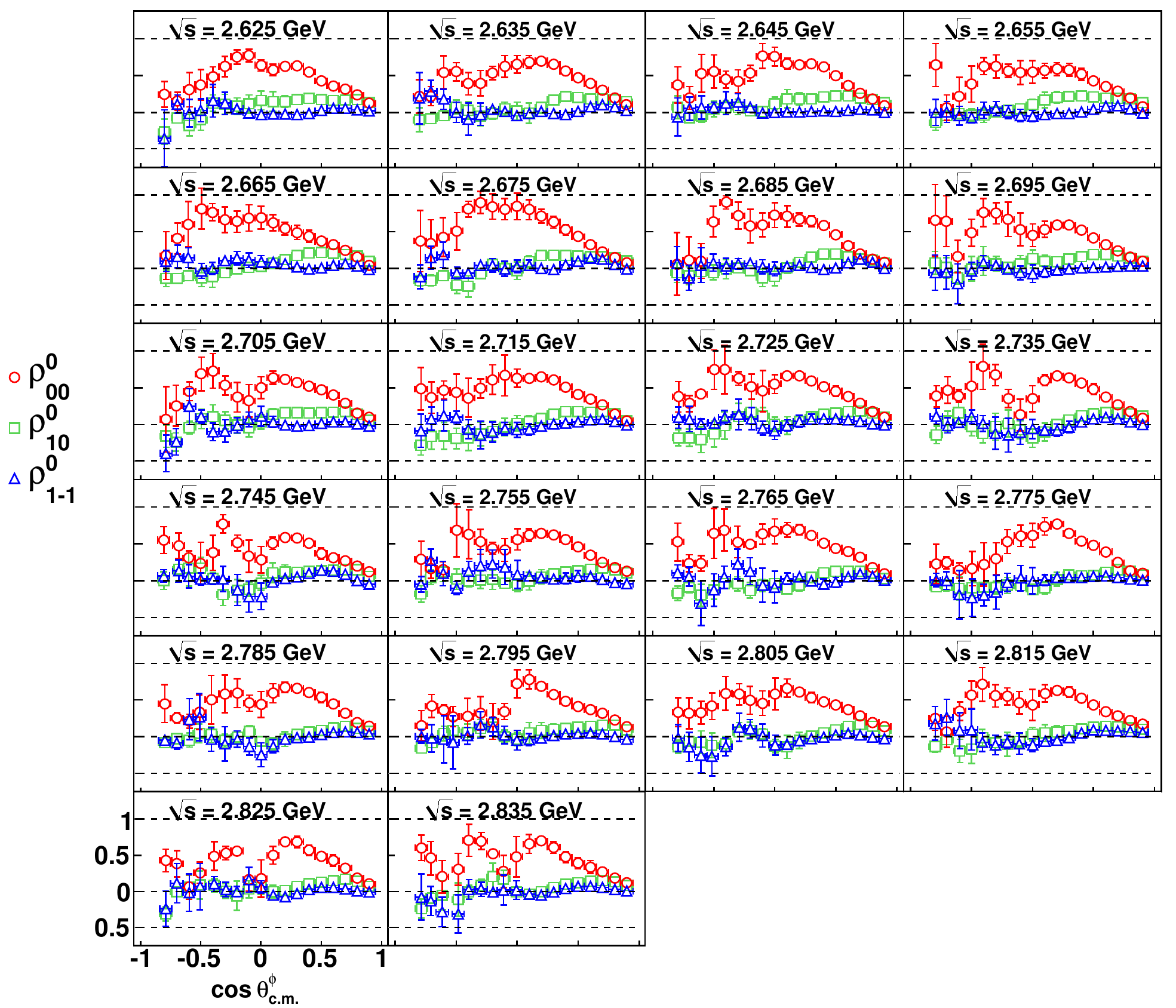} 
\caption[]{\label{fig:sdme_charged_15mev_2}
  (Color online) SDME {\em vs.} $\cmangle$: spin density matrix elements in the Adair frame for the charged-mode (with $\Lambda^\ast$ cuts) topology in the energy range 2.62~GeV~$\leq \sqrt{s} <$~2.84~GeV. The centroid of each 10-MeV-wide bin is printed on the plots. All error bars represent statistical uncertainties only.}
\end{figure*}

\begin{figure*}
    \includegraphics[width=7in]{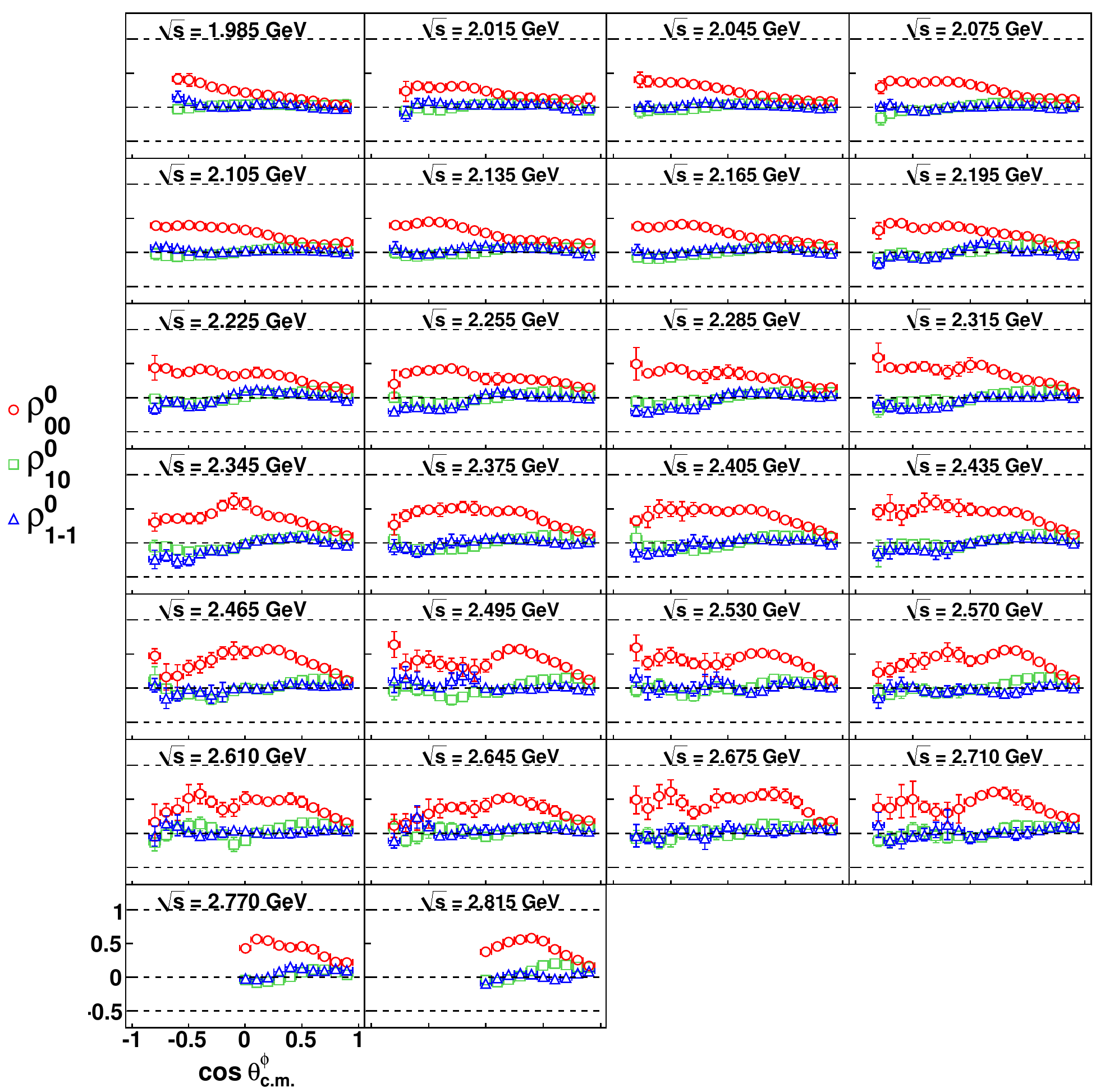} 
\caption[]{\label{fig:sdme_neutral}
  (Color online) SDME {\em vs.} $\cmangle$: spin density matrix elements in the Adair frame for the neutral-mode topology. The minimum bin-width is 30-MeV in $\sqrt{s}$ and the centroid of each bin is printed on the plots. All error bars represent statistical uncertainties only.}
\end{figure*}

\begin{figure*}
   \includegraphics[width=7in]{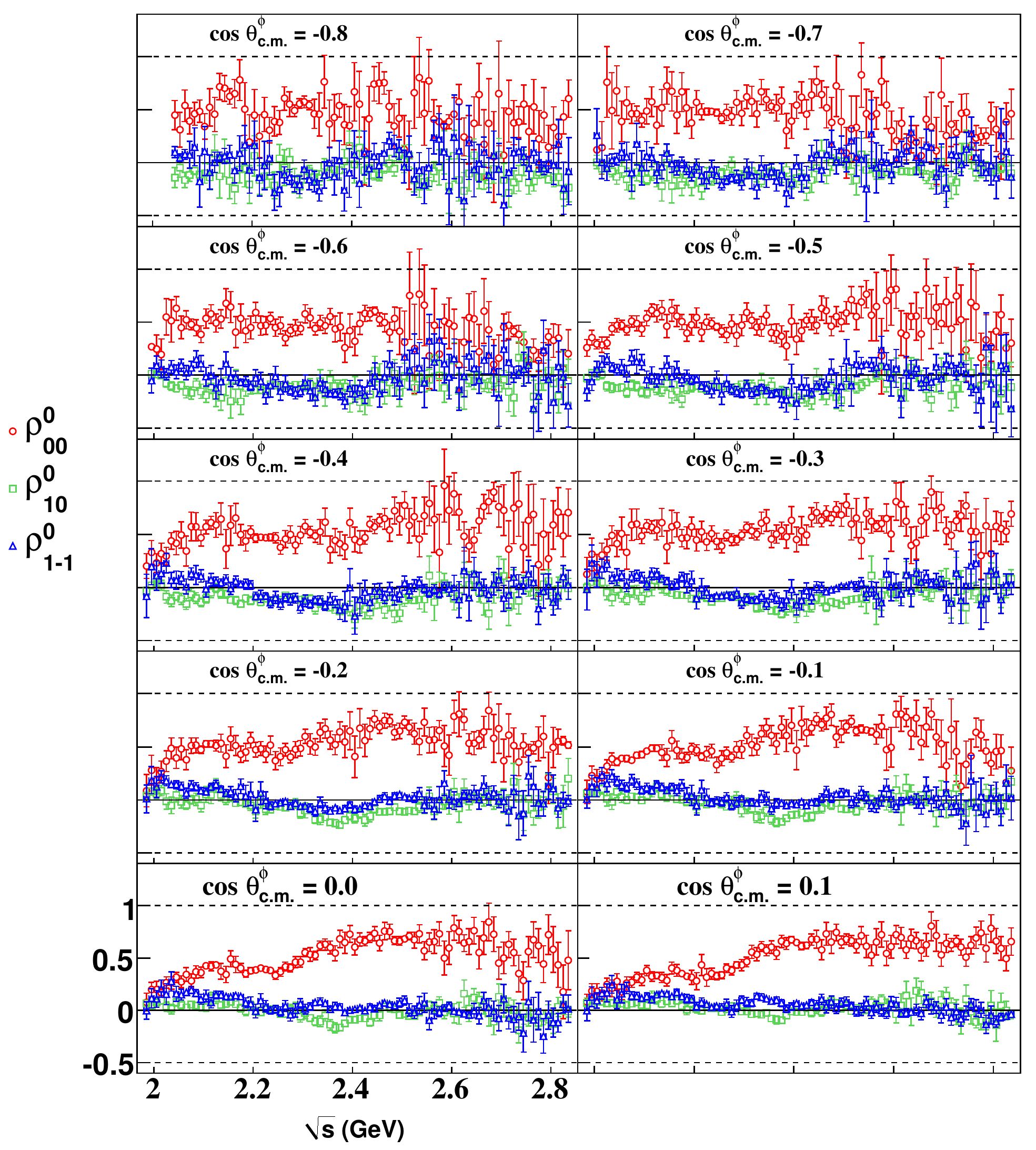} 
\caption[]{\label{fig:sdme_charged_angular_0}
  (Color online)
  The energy dependence of SDME's (Adair frame) in the backward-angle bins for the charged-mode (with $\Lambda^\ast$ cuts) topology. All error bars represent statistical uncertainties only.
}
\end{figure*}

\begin{figure*}
   \includegraphics[width=7in]{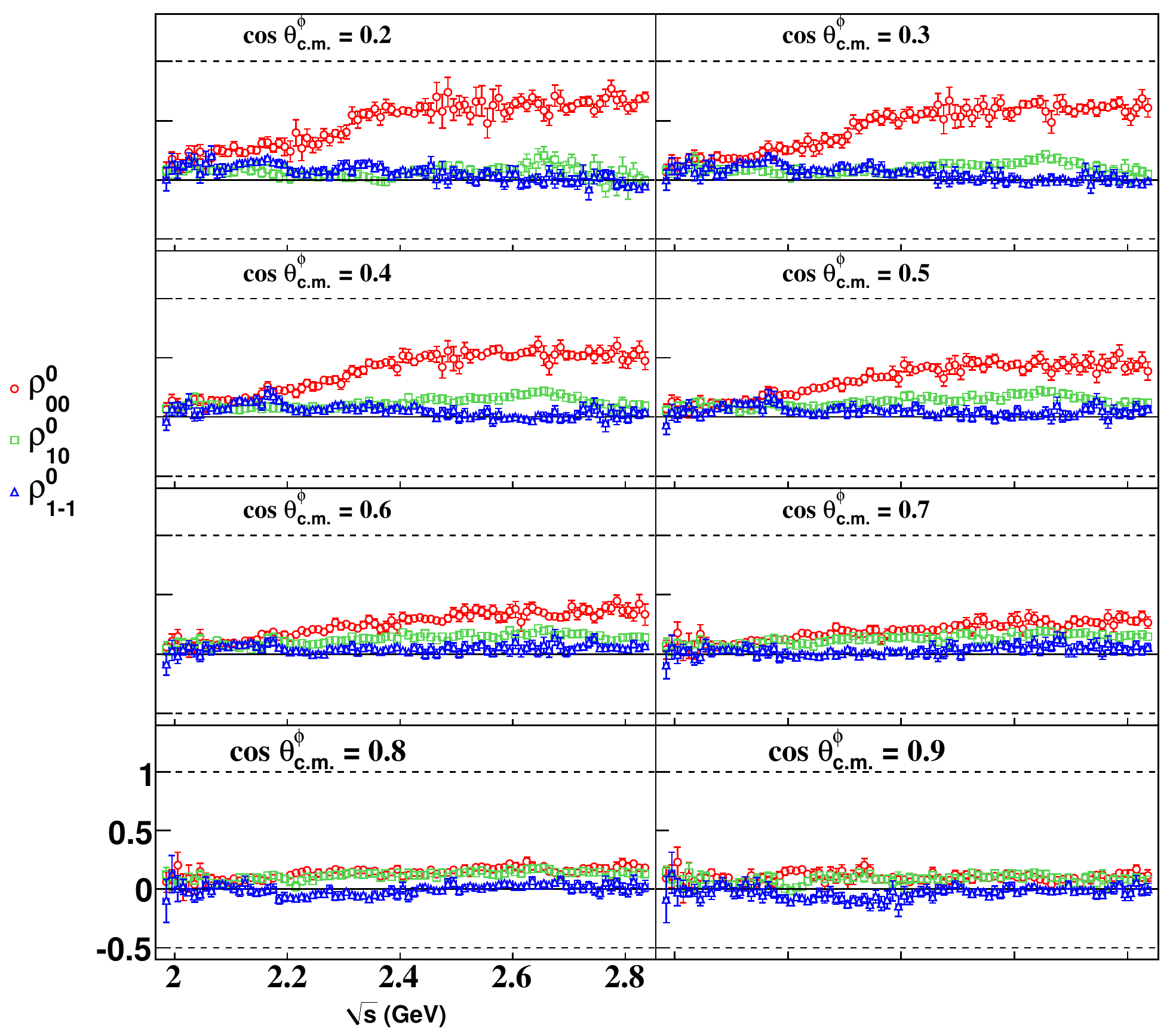} 
\caption[]{\label{fig:sdme_charged_angular_1}
  (Color online)
  The energy dependence of SDME's (Adair frame) in the forward-angle bins for the charged-mode (with $\Lambda^\ast$ cuts) topology. All error bars represent statistical uncertainties only.
}
\end{figure*}

\begin{figure*}
  \includegraphics[width=7in]{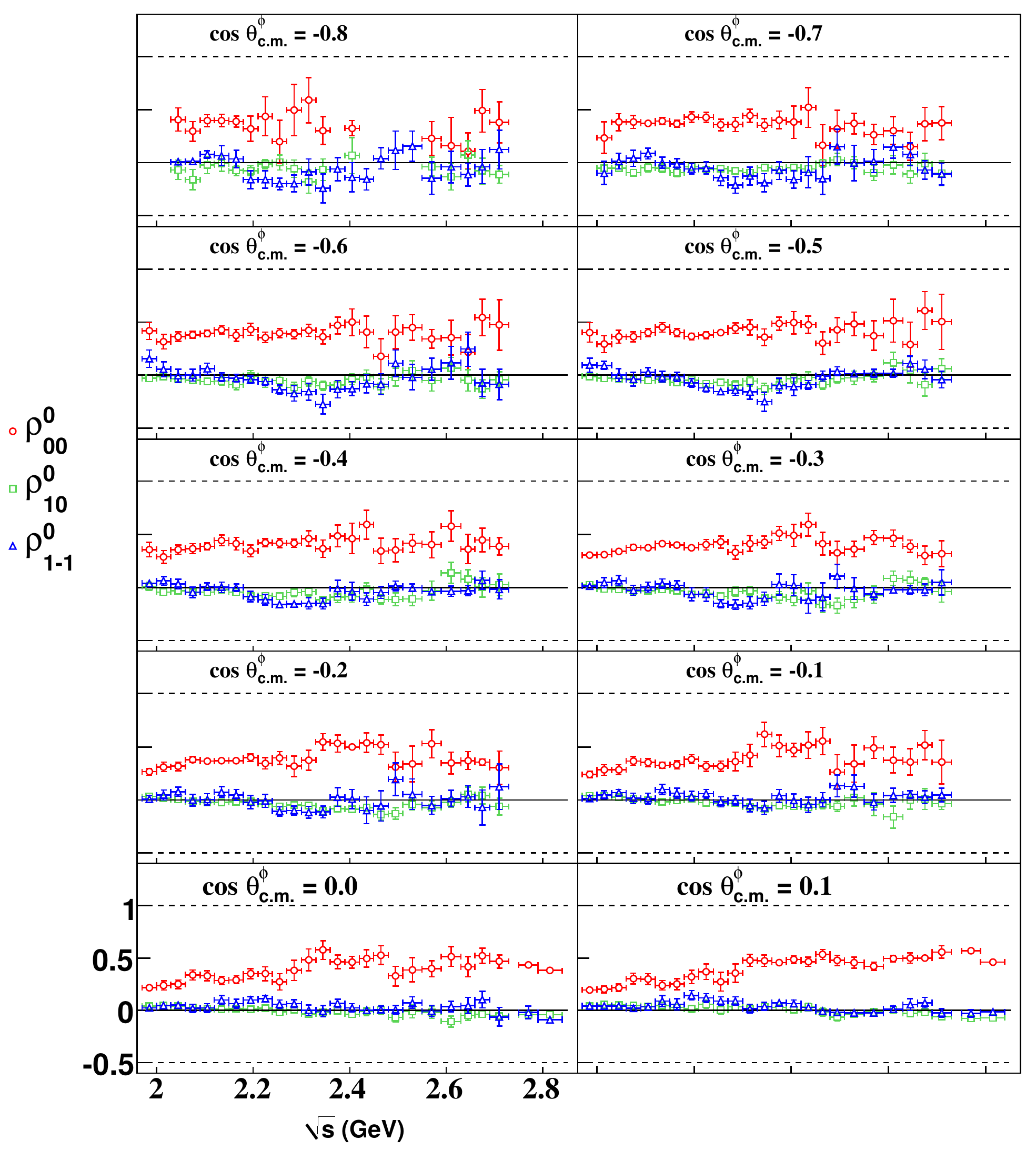} 
\caption[]{\label{fig:sdme_neutral_angular_0}
  (Color online)
  The energy dependence of SDME's (Adair frame) in the backward-angle bins for the neutral-mode topology. All error bars represent statistical uncertainties only.
}
\end{figure*}

\begin{figure*}
\includegraphics[width=7in]{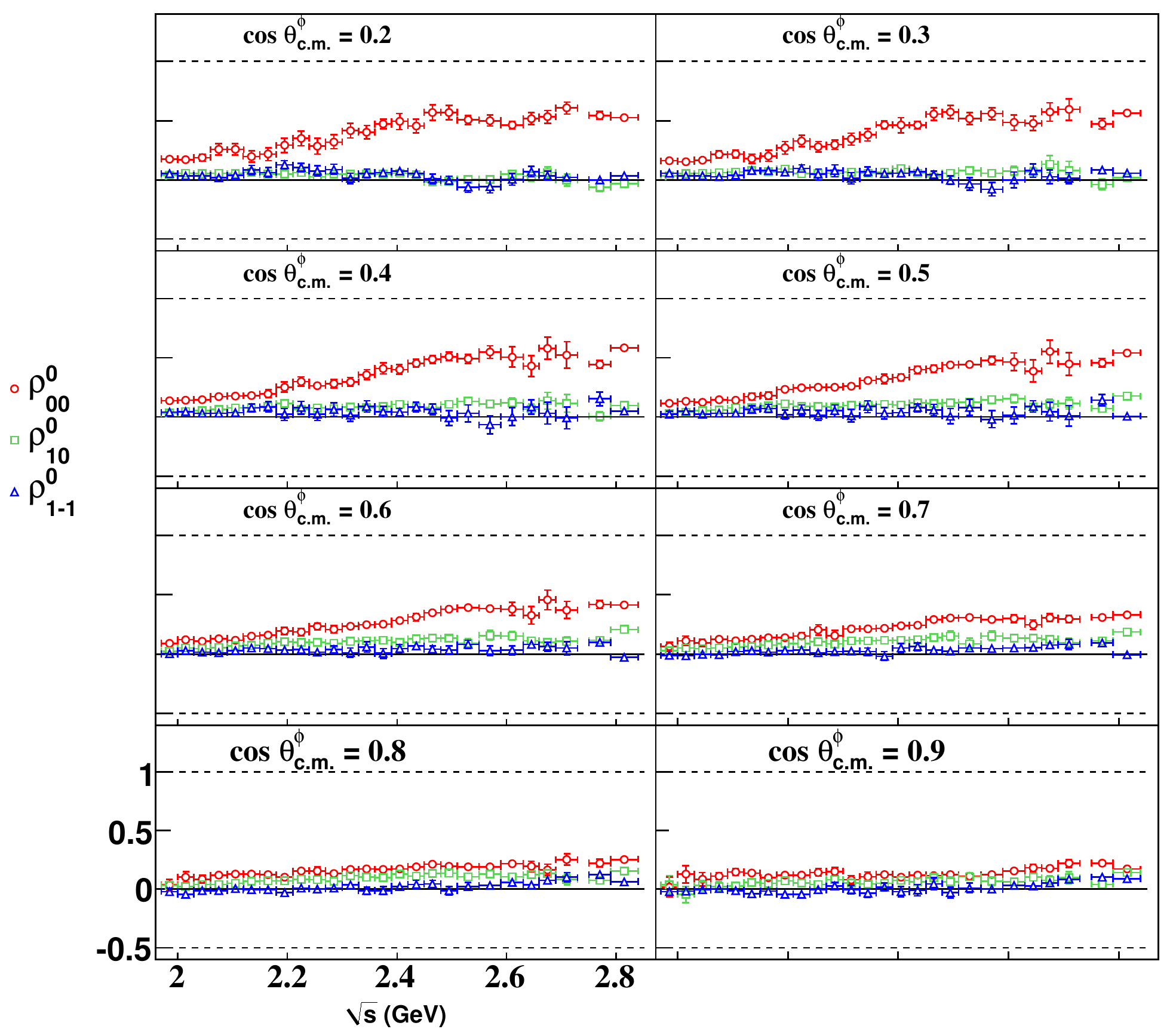}  
\caption[]{\label{fig:sdme_neutral_angular_1}
  (Color online)
  The energy dependence of SDME's (Adair frame) in the mid- and forward-angle bins for the neutral-mode topology. All error bars represent statistical uncertainties only.
}
\end{figure*}

Since polarization measurements are sensitive to interference between amplitudes and require enhanced statistics compared to cross sections which measure sum of the squared amplitudes, we retain a uniform 0.1-unit $\cmangle$ binning for the SDME results.  Figs.~\ref{fig:sdme_charged_15mev_0}-\ref{fig:sdme_charged_15mev_2} show the SDME's for the charged-mode topology in the Adair frame. The most prominent feature is the large value of $\rho^0_{00}$, while $\rho^0_{10}$ and $\rho^0_{1-1}$ are small, but non-zero. There is a similarity with the corresponding results for the $\omega$ channel~\cite{omega_prc} in a ``hump-like'' structure, followed by a ``dip'', for the $\rho^0_{00}$ element. At high $\sqrt{s}$, the $\omega$ results had a distinct ``dip'' for $\rho^0_{00}$ in the mid-forward angles. For the $\phi$, there are indications of a ``dip'' for $\rho^0_{00}$ in the mid- to mid-backward angles, though the structure is much less well-defined due to statistical limitations at high $\sqrt{s}$ and $\cmangle < 0$. Fig.~\ref{fig:sdme_neutral} shows the SDME's for the neutral-mode topology in the Adair frame. The energy bins are at least 30-MeV-wide in $\sqrt{s}$. Figs.~\ref{fig:sdme_charged_angular_0}-\ref{fig:sdme_charged_angular_1} and~\ref{fig:sdme_neutral_angular_0}-\ref{fig:sdme_neutral_angular_1} show the energy dependence of the Adair frame SDME's in different angular bins for the charged- and neutral-mode, respectively.

For systematic uncertainties, we adopt the results from our previous $\omega$ analysis~\cite{omega_prc} where the maximal effect of incorrect acceptance on the extracted SDME's was studied by distorting the decay distributions in Eq.~\ref{eqn:schilling} by the uncertainties in our acceptance calculation. The SDME systematic uncertainties from this study were $\delta (\rho^0_{00} )= 0.0175$, $\delta (\rho^0_{1-1}) = 0.0125$ and $\delta(\rho^0_{10}) = 0.01$ and we quote these for the present $\phi$ analysis as well, since the underlying assumption is only the Schilling's equation for vector mesons.

\subsection{Comparison with previous world data for SDME}

Previous (pre-2010) world data on the $\phi$ SDME's are extremely limited. Although many of the older papers did report a few results, the overall general conclusion was that the $\rho^0$ SDME's in the helicity frame were all near-zero. An important drawback in the SDME extraction method employed in these older data was that instead of fits to the full Schilling's equation as given by Eq.~\ref{eqn:schilling}, fits were performed to integrated intensity profiles. There are several problems that arose due to this. First, the SDME's are functions of both $\sqrt{s}$ and the production angle $\cmangle$. Second, the detector acceptance is a function of every independent kinematic variable ($\{\sqrt{s}, \cmangle, \zeta, \varphi\}$ in Eq.~\ref{eqn:schilling}) and therefore such integrated distributions are not properly acceptance corrected. Furthermore, it was claimed that the only way to obtain a good fit was to incorporate $S$-$P$-wave interference effects (see McCormick {\em et al.}~\cite{mccormick}). As we have stressed earlier in Sec.~\ref{sec:sig_bkgd}, a certain amount of $S$-$P$-wave interference effect indeed must occur. However, claims to the degree of this effect based on such integrated intensity fits have to be considered with caution. In keeping with these facts, and the very limited physics conclusions from previous $\phi$ SDME measurements, we choose not to compare our present results to any of the older data.

\begin{figure}
  \includegraphics[width=3.3in]{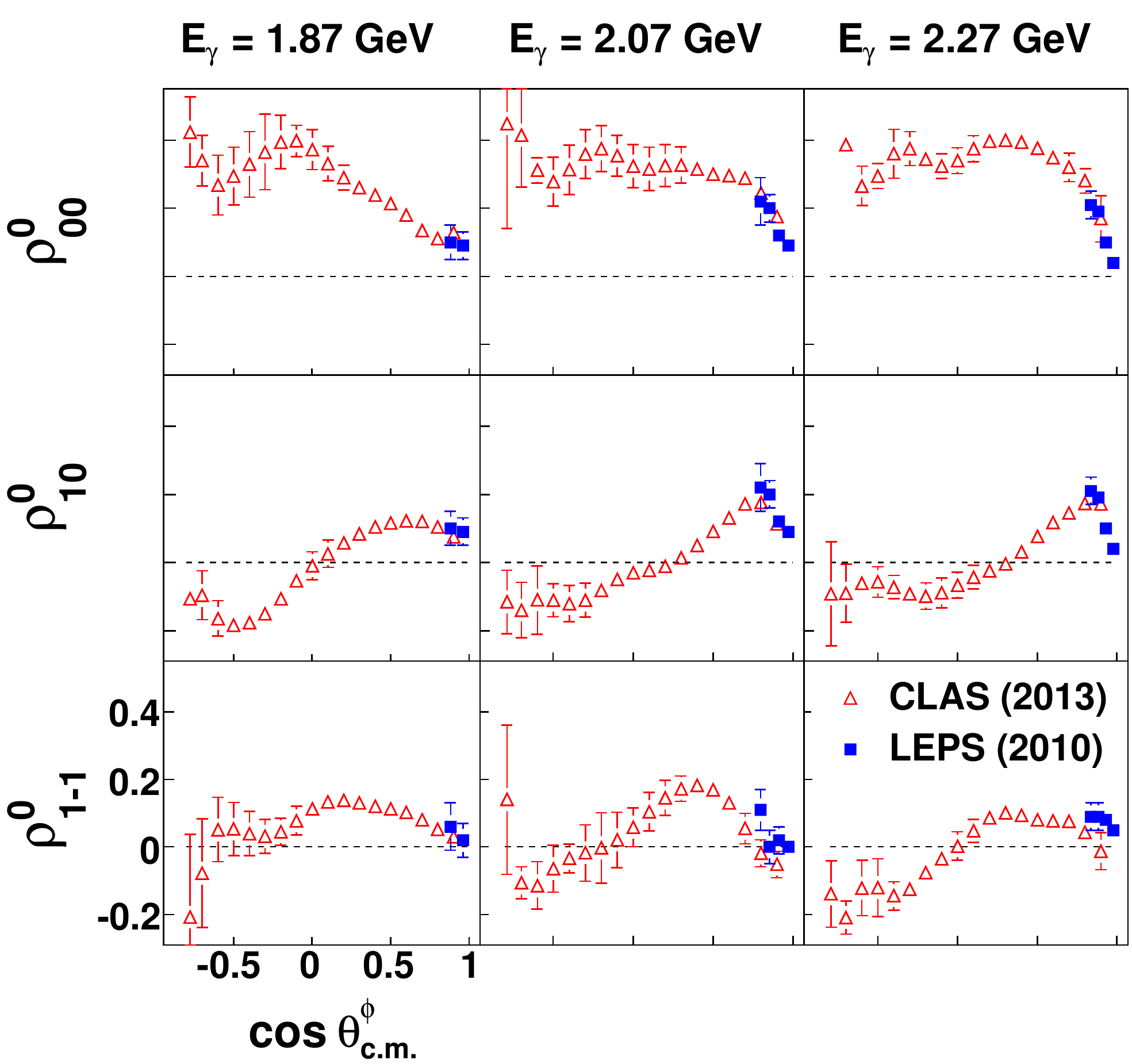} 
\caption[]{\label{fig:sdme_compare_leps}
  (Color online) Comparison of the charged-mode SDME's (Gottfried-Jackson frame) between the current CLAS (red triangles) and forward-angle 2010 LEPS Chang {\em et al.}~\cite{chang_leps_sdme} (blue squares) results. The CLAS results ($\Delta (\sqrt{s}) = 10$-MeV-wide bins) are shown at the approximate LEPS ($\Delta (E_\gamma) = 200$-MeV-wide bins) bin-centers, printed on the top of each column. Note that the LEPS data points also involve an energy-dependent conversion from $|t-t_0|$ to $\cmangle$, which has an intrinsic approximation due to the wide energy bins. All error bars represent statistical uncertainties only.
}
\end{figure}

Recently, however, the LEPS Collaboration~\cite{chang_leps_sdme} has published $\phi$ SDME's by performing fits to the full Schilling's expression and properly accounting for acceptance as well. Fig.~\ref{fig:sdme_compare_leps} shows the comparisons between the LEPS and the current CLAS results. The LEPS binning was 200-MeV in $E_\gamma$, while our bins are much finer. Therefore, the CLAS results are shown at the approximate LEPS bin-centers. Also, the LEPS results were quoted as functions of the variable $|t - t_0|$. The conversion of $|t - t_0|$ to $\cmangle$ depends on the energy $\sqrt{s}$. Since the LEPS energy bins were wide, this conversion process brings an extra degree of uncertainty into the comparisons. Keeping in mind these approximations, the CLAS and LEPS results show good agreement.

\section{\label{sec:disc}Physics discussion}

\subsection{Diffractive exchange parameters $B_\phi$ and $C_\phi$}

\begin{figure}
  \centering
  \subfigure[]{
    {\includegraphics[width=2.3in,angle=90]{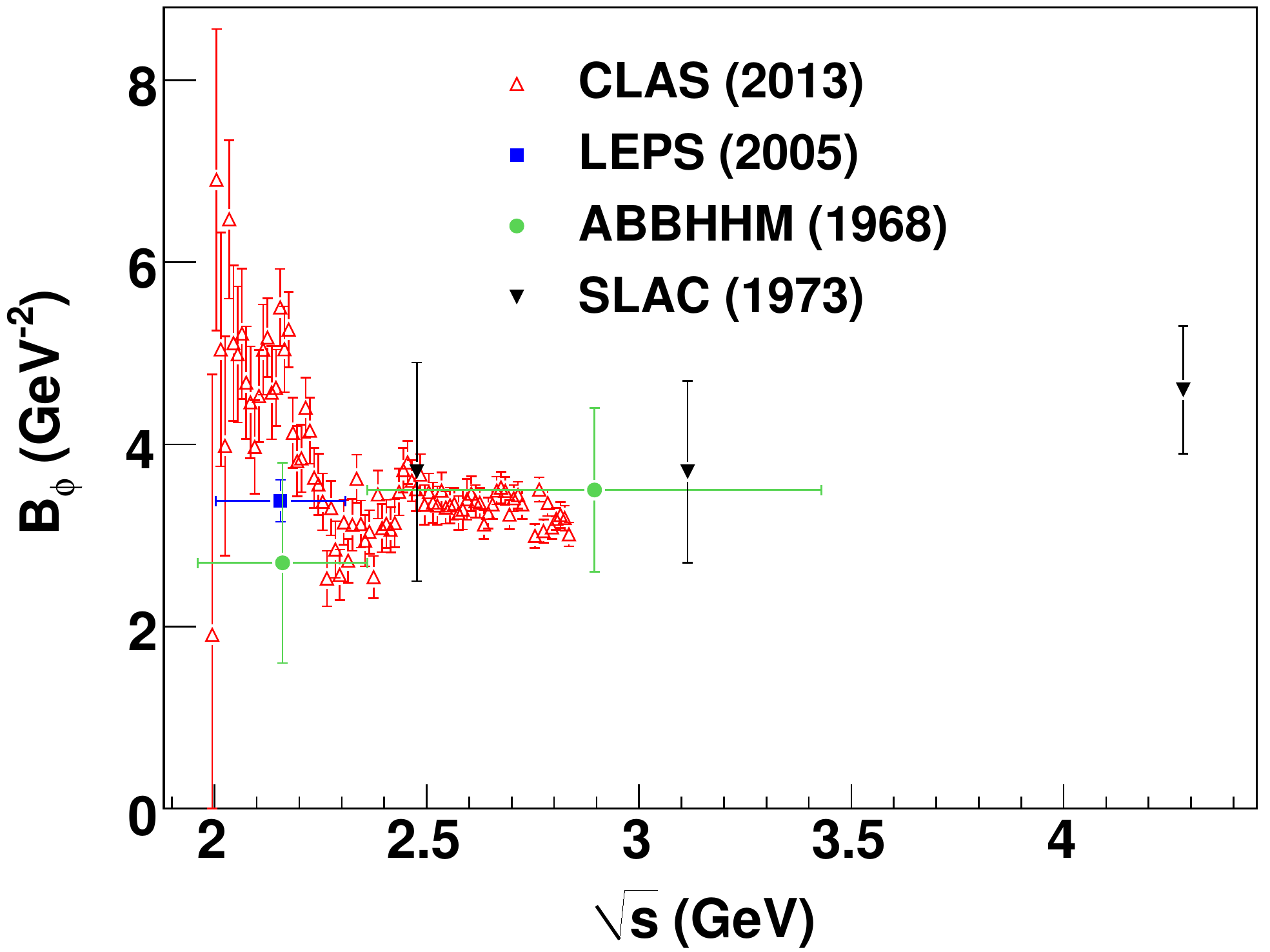}} 
  }
  \subfigure[]{
    {\includegraphics[width=2.3in,angle=90]{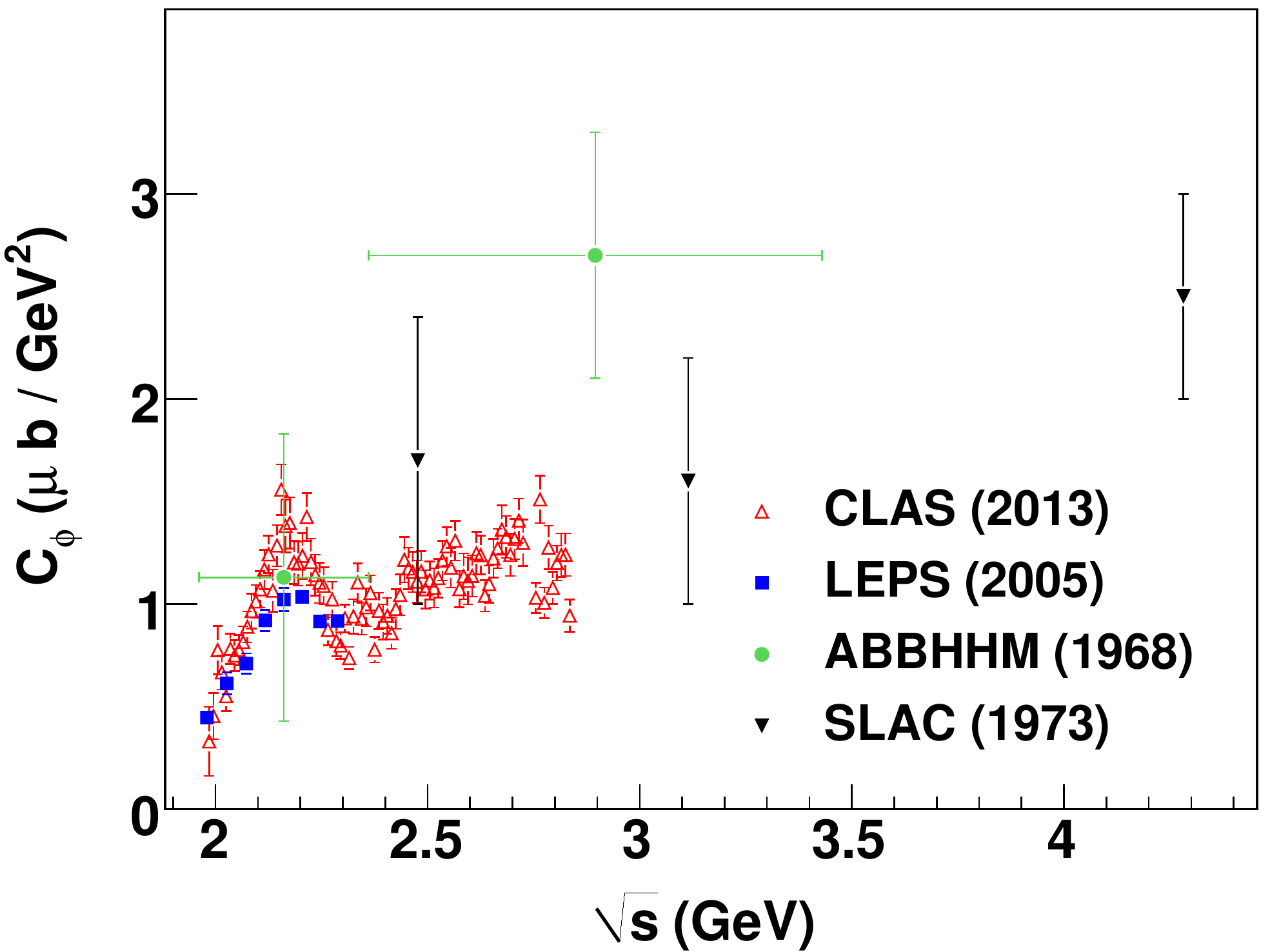}} 
  }
\caption[]{\label{fig:vmd_params}
  (Color online) The variation of the parameters (a) $B_\phi$ and (b) $C_\phi$ from a fit to $d \sigma /dt$ using Eq.~\ref{eqn:vmd_pomeron} for the charged-mode with $\Lambda^\ast$ cuts included, compared to previous world data. Only the forward-angle kinematic points satisfying $0.55 \leq \cmangle \leq 0.95$ were included in the fits for the present analysis. See text for details.
}
\end{figure}

As mentioned in the introductory section, the $\phi$ photoproduction channel is ideally suited to study the phenomenology of Pomeron exchange in the diffractive limit of $t \to t_0$, where ${t_0 = |t|_{min}}$ corresponds to $\cmangle  = 1$ for a given $\sqrt{s}$. The Pomeron Regge trajectory is approximately given by ${\alpha(t) \approx 1.08 + 0.25t}$, where $d \sigma /dt$ scales as $\sim (\beta(t) s^{\alpha(t)})^2/s^2 $ and $\beta(t)$ is the Regge residue that behaves like a form-factor. Therefore, in the diffractive limit of small $t$ and large $s$, $d \sigma /dt$ should show a very slow variation with $s$. 

Such diffractive Pomeron exchanges are expected to occur for all three vector mesons $\rho$, $\omega$ and $\phi$. However, for the $\rho$ and $\omega$, additional meson exchanges occur as well. For the $\omega$, $t$-channel $\pi$ exchanges are thought to have a more dominant contribution than the Pomeron. Since the $\phi$ is almost purely $|s\bar{s}\rangle$, such light-quark $\pi$ exchanges are suppressed and diffractive Pomeron exchange is the dominant contribution to the production amplitude. In the simplest VMD model~\cite{donnachie_book}, the photoproduction cross section can be related to the elastic $\phi p$ scattering cross section as
\begin{equation}
\frac{d \sigma}{dt}(\gamma p \to \phi p) \approx \frac{4\pi\alpha}{\gamma_\phi^2} \left( \frac{|\vec{p}_\phi|}{E_\gamma} \right)^2_{\mbox{\scriptsize c.m.}} \frac{d \sigma}{dt}(\phi p \to \phi p), 
\label{eqn:vmd}
\end{equation}
where $\gamma_\phi$ is the photon-$\phi$ coupling constant from Eq.~\ref{eqn:vmd_amp}. For Pomeron exchanges, the form-factor $\beta(t)$ is generally taken as $\exp(-B_\phi(|t - t_0|)/2)$ and Eq.~\ref{eqn:vmd} can be recast in the form
\begin{equation}
\frac{d \sigma}{dt}(\gamma p \to \phi p) = C_\phi \exp(-B_\phi(|t - t_0|)),
\label{eqn:vmd_pomeron}
\end{equation}
with the parameters $B_\phi$ and $C_\phi$ as the slope and forward-angle cross-sections, respectively. Fig.~\ref{fig:vmd_params} shows the variation of the extracted $B_\phi$ and $C_\phi$ from the charged-mode for this analysis, in comparison with previous world data. It is important to note here that our fits to Eq.~\ref{eqn:vmd_pomeron} included data points with $0.55 \leq \cmangle \leq 0.95$, since it is known that the slope shows a strong $t$-dependence as well~\cite{behrend}. The overall trend in Fig.~\ref{fig:vmd_params} shows only a slow rise of both $B_\phi$ and $C_\phi$ with energy, the signature for diffractive phenomenology.

\subsection{Forward-angle structure at $\sqrt{s} \approx 2.1$~GeV}
\label{sec:fwd_angle_bump}

\begin{figure}
  \centering
   \includegraphics[width=3in]{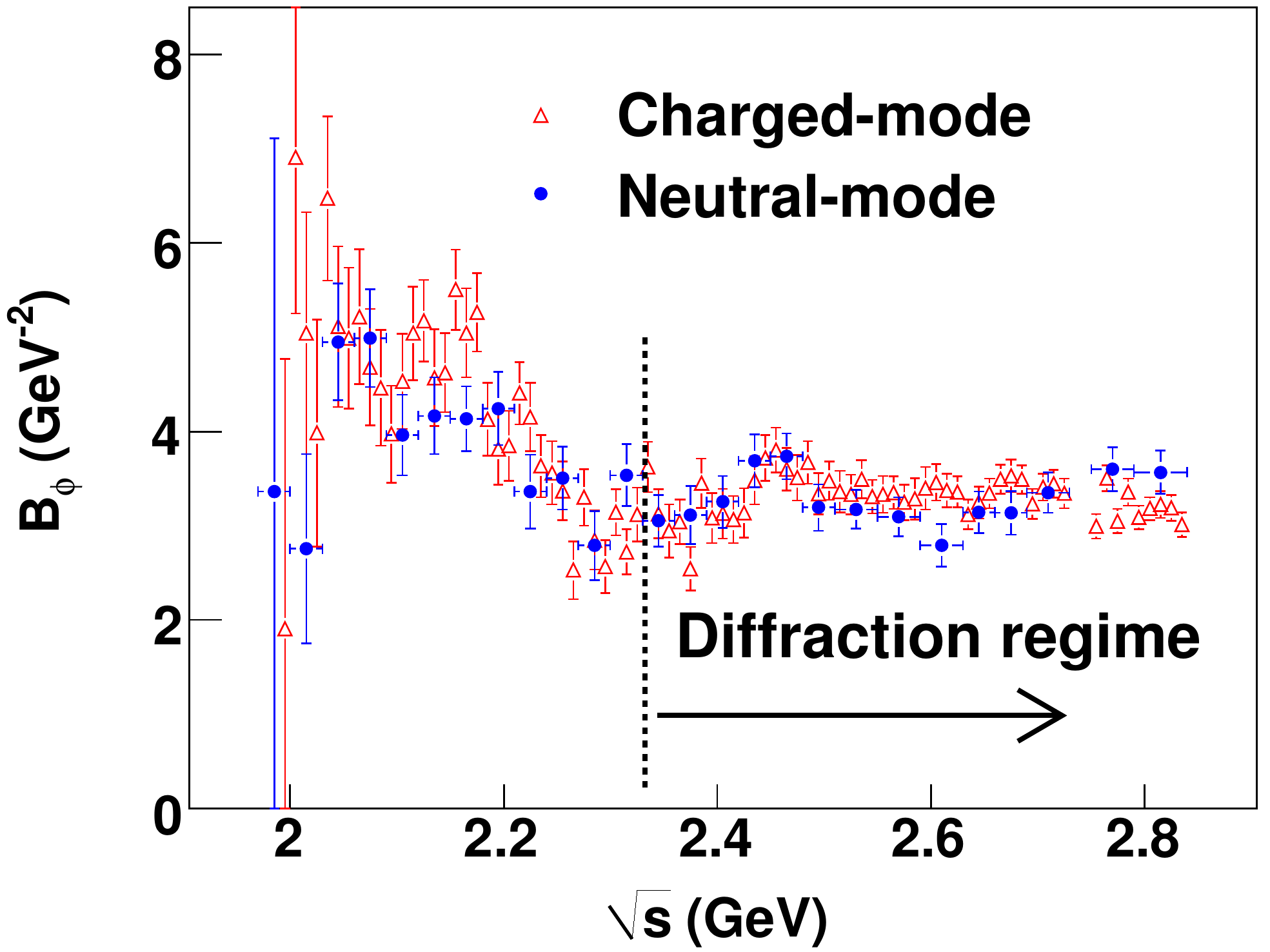} 
\caption[]{\label{fig:slope_compare_charged_neutral}
  (Color online) Comparison between the charged- and neutral-mode slope parameter $B_\phi$. Below $\sqrt{s} \approx 2.3$~GeV, the production mechanism is no longer that of a simple diffractive Pomeron exchange.}
\end{figure}

\begin{figure}
  \centering
   \includegraphics[width=2.4in,angle=90]{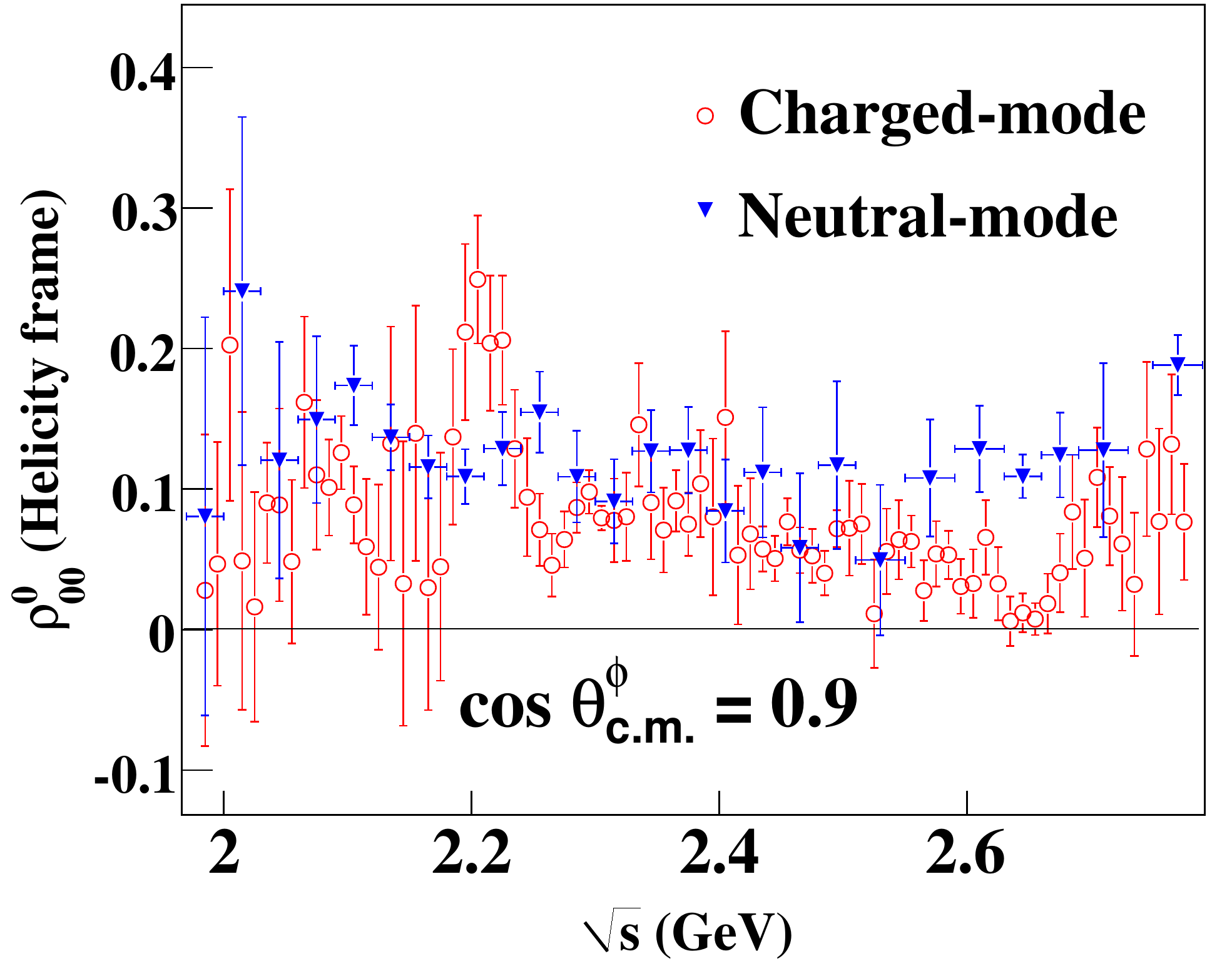} 
\caption[]{\label{fig:rho00_compare_charged_neutral}
  (Color online) Comparison between the charged- (with $\Lambda^\ast$ cuts) and neutral-mode $\rho^0_{00}$ results in the Helicity frame in a forward-angle region. While there is overall good agreement, the charged-mode shows traces of a local ``structure'' around $\sqrt{s} \approx 2.2$~GeV.}
\end{figure}

\begin{figure}
  \centering
  \includegraphics[width=2.2in,angle=90]{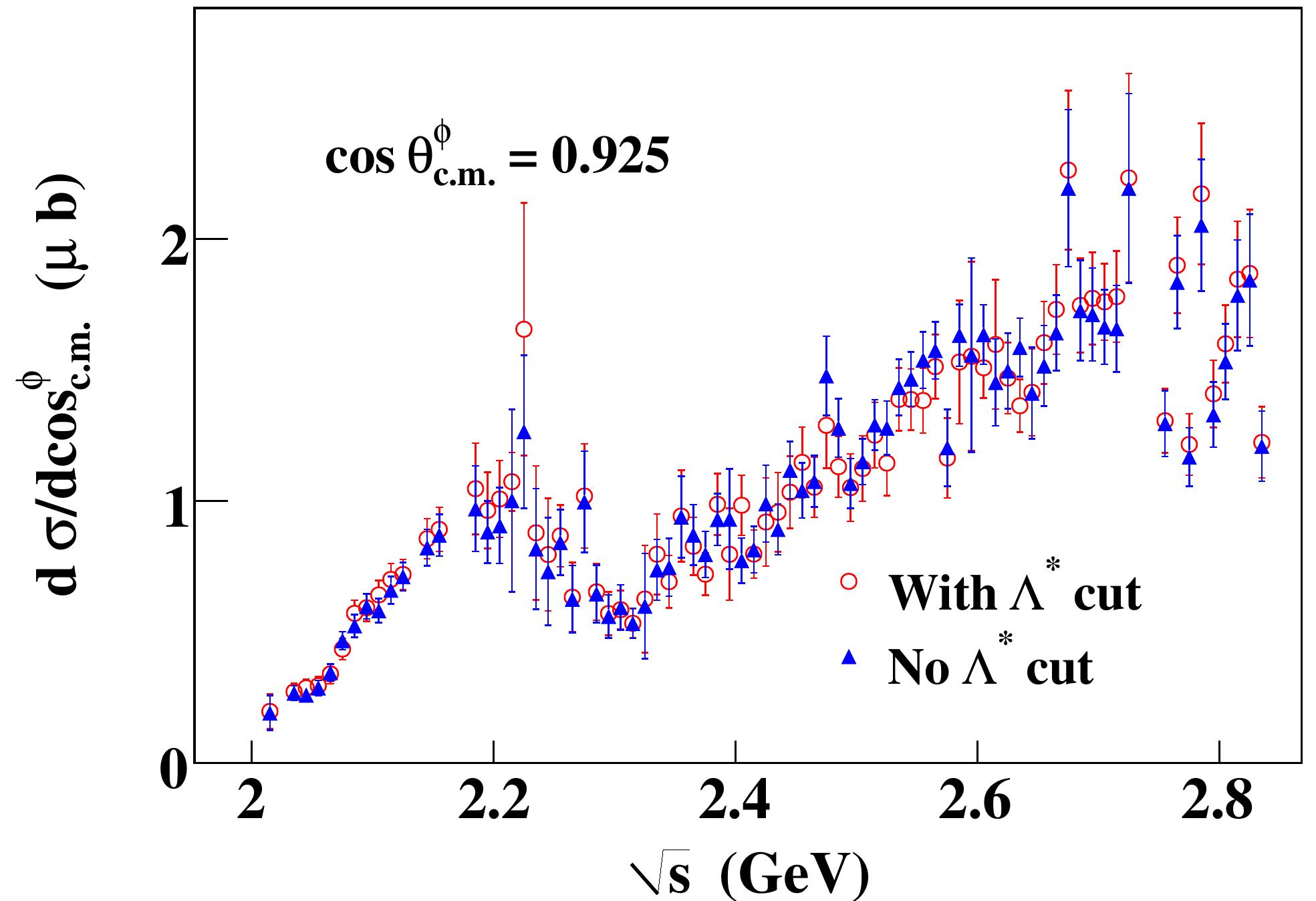} 
\caption[]{\label{fig:phi_lstar_charged2tr_dcs}
  (Color online) $\frac{d\sigma}{d\cmangle}$ ($\mu$b) {\em vs.} $\sqrt{s}$ (GeV) for the charged-mode topology, with or without the $\Lambda^\ast$ cuts from Sec.~\ref{sec:phi_lambda1520_interference}.}
\end{figure}

\begin{figure}
  \centering
  \includegraphics[width=3.4in]{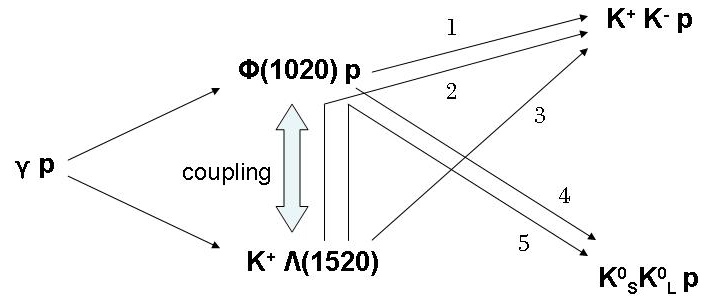}\\ 
\caption[]{\label{fig:phi_lam1520_coupling}
  (Color online) If $\phi$-$\Lambda(1520)$ coupling is allowed, the charged-mode case has three ``paths'' (enumerated as 1, 2 and 3) to its final-state, while the neutral-mode has two ``paths'' (enumerated as 4 and 5) to its final state.}
\end{figure}

As noted earlier in Sec.~\ref{sec:phi_lambda1520_interference}, for $\sqrt{s}$ between 2 and 2.2~GeV, the $\phi p$ and $K^+ \Lambda(1520)$ channels can kinematically overlap in phase space if they have the same $K^+$, $K^-$ and proton final-state particle configuration. For the $\phi$ channel, this corresponds to the charged-mode topology. Therefore any effect of the $K^+ \Lambda(1520)$ channel on $\phi p$ photoproduction might be expected to be enhanced for the charged-mode. Fig.~\ref{fig:slope_compare_charged_neutral} shows the extracted slope parameter $B_\phi$. Above $\sqrt{s} \approx 2.3$~GeV, pure diffraction sets in. However, below $\sqrt{s} \approx 2.2$~GeV, the production mechanism is no longer that of a simple diffractive Pomeron exchange. The slopes extracted from the two modes also show some slight difference here. Similarly, Fig.~\ref{fig:rho00_compare_charged_neutral} shows a ``structure'' around $\sqrt{s} \approx 2.2$~GeV in $\rho^0_{00}$ (Helicity frame). The ``structure'' is also noticeably enhanced for the charged-mode. Fig.~\ref{fig:phi_lstar_charged2tr_dcs} shows a comparison between results including or excluding the hard $\Lambda^\ast$ cuts from Sec.~\ref{sec:phi_lambda1520_interference}. No significant deviation between the two set of results are found in the $\sqrt{s} \approx 2.2$~GeV region.

It is therefore possible that there are two separate phenomena occurring here. First, the $\phi$-$\Lambda(1520)$ re-scatters~\cite{ozaki,ryu} due to kinematic overlap in phase-space and this should affect the neutral-mode as well. Second, there is an interference effect between the $K^+ \Lambda(1520)$ ($pK^-$ mode) and $\phi p$ (charged-mode) when the final states are the same. Therefore, we explicitly distinguish between the terms ``interference'' and ``overlap'', though it is possible that the two phenomena mix in some fashion. Fig.~\ref{fig:phi_lam1520_coupling} illustrates the effect of the $\Lambda(1520)$ on the two topologies. Consider the process $\gamma p \to X \to \kkb$, where the the $\kkb$ refers to either $K^+K^-$ (charged-mode) or $K^0_SK^0_L$ (neutral-mode), and $X$ refers to a generic intermediate state comprising of $\phi p$ and $K^+\Lambda(1520)$. If $\phi$-$\Lambda(1520)$ coupling is allowed, the charged-mode case has three ``paths'' to the final-state, while the neutral-mode has two ``paths'' and this can account for the mild remnant differences in the two results.

A deeper understanding of any possible rescattering effects will require data on $K^+ \Lambda(1520)$ photoproduction (both cross sections and polarizations), in both $p\overline{K}$ and $\Sigma \pi$ decay modes of the $\Lambda(1520)$. Using the same dataset as the present analysis, the CLAS Collaboration has recently published cross section results on the $\Sigma \pi$~\cite{kei-thesis,kei_l1520}, and data on the $p\overline{K}$ mode are anticipated as well.

\begin{figure*}
\begin{center}
\subfigure[]{
{\includegraphics[width=2.1in]{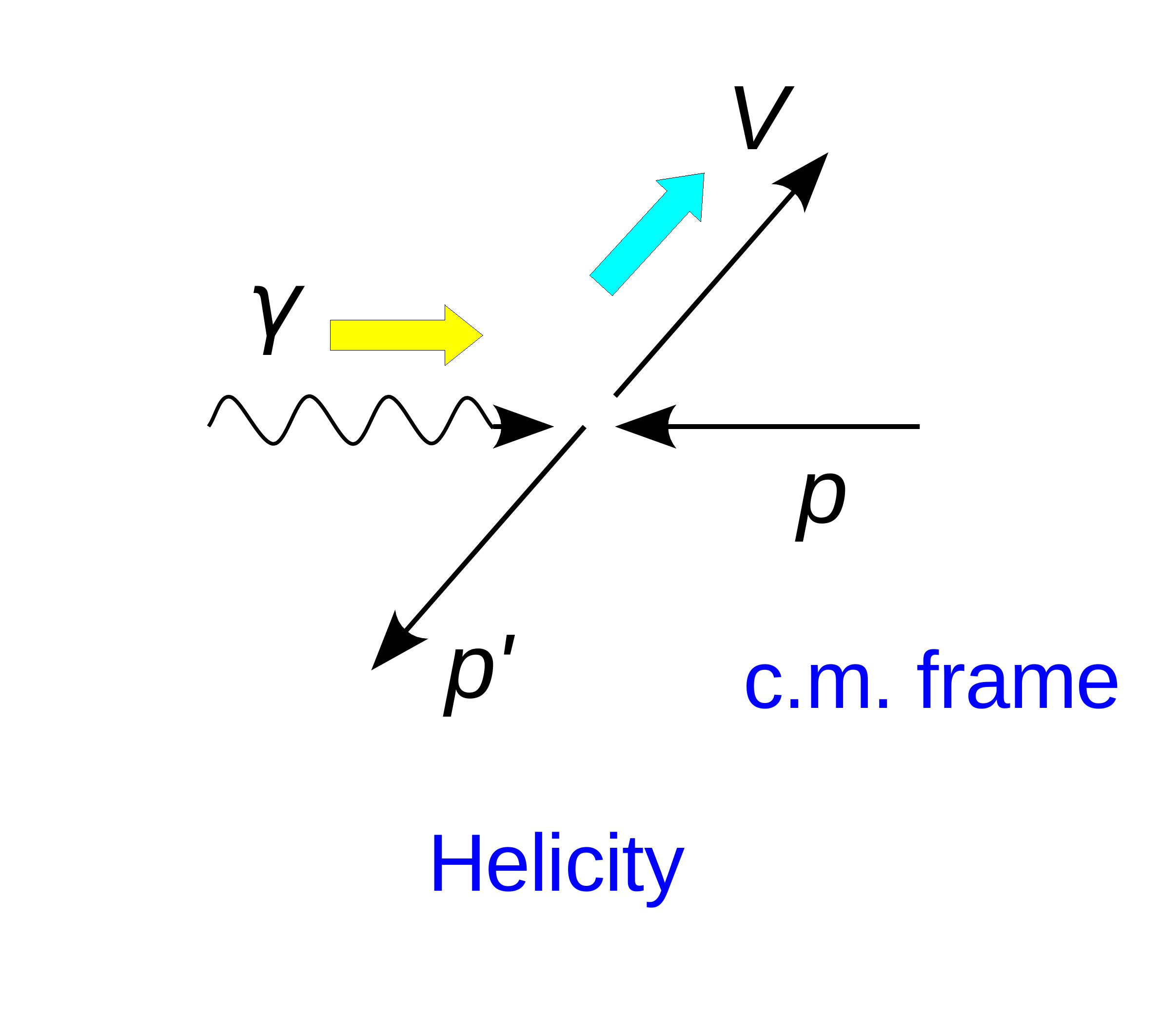}} 
}
\subfigure[]{
{\includegraphics[width=2.1in]{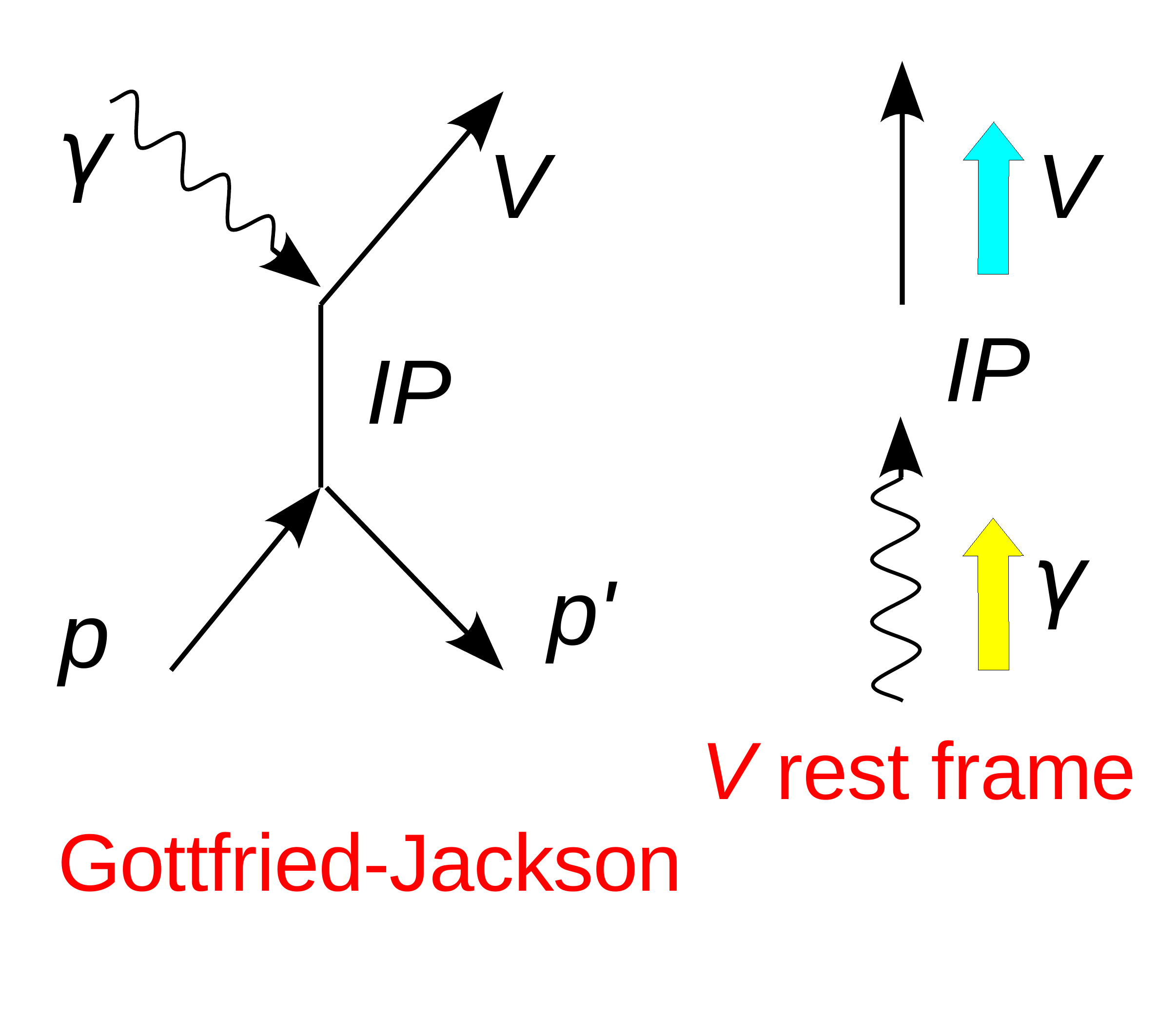}} 
}
\subfigure[]{
{\includegraphics[width=2.0in,angle=90]{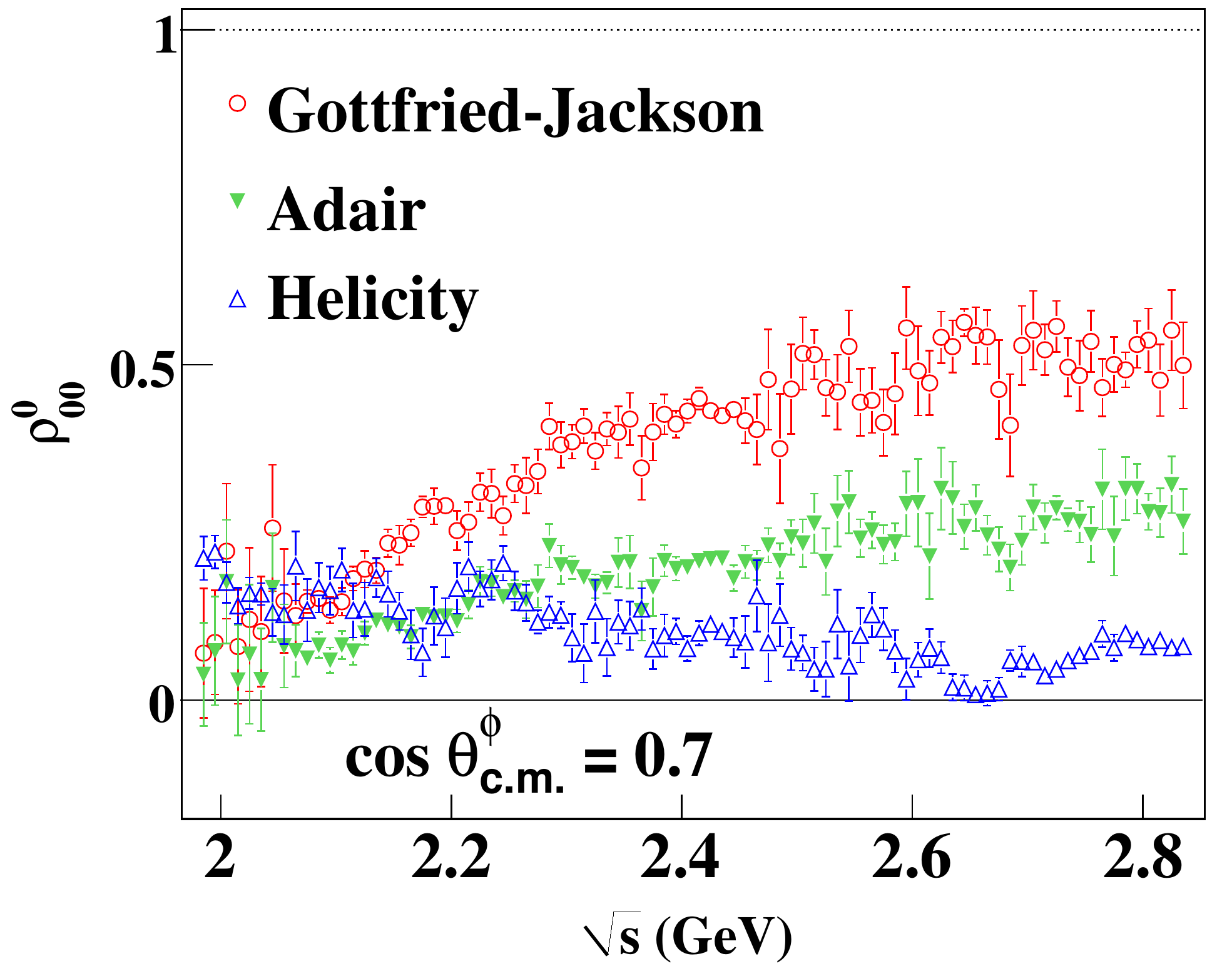}} 
}
\caption[]{(Color online) Helicity conservation in the process $\gamma p \to V p'$, where $V \in \{\rho, \omega, \phi, J/\psi, ...\}$ is a generic vector meson: (a) $s$-channel (SCHC in Helicity frame) (b) $t$-channel (TCHC in the Gottfried-Jackson frame). If the $\Pom$ couples like a $0^+$ object in (b), one would expect TCHC to hold. The $V = \phi$ data in (c) exhibits strong deviation from TCHC since $\rho^0_{00} \neq 0$, implying non-zero helicity flips. The filled arrows in (a) and (b) depict the spins of the incoming and outgoing vector particles.}
\label{fig:disc_tchc_schc}
\end{center}
\end{figure*}

\subsection{Helicity non-conservation and $\rho^0_{00}$}
\label{sec:disc_tchc_schc}

If we ignore the spin-indices of the target- and outgoing-proton (whose polarizations we do not measure) in Eq.~\ref{eqn:sdme_adair}, the definition of $\rho^0_{00}$ becomes
\begin{equation}
\rho^0_{00} \propto |\mathcal{M}_{\lambda_\phi=0, \lambda_\gamma = 1}|^2 + |\mathcal{M}_{\lambda_\phi=0, \lambda_\gamma = -1}|^2.
\end{equation}
Therefore, a non-zero value of $\rho^0_{00}$ is a direct measure of helicity flips between the incoming and outgoing vector particles. Although we have presented most of our SDME results in the Adair frame in Sec.~\ref{sec:results}B, these can easily be converted into the Helicity and Gottfried-Jackson frames by applying Wigner rotations, as described in Sec.~\ref{sec:wigner_rotations}. Doing so, one finds that $\rho^0_{00}$ is distinctly non-zero at all kinematics, in all the three reference frames. For a long time, it was believed that diffractive vector meson photoproduction proceeds via helicity conservation in the $s$-channel~\cite{gilman,donnachie_book}. That is, the $\rho^0$ elements are very small in the Helicity frame. It is indeed puzzling as to why a $t$-channel process (Pomeron exchange) should conserve helicity in the $s$-channel. In Ref.~\cite{gilman}, Gilman {\em et al.} gave some phenomenological arguments for SCHC. The problem boils down to how the Pomeron ($\Pom$) couples (note that Regge theory only gives the overall energy behavior). In the Donnachie-Landshoff (DL) model, the Pomeron couples to partons via a $C = +1$ isoscalar-photon-like $\gamma^\mu$ coupling~\cite{donnachie_book,donnachie_landshoff}. At very high energies where the parton masses can be neglected, the left- and right-handed sectors remain decoupled during a $\gamma^\mu$ coupling and no helicity flips occur in the $s$-channel, as shown in Fig.~\ref{fig:disc_tchc_schc}a. However, the DL model is a phenomenological model after all, and there are no fundamental reasons to expect either SCHC or TCHC. In fact, naively, one would assign a $0^{++}$ $t$-channel exchange-like behavior to the Pomeron (which can only exchange the quantum numbers of the vacuum), and this in turn would lead to TCHC, as Fig.~\ref{fig:disc_tchc_schc}b. Others authors~\cite{titov_lee} have postulated a $2^{++}$ tensor-like coupling as well.

The earlier CLAS $\omega$ results~\cite{omega_prc} already corroborated violation of SCHC for the $\omega$ channel and Fig.~\ref{fig:disc_tchc_schc}c shows that $\rho^0_{00}$ is non-zero in all three frames (Adair, Gottfried-Jackson and Helicity) for the $\phi$ even at forward-angles where soft-diffractive processes are generally expected to be dominant. We hope that future partial wave analyses on these new data will shed light on the Pomeron amplitude.

\section{Summary}

We have presented the first extensive data for the $\phi$ vector meson photoproduction covering both the charged and neutral modes of the $\phi \to \kkb$ decay. The high statistics, wide kinematic coverage and fine energy binning of these results give us a detailed picture of the differential cross sections and $\rho^0$ SDME observables. Access to the neutral-mode results will help understand the physics behind the 2.2~GeV forward-angle ``bump'' structure seen in the differential cross sections and any possible coupling between the $\phi p$ and $K^+ \Lambda(1520)$ channels. Our high-precision SDME data shows that both helicity conservation between the incoming photon and outgoing $\phi$ is broken in both the $t$- and $s$-channels. Electronic versions of the numerical data can be obtained from Ref.~\cite{clas_db}.

A very important aspect of this work has been to ensure that systematic issues that were under very little control in previous analyses due to statistical limitations, have been carefully dealt with. In particular, this pertains to a detailed study of the signal-background separation procedure, use of kinematic fitting and data-driven acceptance calculations. We also note that any further theory model fits to these data should incorporate both the charged- and neutral-mode results as a single dataset, and not as independent analyses, since they were not processed blind to each other. In particular, any point-by-point difference between the two sets of results should be taken as an additional systematic uncertainty.

There is an enormous amount of physics information in these data, in conjunction with the $\omega$~\cite{omega_prc} results published previously. With a wide angular coverage, these latest CLAS results should lead to a better understanding of the transition between the soft and hard Pomeron exchanges.

\begin{acknowledgments}
The authors thank the staff and administration of the Thomas Jefferson National Accelerator Facility who made this experiment possible. B.~Dey thanks Bill Dunwoodie for helpful discussions on the $\phi$ lineshape. This work was supported in part by the U.S. Department of Energy (under grant No. DE-FG02-87ER40315); the National Science Foundation; the Italian Istituto Nazionale di Fisica Nucleare; the French Centre National de la Recherche Scientifique; the French Commissariat \`{a} l'Energie Atomique; the U.K. Research Council, S.T.F.C.; the National Research Foundation of Korea; and the Chilean CONICYT. The Southeastern Universities Research Association (SURA) operated Jefferson Lab under United States DOE contract DE-AC05-84ER40150 during this work.
\end{acknowledgments}


\end{document}